\documentclass[fleqn,usenatbib]{mnras}

\usepackage{newtxtext,newtxmath}

\usepackage[T1]{fontenc}

\DeclareRobustCommand{\VAN}[3]{#2}
\let\VANthebibliography\thebibliography
\def\thebibliography{\DeclareRobustCommand{\VAN}[3]{##3}\VANthebibliography}

\usepackage{graphicx}	
\usepackage{amsmath}	
\usepackage[dvipsnames]{xcolor}

\usepackage[normalem]{ulem}

\newcommand{\Msun}{\, \mathrm{M}_{\odot}}
\newcommand{\LCDM}{$\Lambda$CDM}

\title[Modelling baryon expulsion from haloes]{Modelling the expulsion of baryons from haloes: the role of feedback and of the cosmological constant}

\author[O. Veenema et al.]{
Oscar Veenema,$^{1, 2}$\thanks{E-mail: oscar.veenema@physics.ox.ac.uk}
Daniele Sorini,$^{2}$
Sownak Bose$^{2}$
\\
$^{1}$Department of Physics, University of Oxford, Keble Road, Oxford, OX1 3RH, UK\\
$^{2}$Institute for Computational Cosmology, Department of Physics, Durham University, South Road, Durham DH1 3LE, UK
}

\date{Accepted XXX. Received YYY; in original form ZZZ}

\pubyear{\the\year{}}

\begin{document}
\label{firstpage}
\pagerange{\pageref{firstpage}--\pageref{lastpage}}
\maketitle

\begin{abstract}
The extent to which galactic-scale astrophysical processes conspire with the underlying cosmological model to expel baryons from haloes remains a central question in galaxy formation. We present an analytical model for the gas distribution within and beyond haloes, based on the balance between gravitational collapse, hydrostatic pressure, and cosmic expansion. Our model predicts the halo-centric distance enclosing a baryon mass fraction equal to the cosmic value $f_{\rm b} = \Omega_{\rm b}/\Omega_{\rm m}$ (`closure radius') in an arbitrary \LCDM\ cosmology. We compare the predictions with the results of six variants of the EAGLE cosmological, hydrodynamical simulation, encompassing values of the cosmological constant ranging from 0 to 100 times its observed value in our Universe, $\Lambda_0$. Despite its simplicity, our model exhibits excellent agreement with the simulations for haloes with mass \smash{$M_{\rm 200c} > 10^{11} \Msun$} in the redshift range $0<z<3$, suggesting that it captures the key astrophysical processes and highlighting its robustness to the cosmological parameters. Thus, it provides the first physical explanation for the empirical closure radius--halo mass relation previously observed in simulations. Furthermore, we find that dark energy plays a non-negligible role in baryon evacuation: the simulations reveal that in the fiducial cosmological model, the closure radius at $z<2$ is $\sim 30\%$ larger than in an Einstein-de Sitter universe. In cosmologies with $\Lambda \geq 10 \Lambda_0$, dark energy emerges as the dominant factor in this process -- suggesting that, as our Universe transitions towards $\Lambda$-domination, dark energy eventually becomes the primary driver of baryon evacuation from massive haloes.
\end{abstract}




\section{Introduction}

Despite recent tensions \citep{hu2023hubble, bargiacchi2023tensions, adil2024s}, the \LCDM\ model is still the standard cosmological paradigm. Within this framework, dark matter (DM) haloes arise from hierarchical structure formation driven by gravitational collapse and subsequent mergers \citep{Lacey_1994, article}. This process is well understood thanks to successful early analytical models \citep{1991ApJ...379...52W}, later validated by numerical simulations. However, an accurate description of the role of baryons in shaping galaxy formation and large-scale structure remains a primary challenge.

Galaxies form within their host DM haloes through the accretion of gas, which subsequently forms stars after undergoing several cooling processes. This essential picture constitutes the basis of initial analytical models of star formation \citep{Madau1996, Ellis1997, Hernquist_2003, dave2012analytic}. However, the physics is more complex than that. As star formation proceeds, stellar-driven outflows such as radiative winds and supernovae explosions become more important. Similarly, active galactic nuclei (AGN) at the centre of massive galaxies eject radiative winds and jets. These phenomena, collectively labelled `feedback processes', heat up surrounding gas, and may expel some outside the galaxy into the circumgalactic or even intergalactic medium (CGM and IGM, respectively). 

Due to the complexity and multi-scale nature of feedback processes, it is challenging to accurately describe their action on cosmic star formation and large-scale structure from first principles. Often, theoretical work aims at capturing their average effects on the baryons, rather than focussing on the detailed physics. For instance, analytical models of cosmic star formation typically rely on empirical relationships or idealised physical considerations for feedback processes. Similarly, semi-analytical models may include analytical recipes for baryonic-driven physics, imposed on top of numerical simulations \citep{monaco2014semi}. Fully hydrodynamical cosmological simulations implement feedback processes through various numerical prescriptions \citep{dale2015modelling}, which vary from code to code \citep{chisari2019modelling}. It is therefore paramount to investigate the effect of feedback on as many observables as possible, to constrain our models and to improve our understanding of the phenomena \citep{Bigwood:2024, LaPosta:2024, wayland2025calibrating}.

A principal observable that can provide us with great insight on feedback is the baryon mass fraction in haloes, $f_{\text{b-halo}}$. In galaxy clusters, this is compatible with the cosmic baryon mass fraction $f_{\text{b-cosmic}}$ = $\Omega_{\text{b}} / \Omega_{\text{m}}$, where $\Omega_{\text{b}}$ and $\Omega_{\text{m}}$ are the baryonic and total matter density parameters of the Universe, respectively. However, haloes below a mass scale of order $ 10^{14} \mathrm{M}_{\odot}$ typically enclose a lower baryon mass fraction, which is correlated with the total halo mass  \citep{akino2022hsc}. The discrepancy between the baryon density inferred from early-Universe measurements, such as Big Bang nucleosynthesis (BBN) and the cosmic microwave background (CMB), and the baryonic content directly detected in the local Universe became known as the `missing baryon problem' \citep{Cen_1999, 2010ApJ...708L..14M}. Simulations suggested that a large fraction of these missing baryons reside in diffuse, warm--hot gas phases that are difficult to detect observationally, particularly the warm--hot intergalactic medium (WHIM), helping to explain the apparent deficit \citep{Cen_1999, 2007ARA&A..45..221B}. It is now widely accepted that the baryons within haloes are truly `missing', and not simply undetected, since numerical simulations showed that feedback processes can push baryons far beyond typical halo boundaries \citep{Haider_2016}. Crucially, advancements in measurement techniques have progressively enabled a complete census of baryons across different environments, from within haloes to the IGM. Examples include measurement of hot gas through X-ray emission \citep{allen2002cosmological, simionescu2011baryons}, detection of the warm-hot intergalactic medium X-ray absorption \citep{Yao_2012}, mapping hot gas within clusters via the Sunyaev-Zel'dovich Effect \citep{Grego_2001} (which can also be combined with studies of galaxy-galaxy lensing, e.g. \citealt{Sunseri2026}), analysis of the Lyman-$\alpha$ forest in intergalactic gas absorption spectra \citep{2012ApJ...759...23S}, and, most recently, probing the low-redshift baryon content in the IGM via the dispersion measure of Fast Radio Bursts (FRBs) signals \citep{Macquart_2020, Reischke:2023, Khrykin_2024, Wayland2026}. Furthermore, the deployment of new and improved science instruments such as the JWST \citep{rigby2023science}, coupled with high-redshift galaxy surveys \citep{glazebrook2024massive, Pintos-Castro_2019, Austin_2023}, has provided data on galaxy composition at unprecedentedly early epochs, up to 13.5 Gyr ago \citep{naidu2025cosmic}.

Alongside such a wealth of observations, a large body of theoretical work has provided insight into the evolution of the baryon distribution under the activation of different feedback mechanisms within haloes. In the context of the NIHAO project, \cite{Tollet_2019} demonstrated that galactic winds inhibit the accretion of gas from cosmic filaments up to a distance of six virial radii, thereby decreasing the mass of galaxies by approximately a factor of $2$ to $4$. In the EAGLE simulation \citep{Schaye}, a critical role in this respect is played by AGN feedback in haloes with total mass above $\sim 10^{12.5} \Msun $ \citep{Davies_2019}, as the increased energy transfer from more massive black holes onto the surrounding gas elements can push them beyond the virial radius \citep[see also][]{Davies_2020, Davies_2021, Davies_2022}, consistent with recent FLAMINGO results showing that black hole growth strongly regulates halo gas fractions across galaxy groups and clusters \citep{Costello2026}. In particular, more than half of the baryonic mass is lost by haloes hosting Milky-Way-size galaxies due to feedback processes -- analogous results were also observed in the IllustrisTNG \citep{IllustrisTNG2018} and Magneticum simulations \citep[e.g.][]{Dolag_2016,Lim_2021}. The Simba suite of simulations \citep{Dav__2019} includes a particularly strong AGN-jet feedback prescription that manages to evacuate up to 80\% of the baryonic mass of haloes by $z=0$ \citep{Appleby_2021}, displacing baryons by even 15 Mpc \citep{Borrow_2020}.

In a systematic study of the impact of the different feedback models in the Simba simulations, \cite{Sorini_2022} showed that stellar winds primarily drive baryon evacuation at early ($z > 2$) times and in smaller haloes, while AGN jets become dominant later on ($z < 2$) and in larger haloes. To quantify the impact of each feedback mode, \cite{Sorini_2022} considered the distance enclosing 90\% of the cosmic baryon mass fraction from the centre of haloes, and studied how this quantity varies with the halo mass and redshift. In the flagship simulation including all types of stellar and AGN feedback, this length scale can be as large as 10-20 virial radii for haloes with total mass $\sim 10^{12.5} \Msun $, decaying both for smaller and larger masses. \cite{Ayromlou_2023} independently confirmed these results and extended this analysis to the IllustrisTNG and EAGLE simulations as well. They found that each simulation predicts a different dependence of the aforementioned length scale (which, upon a slightly different definition, they named the `closure radius') on the halo baryon mass fraction and redshift and provided an analytical fitting formula for this empirical correlation. A similar approach was undertaken by \cite{Angelinelli_2022}, who focused on the impact of feedback on the distance where groups and clusters contain their cosmic share of baryons in the Magneticum simulation.

However, the numerical analyses mentioned above are limited in at least three aspects. First, it is inherently difficult to interpret the results of hydrodynamical cosmological simulations that include several, intertwined physical processes. As these, in turn, rely on a set of code-specific numerical parameters, it becomes even more challenging to make comparisons across the results of different simulations. Second, the relationship between closure radius and baryonic or total halo mass is not yet explained. While extremely valuable on a practical level, the analytical fit proposed by \cite{Ayromlou_2023} is purely empirical. Third, all aforementioned works assume a fixed set of cosmological parameters, leaving the question of the cosmological dependence of the closure radius unaddressed.

The interplay between cosmology and baryonic physics has received increasing attention in recent years, particularly in studies aimed at modelling large-scale structure and cosmological observables. Examples include large hydrodynamical simulation suites such as BAHAMAS \citep{mccarthy2016bahamas}, Magneticum \citep{dolag2017distribution}, FLAMINGO \citep{schaye2023flamingo}, and CAMELS \citep{villaescusa2023camels}, as well as semi-analytic approaches such as the halo model \citep{mead2016accurate} and "baryonification" \citep{schneider2015new, arico2021simultaneous, schneider2025baryonification}. However, galaxy-formation simulations are commonly performed within a fixed cosmological framework, typically in a neighbourhood of the \LCDM\ parameters observed by Planck \citep{aghanim2020planck}. Feedback prescriptions are then calibrated to reproduce selected observables, such as the galaxy stellar mass function. While this strategy has proven highly successful, it raises the question of how robust the resulting baryon distributions remain under variations of the underlying cosmology, especially when considering large deviations from the observed \LCDM~parameters. In particular, the dependence of baryon redistribution beyond halo boundaries, and therefore the closure radius, on cosmological parameters remains comparatively less explored. Addressing this question is important for understanding the extent to which feedback models capture universal aspects of galaxy formation, rather than features that may be specific to the cosmological context in which they were calibrated.

It is in this spirit, aside from the interesting implications for anthropic reasoning \citep{Carter_1974, weinberg1987anthropic}, that \cite{BK_2022} followed the pioneering work by \cite{Nagamine_2004} and investigated the abundance and thermal state of gas in the IGM, and the star formation history, in a suite of ENZO simulations where the cosmological constant, $\Lambda$, is varied between 0 and 100 times the observed value, $\Lambda_0$. Unsurprisingly, a sharp increase in the cosmological constant was found to limit the amount of matter able to collapse into haloes, hence increasing the amount of baryons in the IGM. These results echoed the findings by \cite{Barnes_2018}, who considered a suite of EAGLE simulations with fixed feedback prescriptions, but varying cosmological constant in the range $0 \leq \Lambda / \Lambda_0 \leq 300$. They showed that for high values of $\Lambda$, haloes tend to become more concentrated, with `island galaxies' decoupling from a very rarefied IGM due to the early onset of the accelerated expansion of the universe that slows down gas accretion. This effect was explained from first principles by \cite{Sorini_2024_cosmo}, who applied an earlier analytical model for the baryonic--total mass relationship and the star formation history \citep{Sorini_2021} on \LCDM\ universes with $0 \leq \Lambda / \Lambda_0 \leq 10^5$. All these works thus highlight the conceptual importance of the effect of the cosmological constant on the distribution of baryons in the universe, although practically it appears to be impactful at values of $\Lambda$ well above observational constraints.

To summarise, our understanding of the closure radius would benefit from deeper physical insight on its correlation with halo mass over redshift, and from an analysis of its cosmological dependence. In this paper, we address this research gap. We present an analytical model that describes the time-dependent evacuation of baryons from haloes due to galactic outflows and the cosmological constant, offering a predictive and physically grounded perspective. The details of the model are expounded in Section~\ref{sec:model}, where we also derive a physically motivated relationship between the closure radius and the baryon mass fraction in haloes. In Section~\ref{sec:simulations}, we briefly discuss the main features of a suite of EAGLE simulations encompassing values of the cosmological constant between 0 and $100 \Lambda_0$. These are a subset of the simulations first utilised by \cite{Barnes_2018}. In Section~\ref{sec:results}, we show that the numerical results validate the predictions of our analytical model, hence lending support to its physical soundness. In the same section, we also compare our predictions with the empirical fitting formula provided by \cite{Ayromlou_2023} and find excellent agreement for $\Lambda = \Lambda_0$, underscoring that our model provides a physical explanation for the empirical correlation found in simulations. For $\Lambda \geq \Lambda_0$, \cite{Ayromlou_2023} model progressively deviates from our predictions, which are nonetheless in better agreement with the simulations. This suggests that our model constitutes a generalisation of \cite{Ayromlou_2023} formula. Indeed, in Section~\ref{sec:previous_work}, we demonstrate that under appropriate limits, our analytical model for the closure radius -- baryonic mass fraction relationship reduces to the same functional form found by \cite{Ayromlou_2023}, allowing for a fair comparison between the two approaches. In the same section, we discuss our findings in the broader context of cosmological research, and then present our conclusions and future perspectives in Section~\ref{sec:conclusions}.

\section{Modelling baryons outside haloes}
\label{sec:model}

In this section, we present our analytical model for the distribution of baryons within and outside haloes. Section~\ref{sec:cosmo} describes the cosmological models that we consider, while Section~\ref{sec:definitions} defines key length scales that we will adopt throughout this work. Our formalism to derive the gas density profile around haloes is then presented in Section~\ref{sec:newmodel}. This allows us to obtain a physically motivated analytic expression for the closure radius (Section~\ref{sec:closureradiusmodel}). 

\subsection{Cosmological model}
\label{sec:cosmo}

We consider flat \LCDM\ models with a cosmological constant that varies in the range $0\leq \Lambda / \Lambda_0 \leq 100$. Clearly, the correspondence between the Hubble constant and the scale factor varies greatly in this range, due to the dependence on the cosmological parameters. At sufficiently late times (i.e., when the contribution to the energy density of the universe from radiation becomes negligible), the evolution of the Hubble parameter with the scale factor $a$ is always given by:
\begin{equation}
H = H_0 \sqrt{\Omega_{\text{m}} a^{-3} + \Omega_\Lambda} \, ,
\label{eq:friedmann}
\end{equation}
with the usual definitions of the cosmological parameters.

\begin{figure}
	\includegraphics[width=\columnwidth]{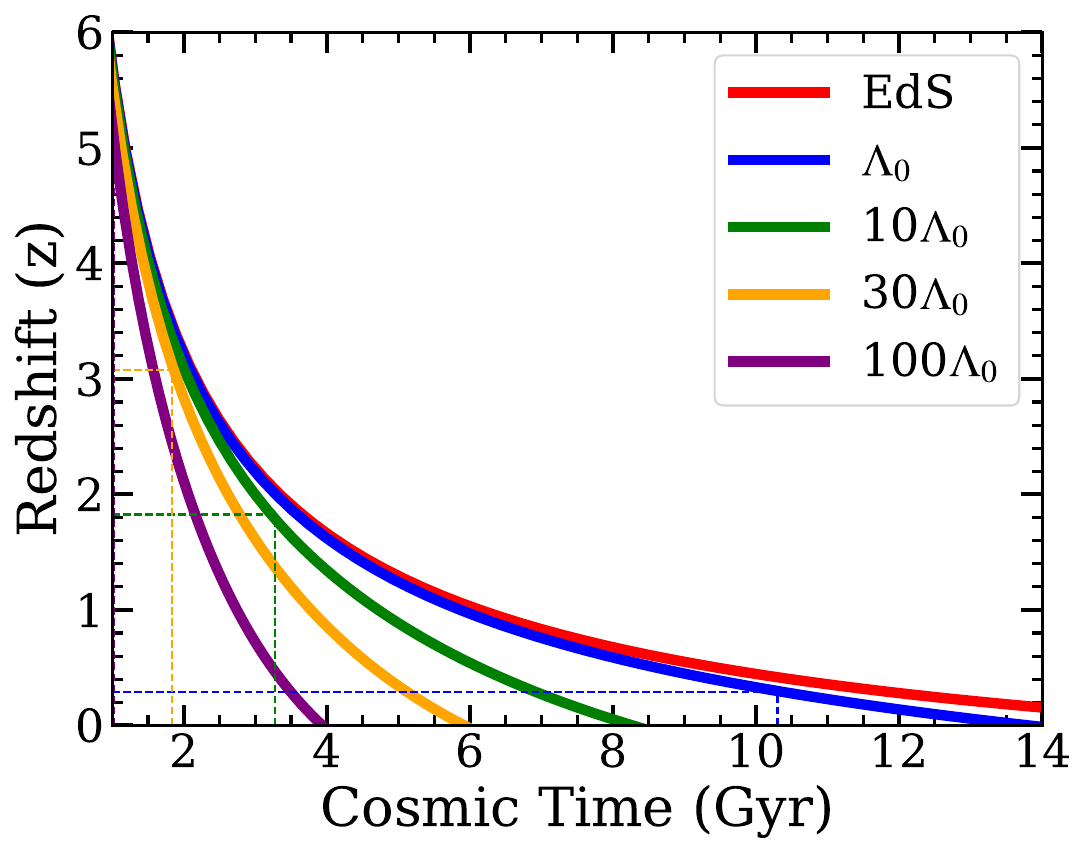}
    \caption{Plot of equation~(\ref{eq:timeevolution}) showing the evolution of cosmological redshift over elapsed cosmic time, highlighting differences in time evolution among universes with varying dark energy. Five cosmologies are shown, including an Einstein-de Sitter universe (EdS) with zero dark energy, a universe with dark energy density corresponding to what is measured in our Universe, $\Lambda_0$, and cosmologies where this density is scaled by factors of 10, 30, and 100. Dashed lines mark the epoch of matter-dark energy equality in each model universe. It is important to bear in mind the different mapping between cosmic time and redshift when comparing results across different cosmological models.}
    \label{fig:timeevolution}
\end{figure}

It follows that the correspondence between the scale factor (and thus redshift) with time is also cosmology dependent. In universes with $\Lambda>0$, equation~\eqref{eq:friedmann} implies that:
\begin{equation}
a(t) = \left( \frac{\Omega_{\text{m}}}{\Omega_{\Lambda}} \right)^{\frac{1}{3}} \sinh \left( \frac{3}{2} \sqrt{\Omega_{\Lambda}} H_0 t \right)^{\frac{2}{3}} \, ,
\label{eq:timeevolution}
\end{equation}
whereas for an EdS universe ($\Lambda=0$) the same equation results in the well known power-law scaling $a(t) \propto t^{2/3}$. In any case, the same cosmic time in different universes corresponds to different values of redshift, and vice versa. This can be clearly seen in Fig.~\ref{fig:timeevolution}, where we plot the redshift -- cosmic time relationship for different values of $\Lambda$ = $3H^{2}\Omega_{\Lambda}/c^2$, where $c$ is the speed of light. Since we are considering only standard \LCDM\ models, we will refer to $\Lambda$ as `cosmological constant' and `dark energy' almost interchangeably throughout this work.

\begin{figure}
	\includegraphics[width=\columnwidth]{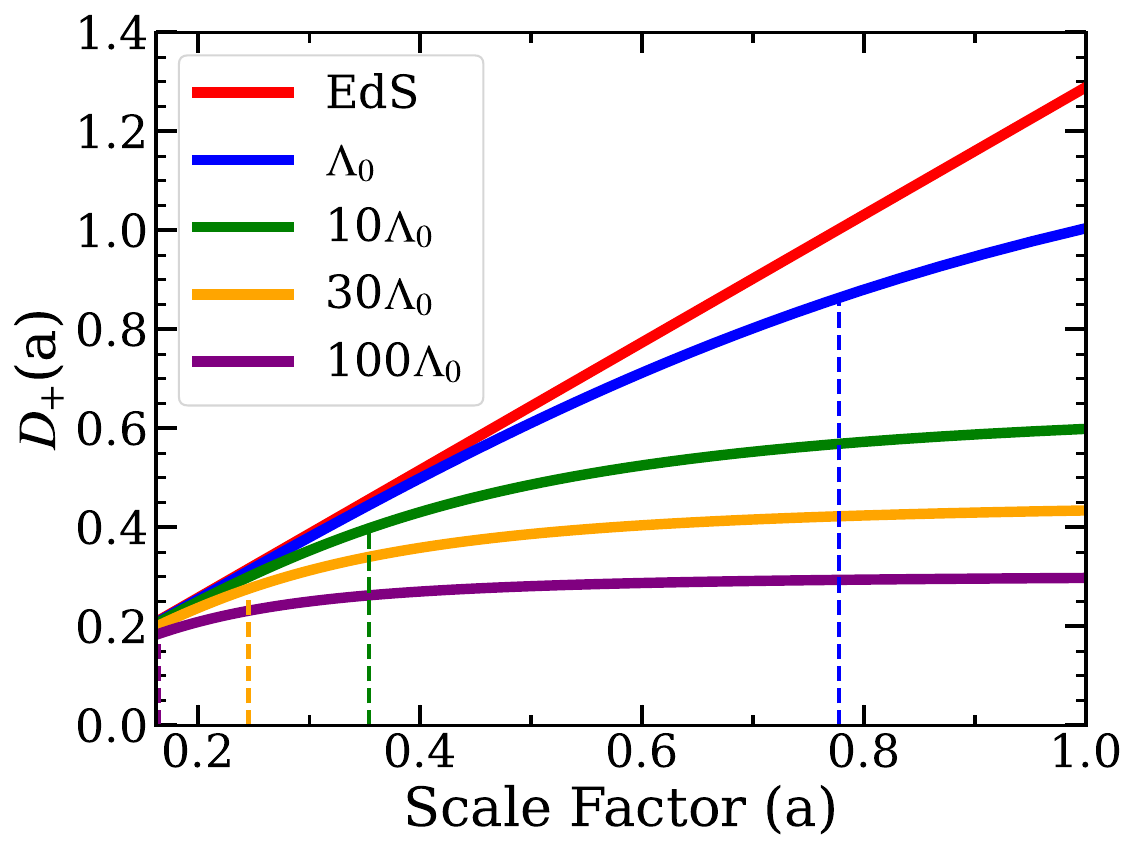}
    \caption{Analytical linear-theory growing-mode evolution for universes with different dark energy contents, showing how the growth of initial matter-density perturbations varies. Dotted lines mark the epoch of matter-dark energy equality in each model universe. In cosmologies with $\Lambda >0$, the linear growth factor asymptotically reaches a plateau as the scale factor increases. In an EdS universe, it grows indefinitely.}
    \label{fig:lineargrowth}
\end{figure}

Dark energy fundamentally affects large-scale structure. Higher dark energy densities result in earlier $\Lambda$-domination and subsequent accelerated expansion. This slows down the growth rate of overdensities sooner, leading to `freezing out' of large-scale structure growth. In a universe with positive $\Lambda$, this phenomenon prevents dark matter haloes from accreting mass indefinitely \citep{Hernquist_2003}. The time evolution of dark matter overdensities (with fractional overdensity, $\delta$) in these universes is governed by the following differential equation from linear growth theory \citep{1980lssu.book.....P}:
\begin{equation}
\frac{d^2 \delta}{dt^2} +2 \frac{\dot{a}}{a} \frac{d\delta}{dt} - 4 \pi G \Bar{\rho}(t) \delta = 0 \, ,
\label{eq:masterequation}
\end{equation}
where $\Bar{\rho}(t)$ is the average matter density of the universe. The growing mode solution of equation~(\ref{eq:masterequation}) is:
\begin{equation}
\delta(t) = D_+ (t) \delta(t_0).
\label{eq:growingmode}
\end{equation}
The linear growth factor, $D_+ (t)$, is then given by the following integral:
\begin{equation}
D_+ (t) \propto \frac{\dot{a}}{a} \int_{0}^{t} \frac{dt'}{\dot{a}^2 (t')} ,
\label{eq:growthfactor}
\end{equation}
which ultimately depends on the cosmology of the model universe considered according to equation~(\ref{eq:friedmann}). It is convention to normalise the linear growth factor so that $D_+ (t_0) = 1$ in a standard cosmology (same parameter values as our real Universe), where $t_0 \sim 13.8$ Gyr, is the present time since the Big-Bang. $D_+ (t)$ can also be plotted for model universes with different $\Lambda$ as shown in Fig.~\ref{fig:lineargrowth}.

Fig.~\ref{fig:timeevolution} and equation~(\ref{eq:timeevolution}) provide useful context for comparing results across universes with differing dark energy. In particular, they illustrate how a fixed redshift corresponds to different cosmic times in different cosmologies, which should be kept in mind when comparing evolutionary stages across model universes.
Equations~(\ref{eq:growingmode}) and ~(\ref{eq:growthfactor}), along with Fig.~\ref{fig:lineargrowth}, depict the evolution of the growth of large-scale structure across various epochs in different model universes. This is useful for understanding the difference in the number of haloes expected to form in each cosmology due to the fact that the growth rate of perturbations freezes out more quickly in the presence of a higher dark energy density. By contrast, the linear growth factor in an EdS universe (red line in Fig.~\ref{fig:lineargrowth}) grows indefinitely as time increases, since it does not experience the freezing out of structure formation induced by the cosmological constant. These insights are important for comparison of results from universes with differing cosmologies explored in this work.

\subsection{Definition of key length scales}
\label{sec:definitions}

In our model, we will consider haloes as being spherically symmetric, with their centres corresponding to the minimum of the gravitational potential. We consider the halo radius, $R_{200}$, to be the halo-centric distance where the enclosed total matter density drops to 200 times the critical density of the universe. This is commonly used as a proxy for the `virial radius', though it does not exactly correspond to the equilibrium radius expected from the virial theorem \citep{Mota_2004}. The associated mass, $M_{200}$, is defined as the mass enclosed within $R_{200}$ under the density criterion $\Delta = 200$:
\begin{equation}
M_{200} = \frac{4 \pi \Delta}{3} \rho_{\text{crit}}(z) R_{200}^3 .
\label{eq:virialmass}
\end{equation}

For simplicity, we use the terms `halo radius', `virial radius', and $R_{200}$ interchangeably, and similarly for `halo mass', `virial mass', and $M_{200}$. While we consider $\Delta=200$ throughout this work, we deliberately leave it as a free parameter to make our formalism readily adaptable to different choices (e.g., $\Delta=500$).

\begin{equation}
R_{\text{crit}} = 1.1 \, {\text{Mpc}} \left( \frac{M_{200}}{10^{12} \mathrm{M}_{\odot}} \right)^{\frac{1}{3}} \left( \frac{\Lambda}{\Lambda_0} \right)^{-\frac{1}{3}} .
\label{eq:criticalradius}
\end{equation}

For another useful way to quantify the extent to which baryons are pushed from haloes, we adopt the notion of the `closure radius'. This  is defined for each halo as the radial distance from the gravitational minimum at which the enclosed baryon fraction, $M_{\text{b}}(<r) / M(<r)$, returns to the cosmic baryon fraction, $f_{\text{b-cosmic}}$. Here, $M_{\text{b}} (<r)$ is the enclosed baryonic mass within a sphere of radius $r$, centred on the halo, and $M(<r)$ represents the total mass within the same sphere. Thus, the closure radius, $R_{\text{closure}}$, is formally defined as:
\begin{equation}
M_{\text{b}}(<R_{\text{closure}})/M(<R_{\text{closure}}) = \Omega_{\text{b}}/\Omega_{\text{m}} .
\label{eq:closureradiusdefinition}
\end{equation}
It is useful to think of the closure radius as the distance from a halo out to which all `missing baryons' can be accounted for, making it a valuable concept for quantifying and understanding the extent of halo baryon evacuation due to different physical mechanisms. 

In practice, when estimating the closure radius from numerical simulations, equation~\eqref{eq:closureradiusdefinition} needs to be applied with reasonable flexibility, due to the finite mass resolution. For example, \cite{Sorini_2022} considered the halo-centric distance beyond which the enclosed baryon mass fraction would exceed 90\% of $f_{\rm b-cosmic} = \Omega_{\text{b}}/\Omega_{\text{m}}$, while \cite{Ayromlou_2023} allowed for 5\% tolerance on $f_{\rm b-cosmic}$. Since we are developing an analytical model, we need not worry about these issues for the time being. However, we will need to make analogous considerations when validating our model against simulations (see Section~\ref{sec:simulations}).

\subsection{Radial gas density profile outside haloes}
\label{sec:newmodel}

For our model for the closure radius, we aim to establish the precise influence of gas pressure, dark energy, and astrophysical feedback on the distribution of baryons within and beyond haloes. Due to the complex astrophysics of star formation and feedback processes, we will inevitably need to make simplifying assumptions. At the price of losing some physical realism, this will enable us to obtain a flexible, predictive and physically intuitive model. 

We employ a two-step methodology. First, we need to determine the gas density beyond the virial radius of a halo. Once the gas density profile is established, we can integrate it radially outward until the enclosed baryonic mass matches that of a homogeneous sphere with a density equivalent to the cosmic baryon density. This will yield an equation involving $R_{\text{closure}}$, which establishes the relationship between the closure radius and the baryonic halo mass fraction. 

To derive the gas density profile beyond the virial radius of a halo, we begin by examining the balance of the forces acting on a small parcel of baryonic mass at a radial distance, $r$, from the halo centre. Assuming spherical symmetry, the following equation is easily obtained:
\begin{equation}
- \frac{r^2}{\rho_{\rm gas}(r)} \frac{dP}{dr} + \frac{\Lambda c^2}{3} r^3 = GM(<r) \, ,
\label{eq:forcebalance}
\end{equation}
where $\rho_{\rm gas}(r)$ is the gas density, $P$ is the gas pressure and $M(<r)$ is the total enclosed mass at $r$. The first term on the LHS accounts for hydrodynamic gas pressure (with a negative sign since $P$ decreases with $r$ and so the derivative is negative), the second term accounts for dark energy and the term on the RHS comes from the halo's self gravity.

We note that equation~(\ref{eq:forcebalance}) assumes that pressure support can be described by a single thermal pressure term. In reality, gas in the outskirts of haloes is expected to receive additional support from non-thermal components, including turbulence and bulk gas motions due to inflows and outflows, whose relative contribution generally increases with radius \citep[e.g.][]{Shaw2010}. Neglecting such effects is a simplification of the present model. Inclusion of non-thermal pressure would modify the resulting density profile and could alter the predicted closure radii quantitatively (likely increasing them as higher pressure would shallow the radial baryon density profile). However, previous analytical models of the baryon mass fraction within haloes achieved good agreement with observations with the same simplifying assumption that we are introducing \citep[e.g.][]{Sorini_2021}. Since our aim here is to develop a simple analytic framework that captures the leading-order behaviour of baryon redistribution, we defer a more detailed treatment of non-thermal pressure support to future work.

Equation~(\ref{eq:forcebalance}) also assumes that all baryons exist in the form of gas; such approximation is supported by observations such as \citet{2010ApJ...708L..14M}, which show that only $\sim 5\%$ end up in other forms such as stars and planets. We further assume that the gas is ideal, with a polytropic equation of state:
\begin{equation}
\label{eq:EoS}
    P(r) = w_n \rho_{\rm gas}(r)^{n+1} \, , 
\end{equation}

where $n$ is the polytopic index of the gas and $w_n$ is a proprotionality constant, which is related to the virial temperature of the halo (see Section~\ref{sec:closureradiusmodel}). Substituting the polytropic equation of state into equation~(\ref{eq:forcebalance}) splitting $M(<r)$ into a density integral over a spherical volume out to $r$ gives a new expression which can then be differentiated with respect to (w.r.t) $r$. Further assuming that the gas density traces the total matter density, we obtain the following 2nd order ODE in $\rho_{\rm gas}(r)$:
\begin{equation}
\frac{-w_n (n+1) f_{\rm gas}}{4 \pi G} \left[\frac{2}{r} \rho_{\rm gas}^{n-1} \frac{d\rho_{\rm gas}}{dr} + \frac{d}{dr}\left(\rho_{\rm gas}^{n-1} \frac{d\rho_{\rm gas}}{dr}\right)\right] = \rho_{\rm gas} - \frac{f_{
\rm gas}\Lambda c^2}{4 \pi G}.
\label{eq:generalODE}
\end{equation}

Here $f_{\rm gas}$ is the gas mass fraction of the halo, and the factors of $f_{\rm gas}$ arise from expressing the total mass profile in terms of the gas mass under this assumption, $M_{\rm tot}(<r)=M_{\rm gas}(<r)/f_{\rm gas}$; these factors cancel when seeking power-law solutions in the limit considered below, leaving the resulting normalisation unchanged. Equation~(\ref{eq:generalODE}) is not trivial to solve and an exact analytic solution may or may not exist, however $\rho_{\rm gas}(r)$ can still be found in certain limits. Examining the term on the RHS of equation~(\ref{eq:generalODE}), $\Lambda c^2 / 4 \pi G$ is of order of $\rho_{\text{crit}}$, thus two limits can be considered. Firstly, where $\rho_{\rm gas} \gg \rho_{\text{crit}}$, which occurs when $r - R_{200} \ll R_{200}$ (which is just beyond the virial radius of the halo), or in the EdS universe where $\Lambda = 0$. The other limit is where $\rho_{\rm gas} \ll \rho_{\text{crit}}$ which occurs far from the halo, $r \gg R_{200}$, and only in universes with $\Lambda > 0$. These limits can be solved separately, with the focus here on the second limit, which is most prominent to understanding the impact of $\Lambda$ on the large-scale baryon distribution in our own Universe. The other limit has limited relevance for the main conclusions of this work, and is discussed in Appendix~\ref{app:solutions}.

Assuming then that the 2nd term on the RHS of equation~(\ref{eq:generalODE}) dominates in our Universe far from the halo (where the closure radius is likely to be located), allows us to neglect the $\rho_{\rm gas}$ term on the RHS - making the ODE easily solvable. One solution is a power-law of the form $\rho_{\rm gas}(r) = Ar^{-\eta}$. Power-law density profiles are a common occurrence in analytical models of galaxy formation \citep{10.1046/j.1365-8711.2003.06207.x,Sorini_2021}, and this assumption has been corroborated over a wide range of halo mass and redshift in full hydrodynamical simulations (e.g., \citealt{Sorini_2024}, but see also \citealt{Sorini_2025, Sorini_2026}). This approximation neglects the detailed structure of the outer halo environment, including contributions from correlated neighbouring haloes and filaments. However, it allows us to construct a tractable analytic model that isolates the dominant dependence of baryon redistribution on cosmological parameters. In our analytical framework, the value of $\eta$ is linked to the polytropic index $n$. The exact relationship can be found by substituting the power-law profile into equation~(\ref{eq:generalODE}). In the limit $\rho_{\rm gas} \ll \rho_{\text{crit}}$, one obtains $\eta = -2/n$. $A$ is a constant that can be determined with an analogous strategy.

We note that the power-law density profile is only one particular solution of equation~\eqref{eq:generalODE} in the limit considered. We provide the most general solution in the Appendix~\ref{app:solutions}. Matching it with the solution to equation~\eqref{eq:generalODE} in the opposite regime (i.e., $\rho_{\rm gas} \gg \rho_{\text{crit}}$), provides a more accurate description of the gas density distribution around haloes. However, it also results in a more complicated expression for the closure radius. As we will demonstrate in Section~\ref{sec:newmodelresults}, the predictions for the closure radius arising from the simplest power-law density profile in the $\rho_{\rm gas} \ll \rho_{\text{crit}}$ regime is sufficiently accurate, providing excellent agreement with full hydrodynamical simulations. We shall therefore adhere to this solution hereafter.

\subsection{Modelling the closure radius}
\label{sec:closureradiusmodel}

With the gas density far from the halo now determined, we can evaluate the volume integral, $V$, out to the closure radius. This equals $f_{\text{b-cosmic}}M_{200}$ at the closure radius by definition (equation~\ref{eq:closureradiusdefinition}), giving us the following integral:
\begin{equation}
\int^{r = R_{\text{closure}}}_{r = 0} \rho_{\rm gas}(r) dV = f_{\text{b-cosmic}} M_{200} .
\label{eq:isolatedhalogeneralintegral}
\end{equation}

Next, we split the integral in equation~(\ref{eq:isolatedhalogeneralintegral}) into two terms; the exact baryonic mass of the halo, $f_{\text{b-halo}} M_{200}$, and a second integral beyond $R_{200}$ where the newly derived gas density profile is physically valid. Neglecting the $r - R_{200} \ll R_{200}$ limit of equation~(\ref{eq:generalODE}), equation~(\ref{eq:isolatedhalogeneralintegral}) can be rewritten as:
\begin{equation}
f_{\text{b-halo}} M_{200} + 4\pi A \int^{R_{\text{closure}}}_{R_{200}} r^{\frac{2}{n}+2} dr = f_{\text{b-cosmic}} M_{200} .
\label{eq:generalintegral2}
\end{equation}
The value of $A$, found from substituting the power-law solution of equation~(\ref{eq:generalODE}) in the limit $r \gg R_{200}$ can be shown to be:
\begin{equation}
    A = \left[\frac{\Lambda c^2 n}{6 w_n (n+1)}\right]^{\frac{1}{n}} .
\label{eq:Asolution}
\end{equation}

The integral in equation~(\ref{eq:generalintegral2}) can be easily evaluated and rearranged to give the following expression for the closure radius:
\begin{equation}
    R_{\text{closure}} = \left[\xi(w_n, n) \left(f_{\text{b-cosmic}} - f_{\text{b-halo}}\right)M_{200} + R_{200}^{\frac{2}{n}+3}\right]^{\frac{n}{2+3n}} \, ,
\label{eq:model2}
\end{equation}
where several terms have been collected into $\xi(w_n, n)$:
\begin{equation}
    \xi(w_n, n) = \left( \frac{1}{4\pi}\right) \left( \frac{2}{n}+3\right) \left(\frac{n+1}{n}\right)^{\frac{1}{n}} \left(\frac{6 w_n}{\Lambda c^2}\right)^{\frac{1}{n}} \, .
\label{eq:mu}
\end{equation}

In order to progress beyond equations~\eqref{eq:model2}-\eqref{eq:mu}, one must find an explicit expression for $w_n$. This can be worked out by evaluating the polytropic equation of state (equation~\ref{eq:EoS}) at a fixed distance from the halo centre. For an ideal gas, the pressure is proportional to the gas density and the temperature, which we assume be of the order of the virial temperature. This enables us to focus on the gas density at a reference distance from the halo centre, for example the virial radius, and obtain the following expression for $w_n$:
\begin{equation}
\label{eq:wn2}
    w_n = \frac{k T_{\rm gas}(R_{200})}{\mu \rho_{\rm gas}(R_{200})^n} \, .
\end{equation}

The pivot density $\rho_{\rm gas}(R_{200})$ can be readily obtained once an expression for the gas density profile in the range $0<r<R_{200}$ is defined and we assume that the gas temperature $T_{\rm gas}$ at the distance $R_{200}$ is equal to the virial temperature $T$. This assumption is well justified by results from hydrodynamical simulations. In the EAGLE suite, the temperature of hot, non-star-forming gas at the virial radius differs from the virial temperature by a factor in the range $0.4 - 0.8$, depending on the feedback variant considered \citep{Stevens_2017}. Accounting for such a  re-scaling is likely a sub-dominant correction to our idealised modelling of the physics shaping the spread of baryons around haloes. Thus, our choice of setting $T_{\rm gas} (R_{200}) = T$ is justified in the spirit of keeping our formalism as simple as possible, while still capturing the essential physics.

Following the formalism by \cite{Hernquist_2003}, we choose a power law for the gas density within the halo, $\rho_{\rm gas} \propto r^{-\eta}$. This enables us to write:
\begin{equation}
\label{eq:HS03gasdensity}
    \rho_{\text{gas}}(R_{200}) = \frac{(3+\eta)M_{200}f_{\text{gas}}}{4 \pi R_{200}^{3}} \, .
\end{equation}
Substituting the above expression in equation~\eqref{eq:wn2}, and with the definition:
\begin{equation}
\label{eq:virialtemperature}
    T = \frac{\mu}{2 k_B} \left[ \sqrt{\frac{\Delta}{2}} G H(z) M_{200} \right]^{\frac{2}{3}} \, ,
\end{equation}
we can write $w_n$ in terms of $f_{\rm gas}$, the Hubble constant, and the virial mass of the halo:
Substituting this into equation~(\ref{eq:wn2}) gives the complete form of $w_n$:
\begin{equation}
\label{eq:wn_3}
    w_n = \frac{1}{2} \left( \sqrt{\frac{2}{\Delta}} GH(z)M_{200} \right)^{\frac{2}{3}} \left[ \frac{(3-\eta) \Delta f_{\rm gas} H(z)^2}{8 \pi G} \right]^{-n} \, .
\end{equation}

Equation~(\ref{eq:wn_3}) can then be substituted for $w_n$ in equation~(\ref{eq:mu}) and used to update equation~(\ref{eq:model2}):

\begin{equation}
\label{eq:Rc_total}
\begin{aligned}
    \frac{R_{\text{closure}}}{R_{200}} = \left[ 
    \left( 1 + \frac{1}{n} \right)^{\frac{1}{n}} 
    \left( 1 + \frac{\Omega_{m0}}{\Omega_{\Lambda 0}}(1+z)^3 \right)^{\frac{1}{n}} \right. \\
     \times \left. \left( \frac{\Delta}{2} \right)^{\frac{1}{n}} \frac{f_{\text{b-cosmic}}}{f_{\text{gas}}} \left( 1 - \frac{f_{\text{b-halo}}}{f_{\text{b-cosmic}}} \right) + 1 \right]^{\frac{n}{2+3n}} \, .
\end{aligned}
\end{equation}

This is our idealised analytic model for the closure radius. The impact of various parameters on the closure radius can be understood by examining equation~(\ref{eq:Rc_total}). The factor by which $R_{\text{closure}}$ exceeds $R_{200}$ is determined by the first term within the square brackets, which is added to 1. As a result, for all physically reasonable parameter values, $R_{\text{closure}}$ always exceeds $R_{200}$. The dependence of our model on the polytropic index, $n$, or equivalently $\eta = 2 / n$, is highly non trivial. Variations in $\eta$ (or $n$) significantly affect this factor, with a rapid increase observed for $\eta > 1.7$, a more gradual increase for $\eta < 1.7$, and a smooth minimum in between. Within our formalism, the physically allowed range is $0 < \eta < 3$: a negative $\eta$ would correspond to an increasing gas density profile at larger radii, which is unphysical, and $\eta \geq 3$ would imply an infinite gas mass within the halo. Though, the function in equation~\eqref{eq:Rc_total} diverges rapidly for $\eta > 2$. Thus, deviations of $\eta$ from approximately $1.7$ lead to an increase in the closure radius.

Furthermore, as expected, a lower gas fraction, $f_{\text{gas}}$, leads to larger closure radii, while a higher $f_{\text{gas}}$ shrinks them. Similarly, an increase in dark energy contributes to an increase in closure radii, and a lower halo baryon fraction produces the same effect -- both of which align with theoretical expectations. However, the redshift dependence of the closure radius is less straightforward. At first glance, equation~(\ref{eq:Rc_total}) suggests that increasing redshift leads to larger closure radii. This is counter intuitive, as one might expect closure radii to grow with time (i.e., decreasing redshift) instead due to the cumulative effects of feedback and the increasing dominance of $\Lambda$. We again remind ourselves that $n > 2/3$, and so the overall trend is that closure radii do decrease with redshift. Moreover, not all redshift dependencies are explicitly encoded in equation~(\ref{eq:Rc_total}), as we make no prior assumptions regarding the redshift evolution of $\eta$ or $f_{\text{gas}}$. In Section~\ref{sec:newmodelresults}, we explore a possible redshift dependence of these parameters and find evidence that they do vary simply with redshift ($\propto (1+z)^\gamma$ for some $\gamma)$. However, deriving an explicit functional form for these variations from theory is beyond the scope of this study.

\section{Simulations}
\label{sec:simulations}

\begin{table*}
    \centering
    \caption{The column headings are, from left to right: Name of simulation, comoving simulation box side length, number of particles (split evenly between baryons and dark matter), matter density parameter, baryon density parameter, vacuum energy density parameter (all at $z = 0$), rms linear fluctuation in the mass distribution on scales of $8h^{-1}$ Mpc, reduced Hubble constant, $H_0/(100 \, \mathrm{km \, s^{-1} \, Mpc^{-1}})$, and vacuum energy strength in units of $\Lambda_0$.}
    \label{tab:simulationparameters}
    \begin{tabular}{|c|c|c|c|c|c|c|c|c|c|}
        \hline
        \textbf{Simulation} & \textbf{$L / \mathrm{Mpc}$} & \textbf{$N$} & \textbf{$\Omega_{\text{m}}$} & \textbf{$\Omega_{\text{b}}$} & \textbf{$\Omega_\Lambda$} & \textbf{$\sigma_8$} & \textbf{$h$} & \textbf{$\Lambda/\Lambda_{0}$} \\
        \hline
        L0050N0752 & 50 & $2 \times 752^3$ & 0.307 & 0.0483 & 0.693 & 0.829 & 0.678 & 1 \\
        \hline
        L0025N0376/EdS & 25 & $2 \times 376^3$ & 1 & 0.157 & 0 &  0.683 & 0.375 & 0 \\
        \hline
        L0025N0376/L1 & 25 & $2 \times 376^3$ & 0.307 & 0.0483 & 0.693 & 0.829 & 0.678 & 1 \\
        \hline
        L0025N0376/L10 & 25 & $2 \times 376^3$ & 0.0424 & $6.67 \times 10^{-3}$ & 0.958 &  0.896 & 1.82 & 10 \\
        \hline
        L0025N0376/L30 & 25 & $2 \times 376^3$ &  0.0146 & $2.29 \times 10^{-3}$ & 0.985 & 0.843 & 3.11 & 30 \\
        \hline
        L0025N0376/L100 & 25 & $2 \times 376^3$ & $4.41 \times 10^{-3}$ & $6.93 \times 10^{-4}$ & 0.996 &  0.748 & 5.65 & 100 \\
        \hline
    \end{tabular}
\end{table*}

To validate our model across different cosmologies, we will compare its predictions to full hydrodynamical simulations. We consider the suite of simulations presented by \cite{Barnes_2018}. These are based on the EAGLE galaxy formation model and span a wide range of values for the cosmological constant ($0 < \Lambda / \Lambda_0 <300$). Since we verified that the $\Lambda = 300 \Lambda_0$ does not contain a statistically significant sample of haloes for the scope of this work, we will exclude it from our analysis. 

Table~\ref{tab:simulationparameters} shows the cosmological parameters in the runs considered in this work, together with the box size and the number of dark matter and baryonic resolution elements. Parameters that were kept constant across all simulations were the dark matter and gas mass resolution ($m_{\text{DM}} = 9.70 \times 10^6 M_{\odot}$ and $m_{\text{g}} = 1.81 \times 10^6 M_{\odot}$, respectively), the Plummer-equivalent gravitational softening length, ($\epsilon_{\text{com}} = 2.66$ kpc), and maximum proper softening length ($\epsilon_{\text{prop}} = 0.70$ kpc). The same initial conditions, generated at $z=127$ using second-order Lagrangian perturbation theory \citep{Jenkins2010}, which were fit to the angular statistics of the CMB observed in our Universe, were maintained across all runs. The fractional variance of matter density fluctuations in linear theory, computed within spherical regions of radius $8h^{-1} \, \rm Mpc$, $\sigma_8$, was calculated using CAMB \citep{Lewis_2000}. All simulations were evolved from the same initial starting snapshot, hence differences in $\sigma_8$ between simulations are due to differences in the structure growth and the value of $h$. 

Every variant also shares the same astrophysical model of galaxy formation. Since this has been widely described in the literature, we will only summarise the features that are most relevant for our study, while referring the reader to the original EAGLE publications for full details \citep{Schaye, Crain_2015}. The simulations rely on a modified version of the GADGET-3 smoothed particle hydrodynamics (SPH) code \citep{10.1111/j.1365-2966.2005.09655.x}. EAGLE describes dark matter as self-gravitating Lagrangian particles, and gas as SPH particles, which are subject to both gravitational interaction and hydrodynamics. Star particles are created whenever set star formation criteria are satisfied. This is modelled stochastically using the pressure law scheme of \citet{Schaye_2007}, which reformulates the observed Kennicutt–Schmidt relation into a pressure-dependent law. This ensures the simulations reproduce the observed star formation law for any gas equation of state. Star formation is restricted to cold, dense gas above a threshold hydrogen number density, which is metallicity-dependent (lower in metal-rich gas where cooling is more efficient). SMBHs are positioned at the centre of haloes with $M_{200} > 10^{10}$ $\mathrm{M}_{\odot}/h$, and grow via mergers at modified Bondi-Hoyle accretion rates \citep{bondi1952spherically}, adjusted for the angular momentum of surrounding gas as explained in \cite{Schaye}.

Stellar feedback is incorporated by injecting thermal energy from supernovae, stochastically increasing the temperature of surrounding gas by a constant amount \citep{Schaye_2007}. AGN feedback is similarly addressed through a single thermal mode, stochastically increasing the temperature of surrounding gas by a fixed amount proportional to the accretion rate onto the supermassive black hole (SMBH; \citealt{Crain_2015}).  Additionally, EAGLE activates a UV background at $z$ = 11.5 to account for the increased heating of the gas during the epoch of reionisation \citep{10.1111/j.1365-2966.2008.14191.x}. 

The parameters governing the strength of the subgrid models were calibrated using the observed galaxy stellar mass function and the galaxy size-mass relation \citep{Crain_2015} at $z \sim 0.1$. It is  implicitly assumed that this parametrisation is valid at high redshift as well, and for all cosmological models considered. The latter is a challenging assumption to verify experimentally, but it enables focusing on the impact of varying $\Lambda$ only, and can be considered as a `stress test' for the underlying galaxy formation model in an unusual cosmological regime.

EAGLE uses the SUBFIND algorithm \citep{Springel_2001} to categorise particles into gravitationally bound subhaloes. This uses the friends-of-friends (FoF) algorithm \citep{1985ApJ...292..371D}, grouping dark matter particles separated by or closer than 0.2 times the mean particle separation. Baryons are then assigned to the same FoF halo as their closest dark matter particle. In each FoF halo, overdense regions are identified, partitioned into distinct subhaloes, and any non-gravitationally bound particles are removed \citep{Mcalpine}. The subhalo containing the particle with the lowest gravitational potential is designated as the central galaxy within the FoF halo, with its minimum gravitational potential location serving as the centre of the entire halo. Subsequently, parameters such as $R_{200}$ and $M_{200}$ are calculated for each halo through examining the density at different ranges from the centre and counting particles. The data are then stored in each SUBFIND file for every snapshot.

\section{Results and Discussion}
\label{sec:results}

\subsection{Computing the closure radius}
\label{sec:closureradiuscalc}

To compute the closure radius for each halo in the EAGLE simulations, we first restrict ourselves to haloes where at least one galaxy (substructure) is identified by the SUBFIND algorithm. For each halo, we then read the values of $R_{200}$, $M_{200}$ and the position of the minimum of the gravitational potential (which we consider as the halo centre), directly from the halo catalogue.

We then consider spherical surfaces centred on each halo with radii at every particle position within the query range, which we set as 50$R_{200}$ to ensure the closure radius is always resolved. At each radius, the enclosed baryon fraction, $f_{\text{b-enclosed}}$, is calculated as the sum of all baryonic particle masses divided by the total mass of all particles within the sphere of that radius. In practice, this involves calculating $f_{\text{b-enclosed}}$ at the position of the particle closest to the centre, moving to the next closest particle, recalculating $f_{\text{b-enclosed}}$ at that position, and repeating for every particle within the query range (typically many tens of virial radii, as discussed). 

This method constructs the enclosed baryon fraction radial profile of every halo in each simulation snapshot with the highest precision — more so than if integrating spheres with discrete radii had been used. The profiles are then examined to observe how $f_{\text{b-enclosed}}$ changes at each particle position from the centre of the halo. This is then compared to $f_{\text{b-cosmic}}$ (calculated using every particle in the simulation) and saved as a function of radius in units of $R_{200}$. $R_{\text{closure}}$ is determined by identifying where the enclosed baryon fraction of each halo first reaches $f_{\text{b-cosmic}}$ $\pm$ $\Delta f_{\text{b-cosmic}}$ beyond 0.5$R_{200}$. Here, the error range, $\Delta f_{\text{b-cosmic}}$, is set at 5\%. Essentially, the closure radius is now defined as the halocentric distance where the enclosed baryon fraction reaches 95\% of the cosmic level, with the 5\% error range largely accounting for natural variation in baryon distribution (perhaps due to the presence of other haloes nearby) and chosen to align with other studies on the closure radius, e.g. \citet{Ayromlou_2023}.

\begin{figure}
	\includegraphics[width=\columnwidth]{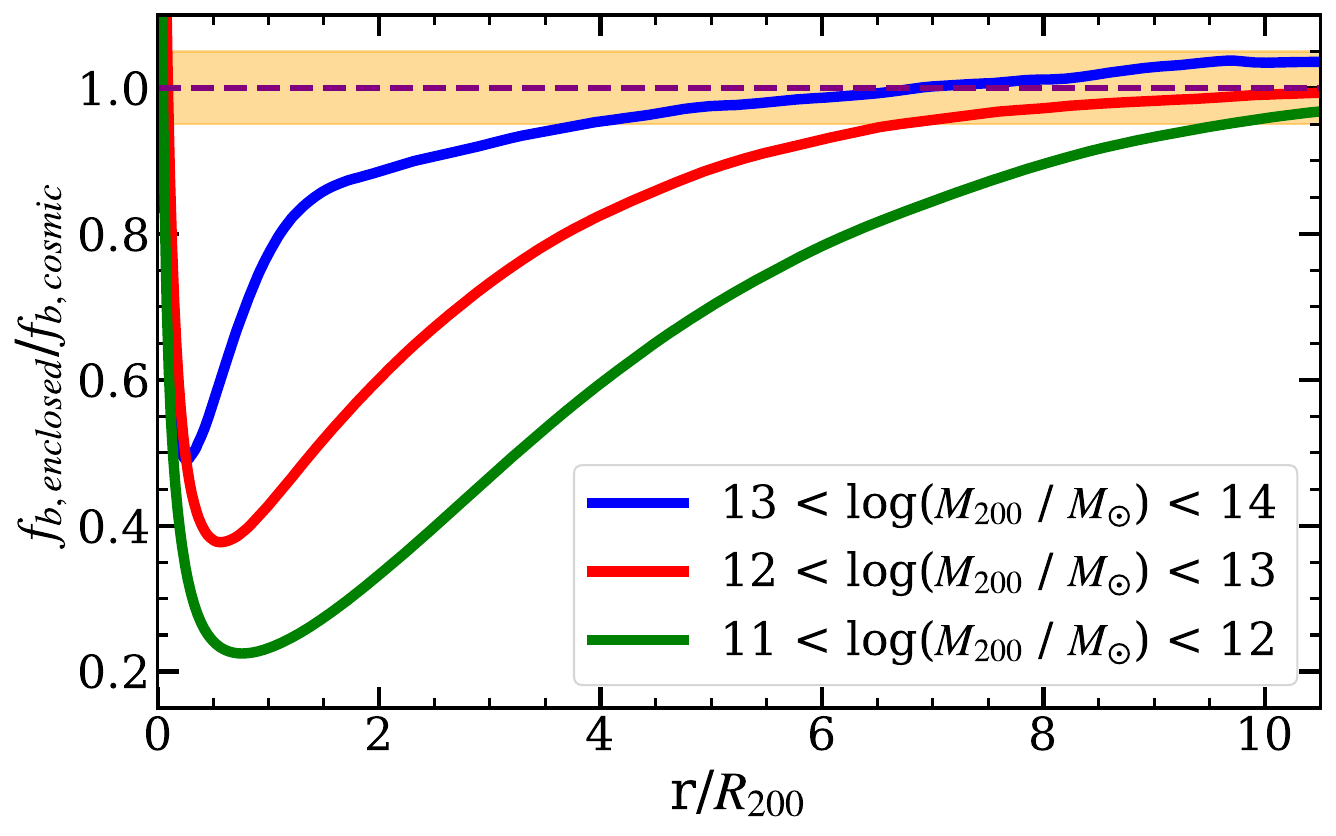}
    \caption{Radial profile of the average enclosed baryon fraction for haloes of different masses in the 50 Mpc simulation at $z$ = 0. The yellow band shows the 5$\%$ error range meaning that the average closure radius of haloes in each mass range would be taken as the point each line intersects the lower bound of the yellow band. The closure radius error is then the difference between that and where the line intersects the purple dashed line at exactly $f_{\text{b-cosmic}}$. The closure radius is generally smaller in larger haloes, owing to the deeper potential wells retaining baryons.}
    \label{fig:radialprofiles}
\end{figure}

Fig.~\ref{fig:radialprofiles} shows the average enclosed baryon fraction profile for haloes of different mass haloes profile in the L0050N0752 simulation ($\Lambda_0$). In the innermost regions, $f_{\text{b-enclosed}}$ is large due to the high baryon fraction resulting from star formation, declines to a minimum around the virial radius, and thereafter increases gradually toward the cosmic value at larger radii. The theoretical concept of the closure radius allows us to quantify the maximum distance from a halo where its missing baryons may accumulate, but its exact location depends strongly on the definition  (and associated error margin) assumed. Adopting a 10\% error range instead would generally reduce closure radii substantially. Thus, the precise localisation of the closure radius in practice is challenging, particularly for lower-mass haloes which have larger errors, resulting in considerable scatter. Additionally, Fig.~\ref{fig:radialprofiles} shows that, on average, larger haloes tend to have smaller closure radii (relative to $R_{200}$), consistent with the notion that their deeper gravitational potential wells allow them to retain their complement of baryons closer to the halo. For instance, haloes within the mass range $10^{13}$ – $10^{14} M_{\odot}$ have $R_{\text{closure}} \sim (3 \pm 2) R_{200}$, whereas those within the $10^{12}$ – $10^{13} M_{\odot}$ mass range have $R_{\text{closure}} \sim (6 \pm 3) R_{200}$.

The profile for haloes with masses $10^{13}$ – $10^{14} M_{\odot}$ does not asymptote to $f_{\text{b-cosmic}}$ within $10R_{200}$ due to the small number ($<10$) of haloes in this range in the simulation, making the average enclosed baryon profile more susceptible to individual halo-to-halo variations. We have verified that their profile asymptotes to $f_{\text{b-cosmic}}$ after $r \sim 20R_{200}$, although this is not plotted for clarity. Massive haloes tend to retain baryons closer to the centre and often have smaller satellite galaxies nearby, from which more baryons may be accreted nearer to the central halo. This may have contributed to the slight overshooting of $f_{\text{b-enclosed}}$ beyond $f_{\text{b-cosmic}}$ for the most massive haloes in Fig.~\ref{fig:radialprofiles}.

\begin{figure}
  \centering
  \includegraphics[width=\columnwidth]{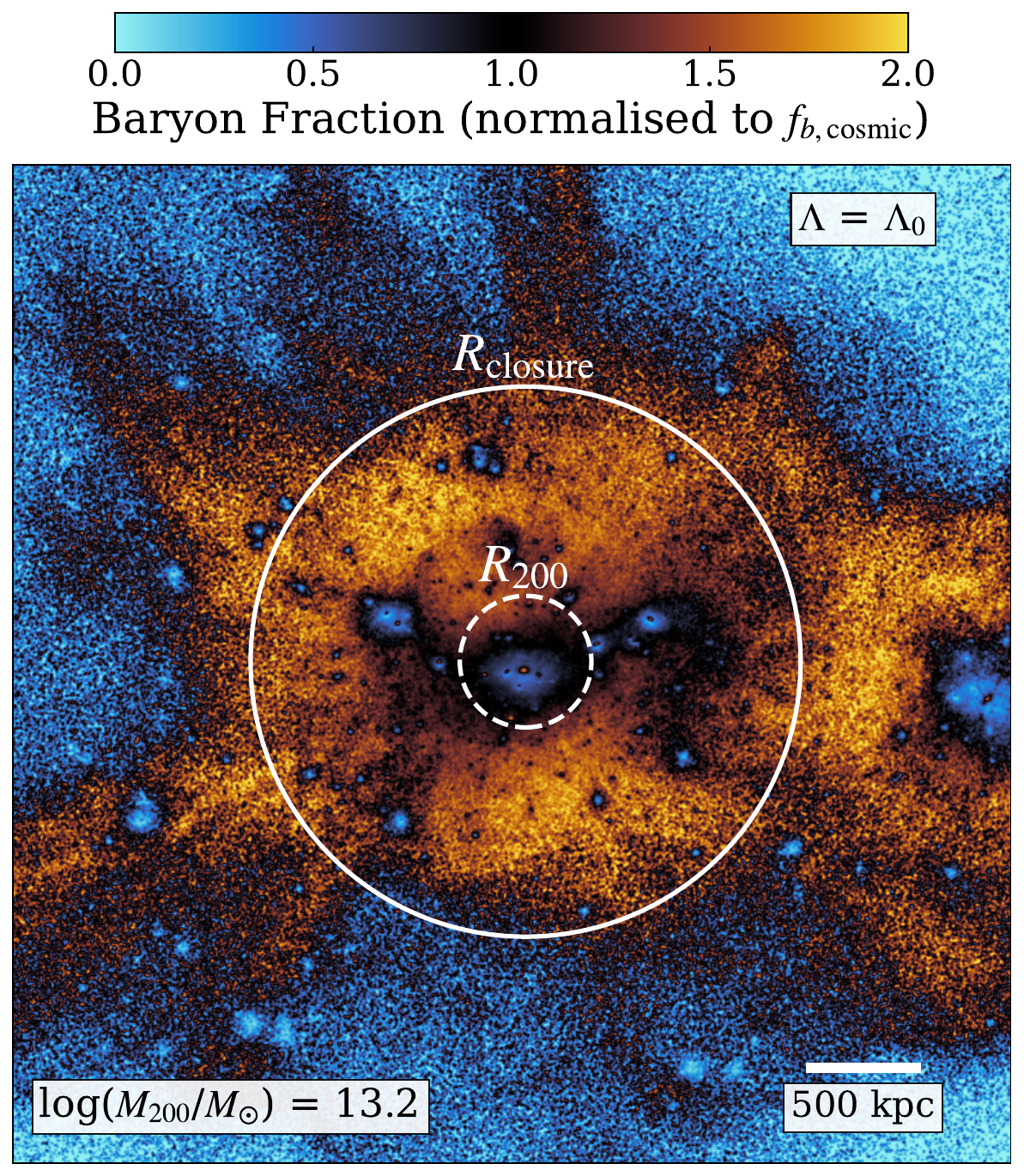}
  \caption{Distribution of baryons in and around a halo in the EAGLE L0050N0752 ($\Lambda_0$) simulation at $z = 0$. Orange and blue regions indicate locations where the baryon fraction is above or below the cosmic value respectively. The halo's characteristic radial scales, $R_{200}$ and $R_{\rm closure}$, are marked by the dashed and solid circles. The figure illustrates how feedback processes evacuate baryons from the central halo region, producing a baryon deficit within $R_{200}$, while redistributing baryons to much larger distances. As a result, the radius at which the enclosed baryon fraction returns to the cosmic value, $R_{\rm closure}$, lies several times beyond $R_{200}$.}
  \label{fig:baryon_fraction_map}
\end{figure}

We illustrate the baryon distribution for an example halo in Fig.~\ref{fig:baryon_fraction_map}, which shows the baryon fraction within and around a halo from the 50 Mpc ($\Lambda_0$) simulation at $z = 0$, with the halo's $R_{200}$ and $R_{\rm closure}$ indicated. Baryons are seen to be expelled from both the central galaxy and its satellites, producing baryon-deficient (blue) regions within the halo. These baryons are redistributed to distances well beyond $R_{200}$, extending into the surrounding circumgalactic and intergalactic medium. This provides a visual counterpart to the behaviour shown in Fig.~\ref{fig:radialprofiles}: haloes are typically baryon deficient within $R_{200}$, while the closure radius lies several times further out, marking the scale at which the baryons originally associated with the halo are fully accounted for.

\begin{figure*}
  \centering
  \includegraphics[width=\linewidth]{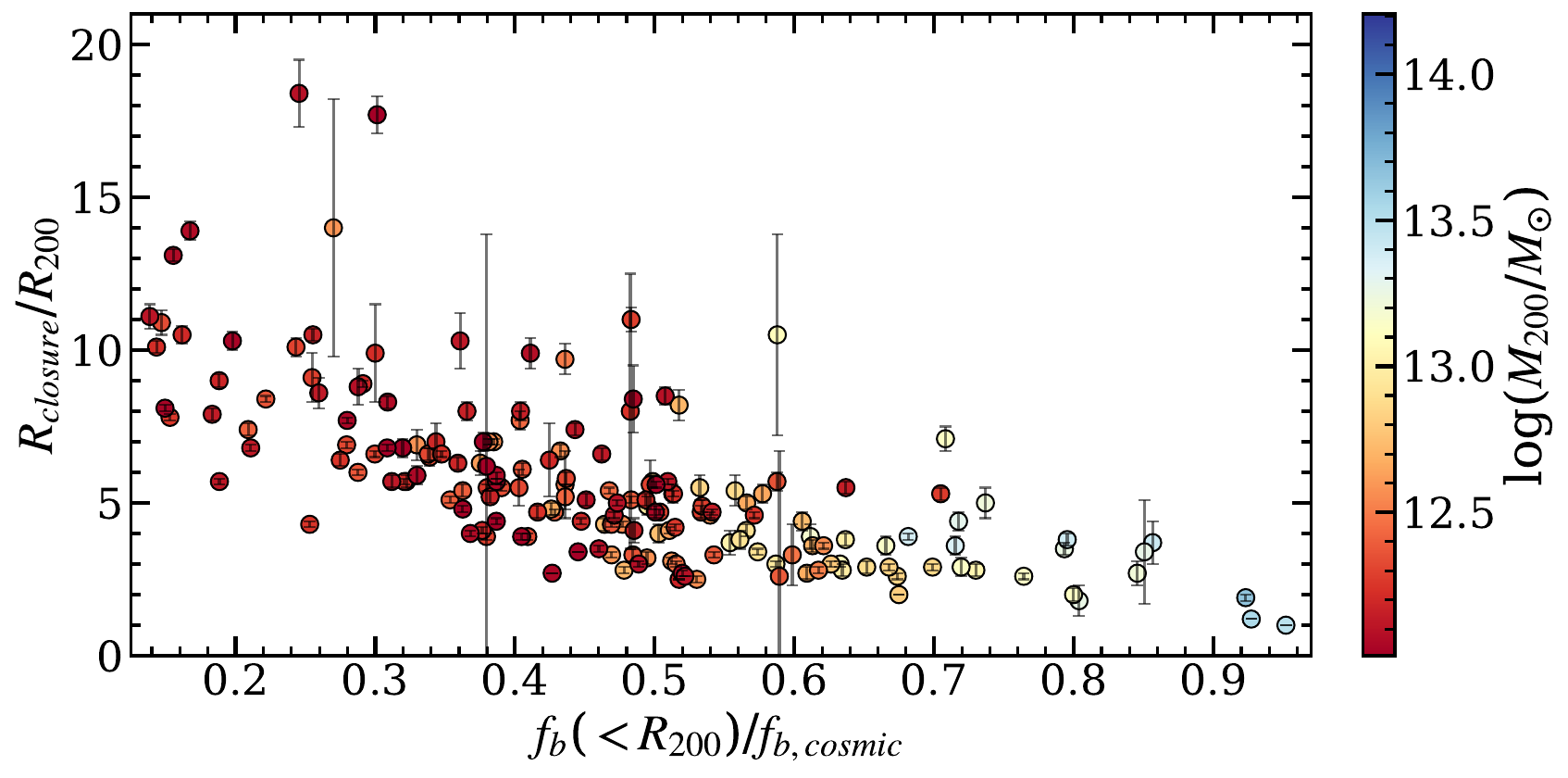}
  \caption{$R_{\text{closure}}/R_{200}$ against $f_{\text{b-halo}}/f_{\text{b-cosmic}}$ for every halo with $M_{200} > 10^{12} \mathrm{M}_{\odot}$ in the 50 Mpc $\Lambda_0$ simulation at $z = 0$. Each halo is colour coded by its mass and has error bars denoting the 5\% error range on its closure radius calculation (see main text for details). As expected, more massive haloes have higher enclosed baryon fractions and smaller relative closure radii on average due to their larger gravitational potential wells making trapping baryons within $R_{200}$ easier. Moreover, haloes with larger $f_{\text{b-halo}}$ tend to have smaller $R_{\text{closure}}/R_{200}$ as they are less void of baryons within $R_{200}$.}
  \label{fig:50MpcLambda0}
\end{figure*}

In Fig.~\ref{fig:50MpcLambda0} we show the closure radii of every halo at $z = 0$ in the 50 Mpc $\Lambda_0$ simulation, with a minimum mass of $\log_{10}\left(M_{200}/M_{\odot}\right) >$ 12, plotted against $f_{\text{b-enclosed}}(<R_{200})$. Importantly, adopting a less restrictive mass range significantly increases the number of haloes for analysis, enhancing the statistical power of best-fit parameter minimisation, crucial for when we come to fit different models. Fig.~\ref{fig:50MpcLambda0} shows that we are able to recover the expected closure radius trend; specifically that haloes with larger halo baryon fractions tend to have smaller relative closure radii.

Fig.~\ref{fig:50MpcLambda0} also highlights notable dispersion in closure radii, even among haloes of similar mass and enclosed baryon fraction. This dispersion complicates precise predictions of the closure radius for individual haloes in simulations, reflecting the complex interplay of baryon physics and the absence of perfectly isolated haloes in a cosmological context. For instance, smaller haloes in Fig.~\ref{fig:50MpcLambda0} often show larger closure radii and lower enclosed baryon fractions. While some of this scatter may be attributed to genuine satellites of even more massive haloes, many lower-mass systems are themselves independent, virialised haloes that may nonetheless lie in the vicinity of larger structures. Their baryon content may therefore be affected by environmental factors such as tidal interactions or large-scale gravitational fields, contributing to the observed diversity in closure radii. However, this question falls outside the scope of the present work, and we intend to pursue it with future investigations.

\subsection{The closure radius -- baryon mass fraction relationship}

\subsubsection{Fitting our model to the EAGLE simulations}
\label{sec:newmodelresults}

Here we apply our model for the closure radius by calculating best-fit $\eta$ and $f_{\text{gas}}$, equation~(\ref{eq:Rc_total}), using chi-squared minimisation. The EdS universe cannot be solved for by our new model as we assumed non-zero dark energy in our formalism. In this case, one should consider the EdS limit of the density profile (discussed in Appendix~\ref{app:solutions}) and repeat our formalism with no $\Lambda$. This may or may not lead to a form for the closure radius similar to equation~(\ref{eq:Rc_total}) and is beyond the scope of this work. 

We also switch to using $\eta$ as our fitting parameter instead of $n$, as negative reciprocal powers of $n$ cause optimisation functions (e.g., \texttt{scipy.optimize} by \citealt{2020SciPy-NMeth} as was used here) to slow down or not converge properly. This simply involves recasting our model in terms of $\eta$ as:
\begin{equation}
\label{eq:Rc_total_eta}
\begin{aligned}
    \frac{R_{\text{closure}}}{R_{200}} = \left[ 
    \left( 1 - \frac{\eta}{2} \right)^{\frac{-\eta}{2}} 
    \left( 1 + \frac{\Omega_{m0}}{\Omega_{\Lambda 0}}(1+z)^3 \right)^{\frac{-\eta}{2}} \right. \times \\
     \left. \left( \frac{\Delta_{200}}{2} \right)^{\frac{-\eta}{2}} \frac{f_{\text{b-cosmic}}}{f_{\text{gas}}} \left( 1 - \frac{f_{\text{b-halo}}}{f_{\text{b-cosmic}}} \right) + 1 \right]^{\frac{1}{3-\eta}}.
\end{aligned}
\end{equation}
We also set priors on the physically allowed values of parameters, notably $0 < \eta < 3$ (see discussion in Section~\ref{sec:closureradiusmodel}) and $0 < f_{\text{gas}} < 1$ -- although most optimisation methods find values within these domains even without setting them a priori.

We extend our analysis to haloes at $z = 0.5, 1.0, 1.5, 2.0, 2.5$, and $3.0$. At higher redshifts most haloes are insufficiently formed, limiting the utility of the closure radius. As shown in Fig.~\ref{fig:50MpcLambda0}, smaller haloes exhibit substantially larger scatter, partly due to finite resolution, which makes their closure radii less reliable and leads to poorly constrained best-fit values of $\eta$ and $f_{\text{gas}}$. We therefore assess how the choice of minimum halo mass affects the stability of the fitted parameters and adopt a mass threshold that ensures robust closure radius measurements. For meaningful comparisons across cosmologies, this threshold is kept fixed across all simulations with different values of $\Lambda$.

\begin{figure}
    \includegraphics[width=\columnwidth]{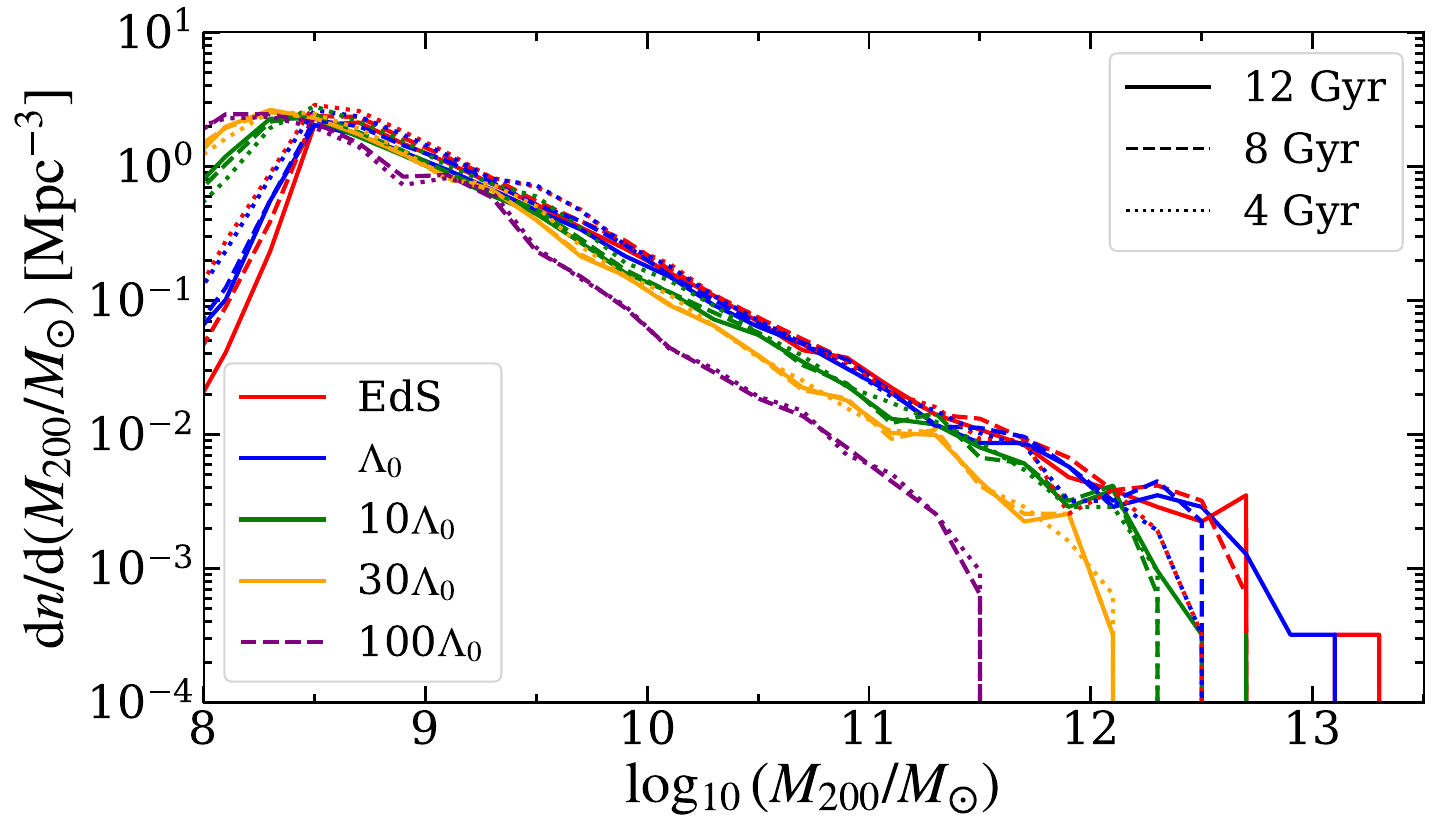}
    \caption{The Halo Mass Function (HMF) in each 25 Mpc alternate cosmology simulation at three cosmic times. This depicts the number density, $n$, of haloes of different masses in each simulation at those times, showing that the higher $\Lambda$ simulations have less haloes overall at every mass. The 100$\Lambda_0$ simulation did not run for as long a cosmic time, so only the 4 and 8 Gyr times are plotted. The HMF was calculated here using mass bins of $\Delta$ log$(M_{200}/\mathrm{M}_{\odot}) = 0.25$. Haloes with mass below $M_{200}\approx 10^{8.5} \, \rm M_{\odot}$ are under-resolved, causing the drop of the HMF at the low-mass end.}
    \label{fig:HMF}
\end{figure}

To determine a consistent threshold across all cosmologies, we use the halo mass function (HMF), which quantifies the number density of haloes per logarithmic interval in mass. Fig.~\ref{fig:HMF} shows the familiar dependence of halo abundance with mass in all cosmologies. Below $M_{200} \lesssim 10^{8.5} \mathrm{M}_{\odot}$, there is a sharp downturn which is due to finite mass resolution. Selecting a threshold of $M_{200} \sim 10^{8.5} \mathrm{M}_{\odot}$ therefore substantially increases the number of haloes available for analysis compared to a higher mass threshold. Fig.~\ref{fig:HMF} also demonstrates the reduced abundance of massive haloes in higher-$\Lambda$ simulations, reflecting the impact of enhanced dark energy on large-scale clustering, consistent with the findings of \citet{BK_2022}. This highlights that a mass threshold suitable for one cosmology may not be directly transferable to another.

To identify an optimal, uniform threshold, we recalculated the minimum reduced chi-squared from applying our model for each cosmology at each redshift, testing discrete thresholds in steps of $\Delta\log_{10}(M_{200}/\mathrm{M}_{\odot}) = 0.25$ over the range $10^{10.5} - 10^{13} \mathrm{M}_{\odot}$. We then compared relative chi-squared values to select a threshold that balances statistical robustness (low chi-squared values) with reliable closure radius measurements. The optimal choice was found to be $M_{200} > 10^{11}\,\mathrm{M}_{\odot}$, which ensures at least ten haloes per snapshot across all simulations and redshifts, while yielding reduced chi-squared values of $\lesssim 20$ in most cases. This threshold is adopted for all subsequent analysis.

The $100\Lambda_0$ simulation is excluded from further discussion as it contains too few high-mass haloes with well-defined closure radii. Future improvements will require larger simulation volumes and higher particle counts to increase the abundance of massive haloes, enabling more stringent mass cuts, reduced parameter uncertainties, and more reliable comparisons across cosmologies, particularly at high $\Lambda$.

\begin{figure*}
    \centering
    \includegraphics[width=0.32\textwidth]{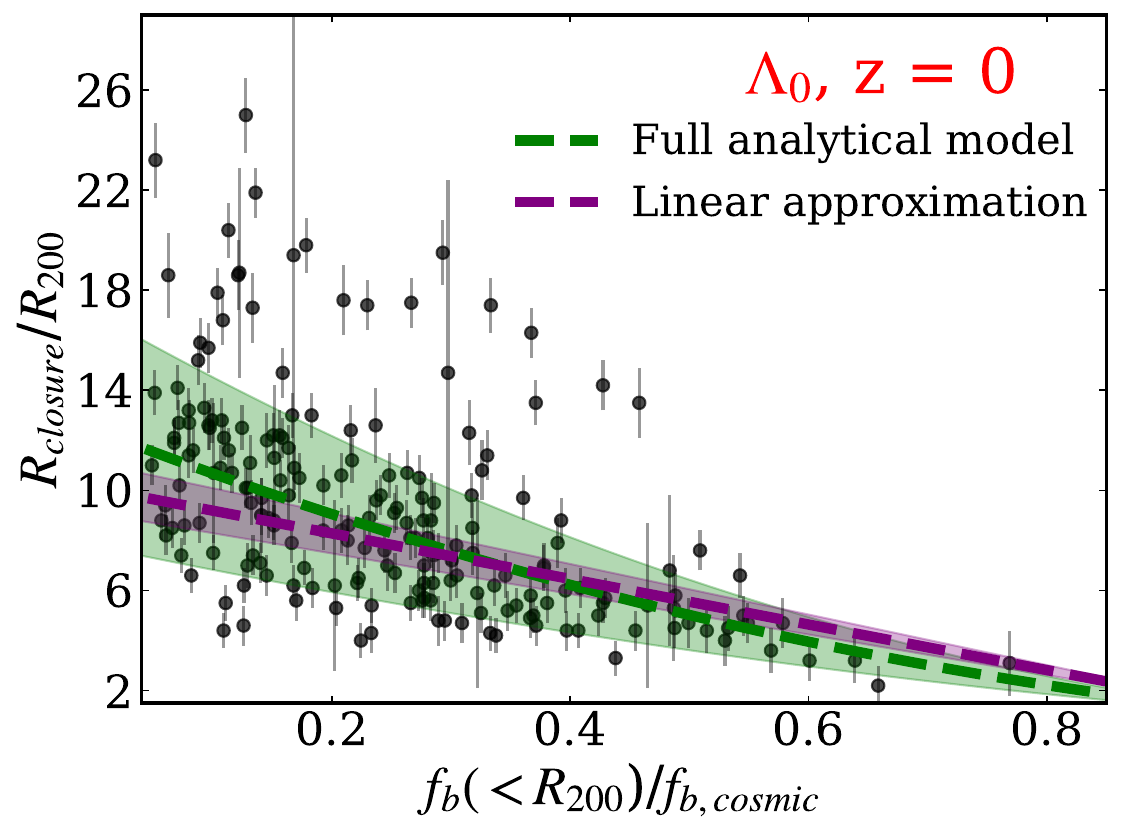}
    \includegraphics[width=0.32\textwidth]{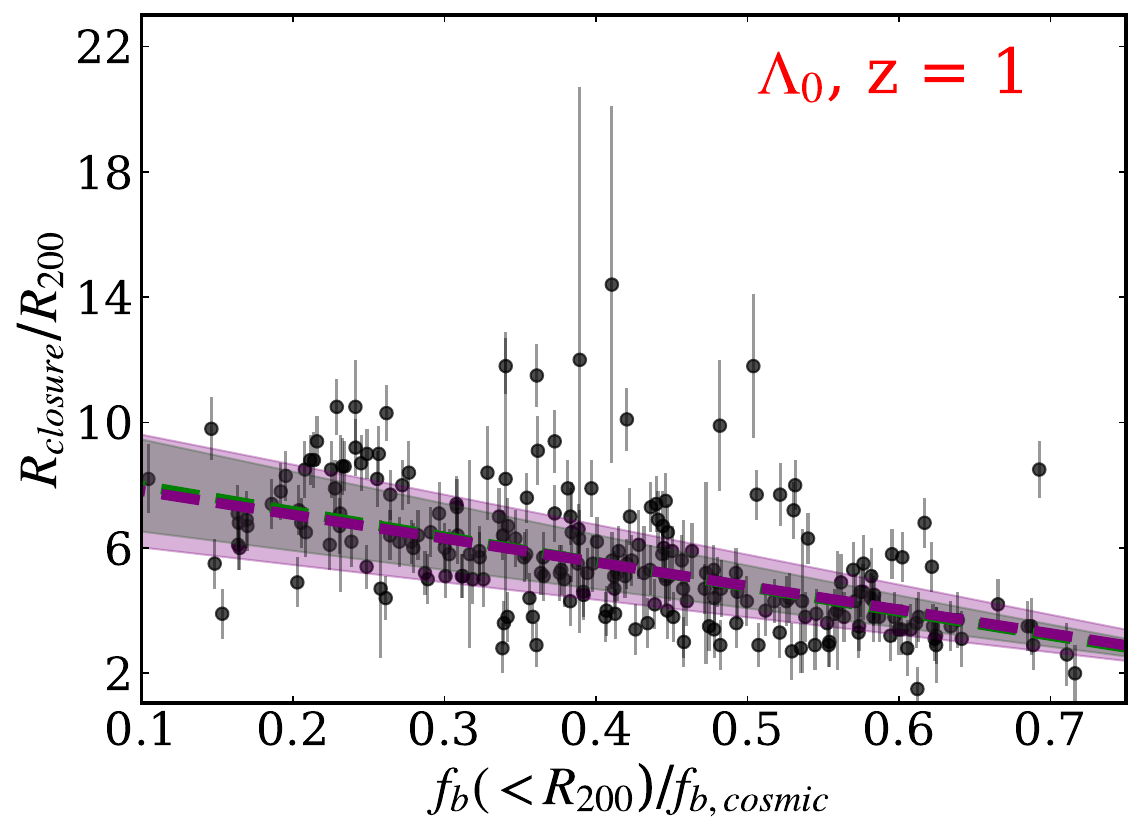}
    \includegraphics[width=0.32\textwidth]{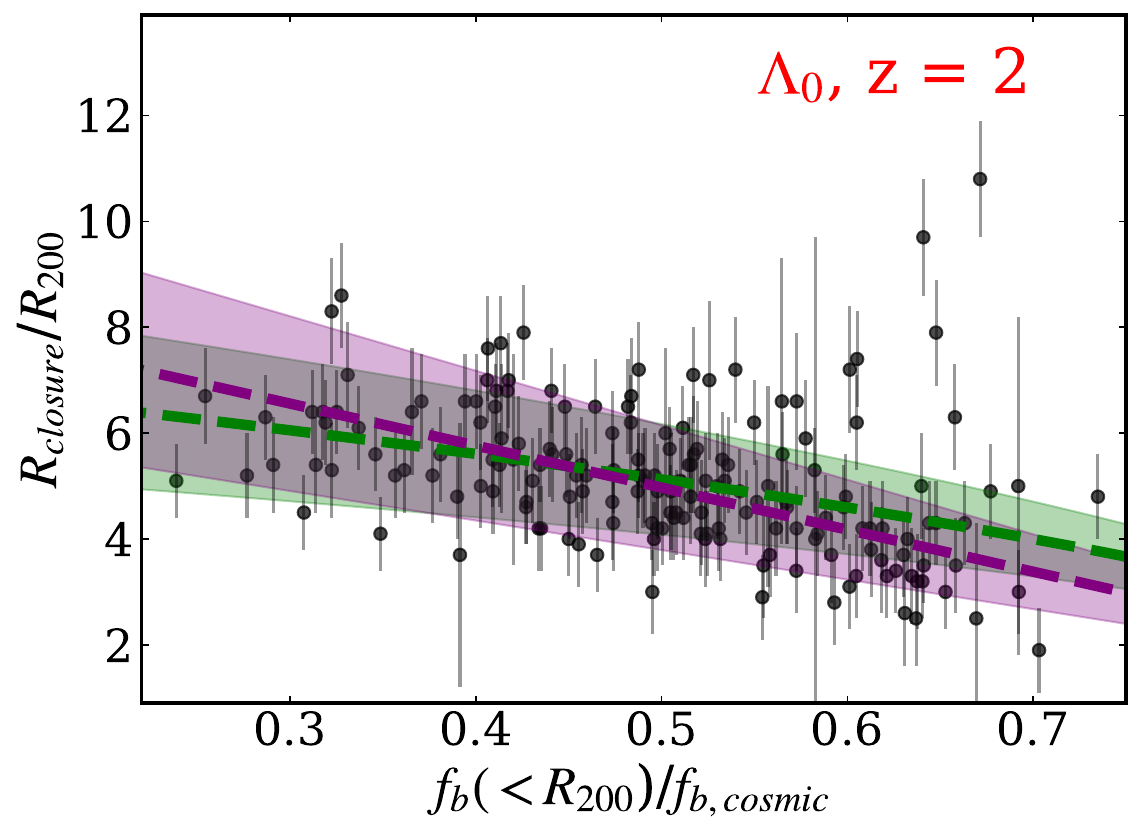} \\
    \includegraphics[width=0.32\textwidth]{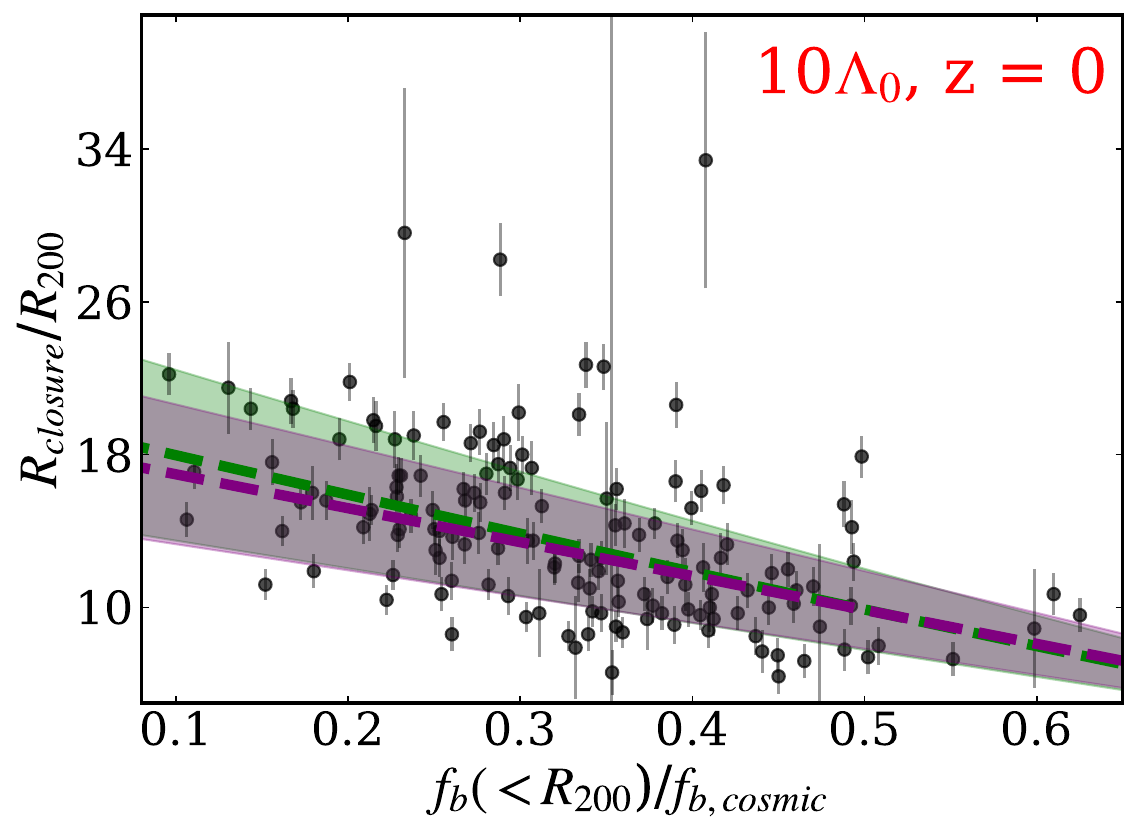}
    \includegraphics[width=0.32\textwidth]{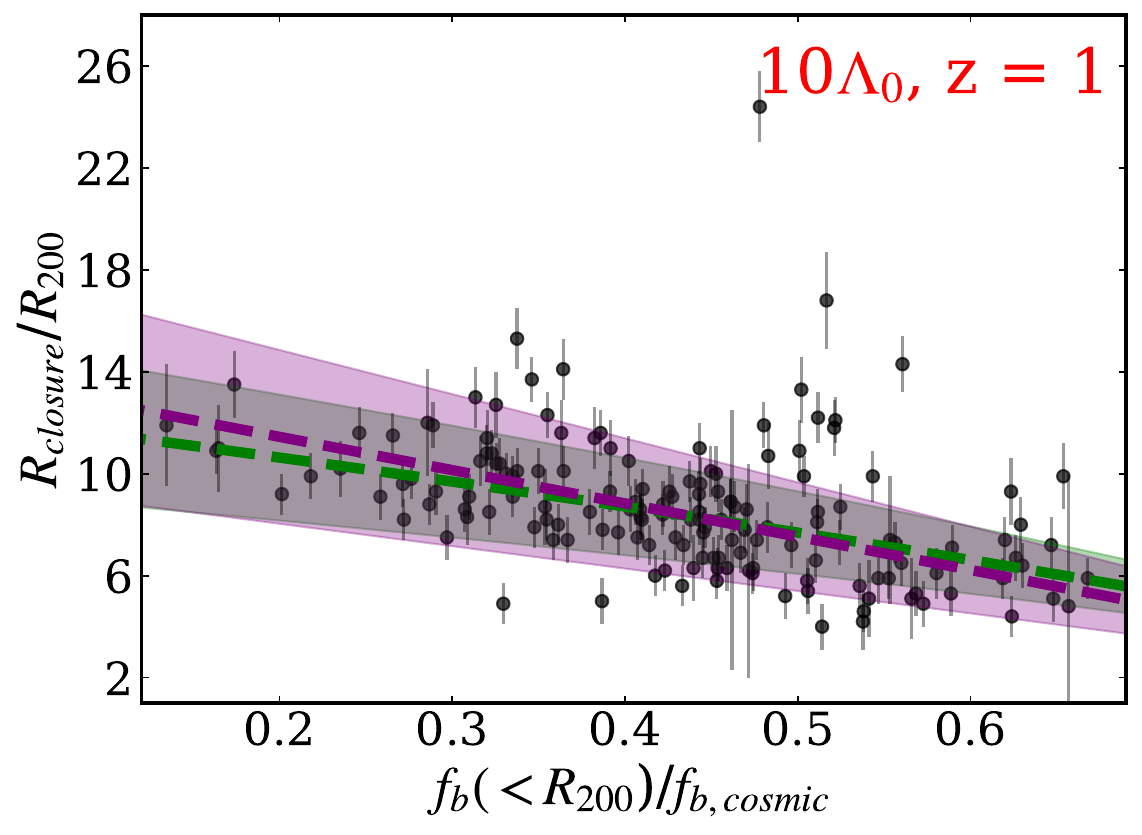}
    \includegraphics[width=0.32\textwidth]{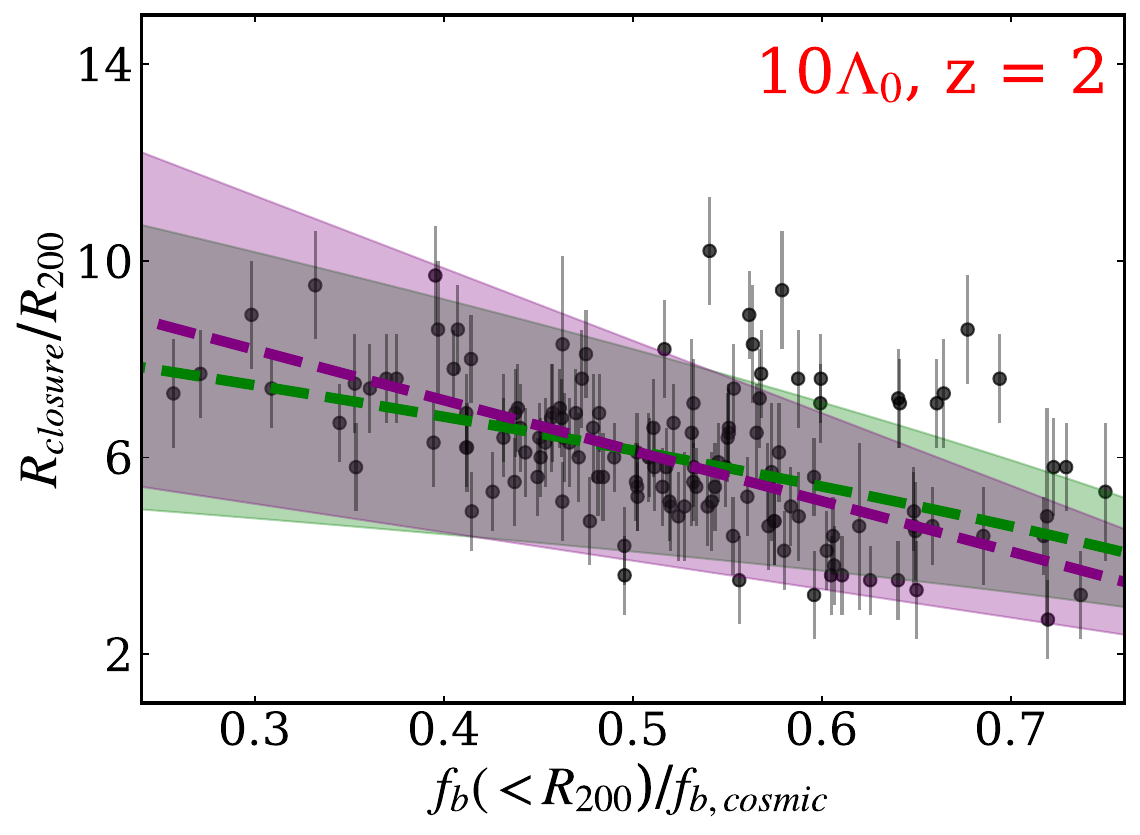} \\
    \includegraphics[width=0.32\textwidth]{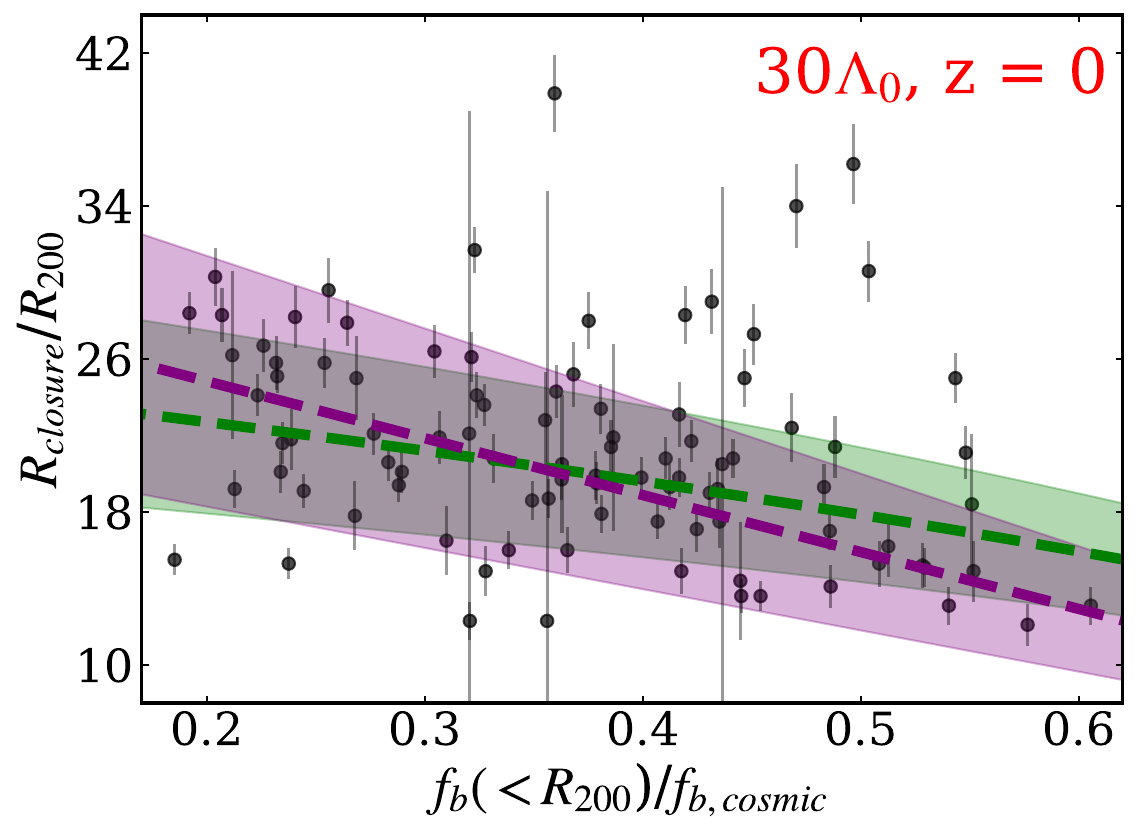}
    \includegraphics[width=0.32\textwidth]{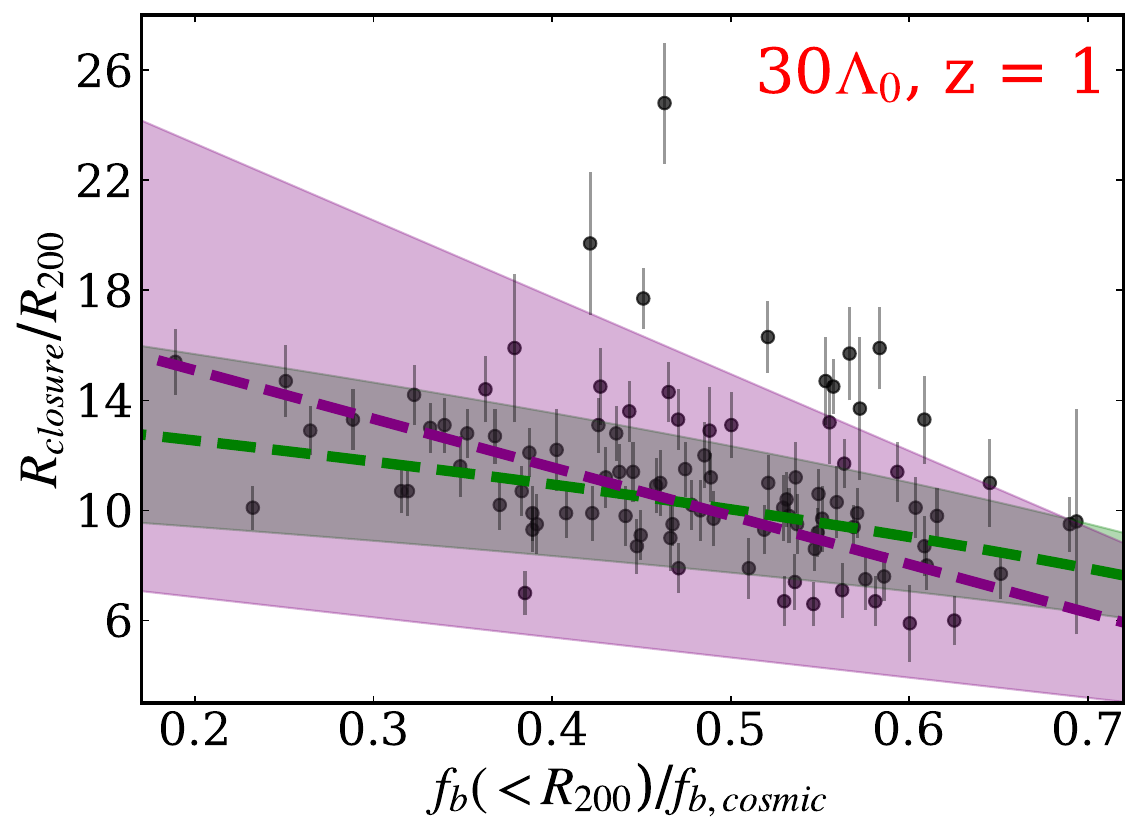}
    \includegraphics[width=0.32\textwidth]{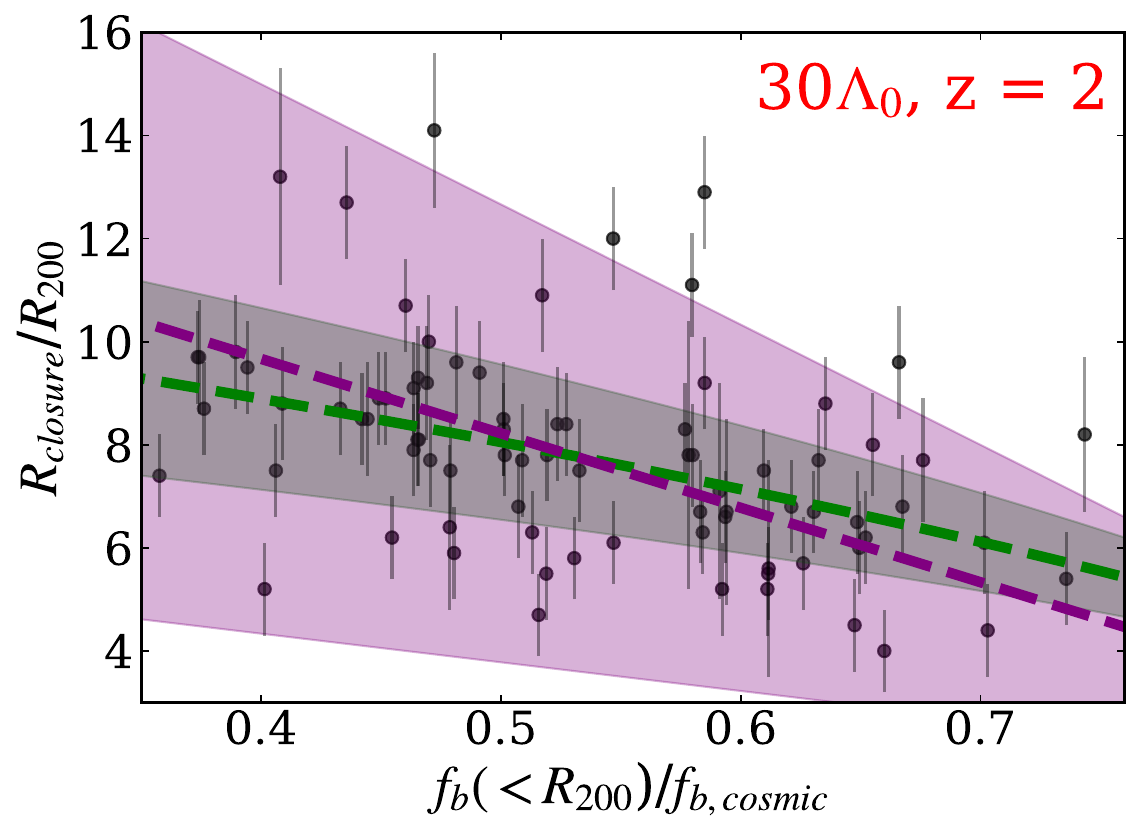}
    \caption{Plots of relative closure radii against halo baryon fraction with the two best-fit models for various cosmologies and redshift. Green shows our new (non-linear) model from this study, equation~(\ref{eq:Rc_total_eta}) and purple shows the approximate model, equation~(\ref{eq:Rc_approx}) (or equivalently the best-fit A23 formula, equation~(\ref{eq:A23model}), which agree completely with each other as they are both linear at fixed $z$), with parameter values as given in Table~\ref{tab:combinedparameters}. Our new model typically has better or just as good $\chi^2$ values as the approximate and A23 formula at every redshift tested in every cosmology.}
    \label{fig:3x3grid}
\end{figure*}

We present the results from our chi-squared minimisation optimisation for each individual redshift snapshot with fixed optimal minimum halo mass threshold ($M_{200} > 10^{11} \mathrm{M}_{\odot}$) in Table~\ref{tab:combinedparameters}. We also show the resulting fits graphically in Fig.~\ref{fig:3x3grid}. Every panel reports the closure radius–baryon mass fraction relationship for a different cosmological model and redshift; the best-fit model to the numerical data given by equation~\eqref{eq:Rc_total_eta} is indicated with the green dashed lines. The green shaded area shows the corresponding $16^{\rm th}-84^{\rm th}$ percentile range following from the errors on the fit parameters. The dashed purple lines refer to the best-fit from a linear approximation of our model, which we shall discuss later (see Section~\ref{sec:approxsolutionresults}). At the current stage, we highlight that our model for the closure radius--baryon mass fraction relationship as defined in equation~\eqref{eq:Rc_total_eta} provides a good description of the numerical data over all cosmological models and redshifts considered.

\begin{table*}
    \centering
    \caption{Best-fit parameter values from individual snapshots in the redshift range $0 < z < 3$ using both our new model, equation~(\ref{eq:Rc_total_eta}), and the approximated model, equation~(\ref{eq:Rc_approx}), along with the A23 formula, equation~(\ref{eq:A23model}). Note: some alternate cosmology `$z = 0$' snapshots have a best-fit $\gamma$ due to their redshift being not exactly equal to 0, making $\gamma$ no longer arbitrary.}
    \label{tab:combinedparameters}
    \begin{tabular}{|c|c|c|c|c|c|c|c|}
        \hline
        \textbf{$\Lambda / \Lambda_0$} & \textbf{$z$} & \textbf{$\eta$} & \textbf{$f_{\text{gas}}$ / $10^{-4}$} & \textbf{$\eta_{\text{approx}}$} & \textbf{$f_{\text{gas - approx}}$ / $10^{-4}$} & \textbf{$\alpha$} & \textbf{$\gamma$} \\
        \hline
        1 & 0 & 1.69 $\pm$ 0.03 & 49 $\pm$ 1 & 1.72 $\pm$ 0.06 & 12 $\pm$ 3 & 9.1 $\pm$ 0.4 & - \\
        \hline
        1 & 0.5 & 1.63 $\pm$ 0.04 & 26 $\pm$ 5 & 1.75 $\pm$ 0.05 & 14 $\pm$ 2 & 9.0 $\pm$ 0.3 & -0.3 $\pm$ 0.2 \\
        \hline
        1 & 1 & 1.39 $\pm$ 0.06 & 8.0 $\pm$ 0.9 & 1.84 $\pm$ 0.04 & 16 $\pm$ 2 & 10.0 $\pm$ 0.4 & -0.4 $\pm$ 0.2 \\
        \hline
        1 & 1.5 & 1.22 $\pm$ 0.09 & 4.3 $\pm$ 0.3 & 1.79 $\pm$ 0.04 & 6 $\pm$ 1 & 9.3 $\pm$ 0.5 & -0.2 $\pm$ 0.2 \\
        \hline
        1 & 2 & 0.7 $\pm$ 0.2 & 3.2 $\pm$ 0.3 & 1.94 $\pm$ 0.03 & 20 $\pm$ 2 & 9.9 $\pm$ 0.5 & -0.2 $\pm$ 0.2 \\
        \hline
        1 & 2.5 & 0.9 $\pm$ 0.2 & 2.5 $\pm$ 0.3 & 1.92 $\pm$ 0.03 & 10 $\pm$ 2 & 10 $\pm$ 1 & -0.1 $\pm$ 0.2 \\
        \hline
        1 & 3 & 0.1 $\pm$ 0.6 & 1.6 $\pm$ 0.2 & 1.48 $\pm$ 0.06 & 3 $\pm$ 1 & 10 $\pm$ 1 & -0.1 $\pm$ 0.2 \\
        \hline
        10 & 0 & 1.41 $\pm$ 0.03 & 11 $\pm$ 1 & 1.82 $\pm$ 0.05 & 10 $\pm$ 1 & 20 $\pm$ 4 & -1.5 $\pm$ 0.3 \\
        \hline
        10 & 0.5 & 1.08 $\pm$ 0.07 & 5.2 $\pm$ 0.8 & 1.88 $\pm$ 0.06 & 9 $\pm$ 2 & 18 $\pm$ 3 & -0.8 $\pm$ 0.3 \\
        \hline
        10 & 1 & 1.08 $\pm$ 0.08 & 7 $\pm$ 2 & 1.81 $\pm$ 0.07 & 13 $\pm$ 3 & 18 $\pm$ 3 & -0.5 $\pm$ 0.2 \\
        \hline
        10 & 1.5 & 1.1 $\pm$ 0.1 & 5.9 $\pm$ 0.9 & 1.76 $\pm$ 0.04 & 13 $\pm$ 2 & 17 $\pm$ 3 & -0.5 $\pm$ 0.2 \\
        \hline
        10 & 2 & 0.9 $\pm$ 0.2 & 4.7 $\pm$ 0.6 & 1.81 $\pm$ 0.06 & 17 $\pm$ 3 & 20 $\pm$ 3 & -0.6 $\pm$ 0.2 \\
        \hline
        10 & 2.5 & 0.2 $\pm$ 0.3 & 3.8 $\pm$ 0.1 & 1.89 $\pm$ 0.05 & 29 $\pm$ 3 & 22 $\pm$ 3 & -0.7 $\pm$ 0.2 \\
        \hline
        10 & 3 & 0.1 $\pm$ 0.6 & 1.1 $\pm$ 0.7 & 1.61 $\pm$ 0.05 & 11 $\pm$ 2 & 22 $\pm$ 3 & -0.6 $\pm$ 0.2 \\
        \hline
        30 & 0 & 0.7 $\pm$ 0.1 & 0.7 $\pm$ 0.2 & 1.86 $\pm$ 0.06 & 7.7 $\pm$ 0.8 & 30 $\pm$ 8 & -0.4 $\pm$ 0.4 \\
        \hline
        30 & 0.5 & 0.8 $\pm$ 0.1 & 1.2 $\pm$ 0.3 & 1.98 $\pm$ 0.03 & 54 $\pm$ 6 & 42 $\pm$ 9 & -1.8 $\pm$ 0.8 \\
        \hline
        30 & 1 & 0.6 $\pm$ 0.2 & 1.7 $\pm$ 0.4 & 1.82 $\pm$ 0.06 & 10 $\pm$ 1 & 50 $\pm$ 10 & -1.5 $\pm$ 0.4 \\
        \hline
        30 & 1.5 & 0.4 $\pm$ 0.2 & 1.6 $\pm$ 0.4 & 1.78 $\pm$ 0.05 & 9 $\pm$ 1 & 34 $\pm$ 8 & -0.9 $\pm$ 0.4 \\
        \hline
        30 & 2 & 0.8 $\pm$ 0.2 & 3.2 $\pm$ 0.7 & 1.40 $\pm$ 0.06 & 6.5 $\pm$ 0.7 & 20 $\pm$ 10 & -0.3 $\pm$ 0.2 \\
        \hline
        30 & 2.5 & 0.6 $\pm$ 0.3 & 3.4 $\pm$ 0.8 & 1.81 $\pm$ 0.04 & 15 $\pm$ 2 & 30 $\pm$ 10 & -0.6 $\pm$ 0.6 \\
        \hline
        30 & 3 & 0.5 $\pm$ 0.5 & 3.6 $\pm$ 0.7 & 1.89 $\pm$ 0.04 & 26 $\pm$ 2 & 30 $\pm$ 10 & -0.6 $\pm$ 0.6 \\
        \hline
    \end{tabular}
\end{table*}

\begin{figure*}
    \centering
    \includegraphics[width=0.42\textwidth]{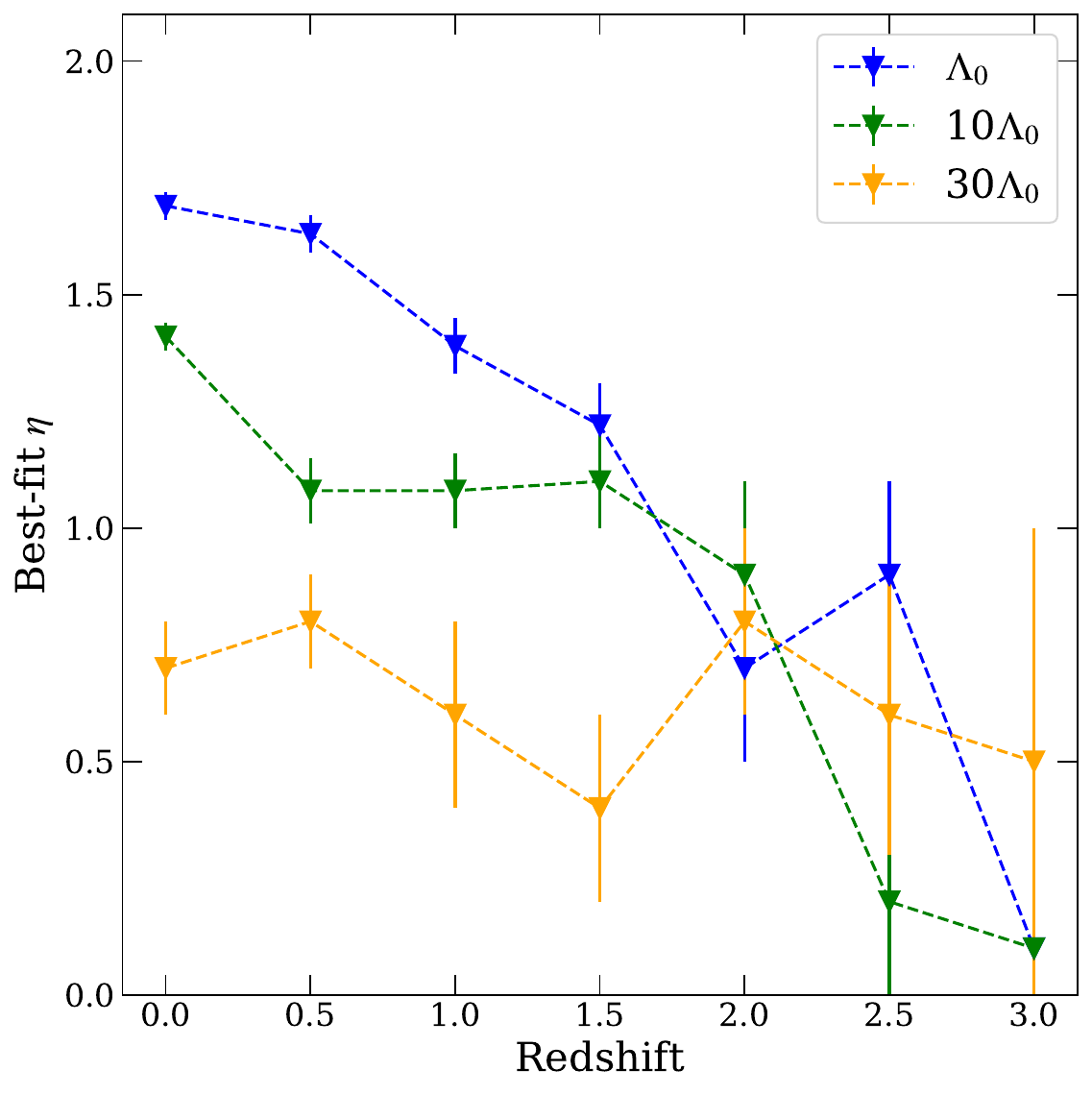}
    \includegraphics[width=0.42\textwidth]{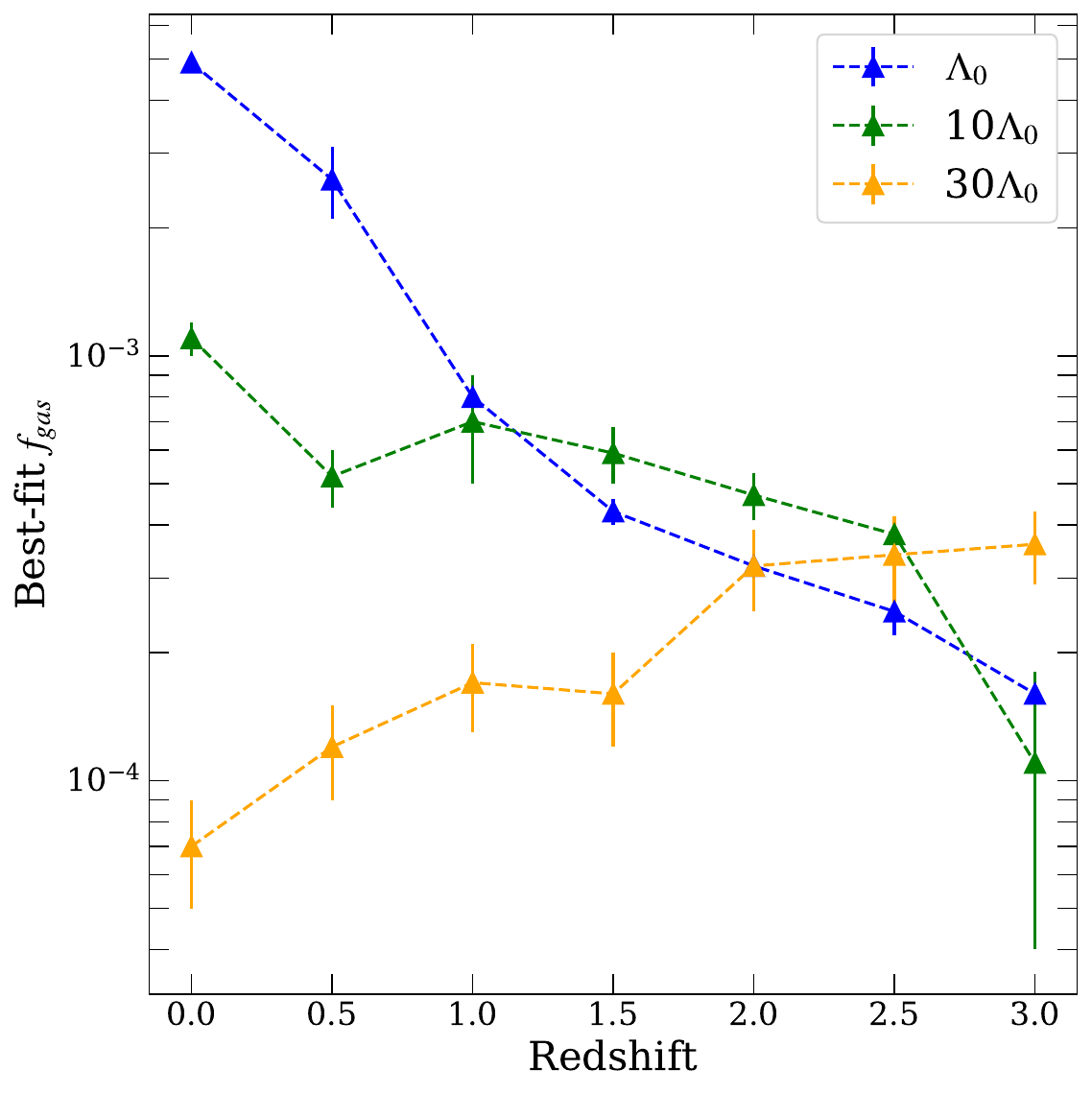} \\
    \includegraphics[width=0.42\textwidth]{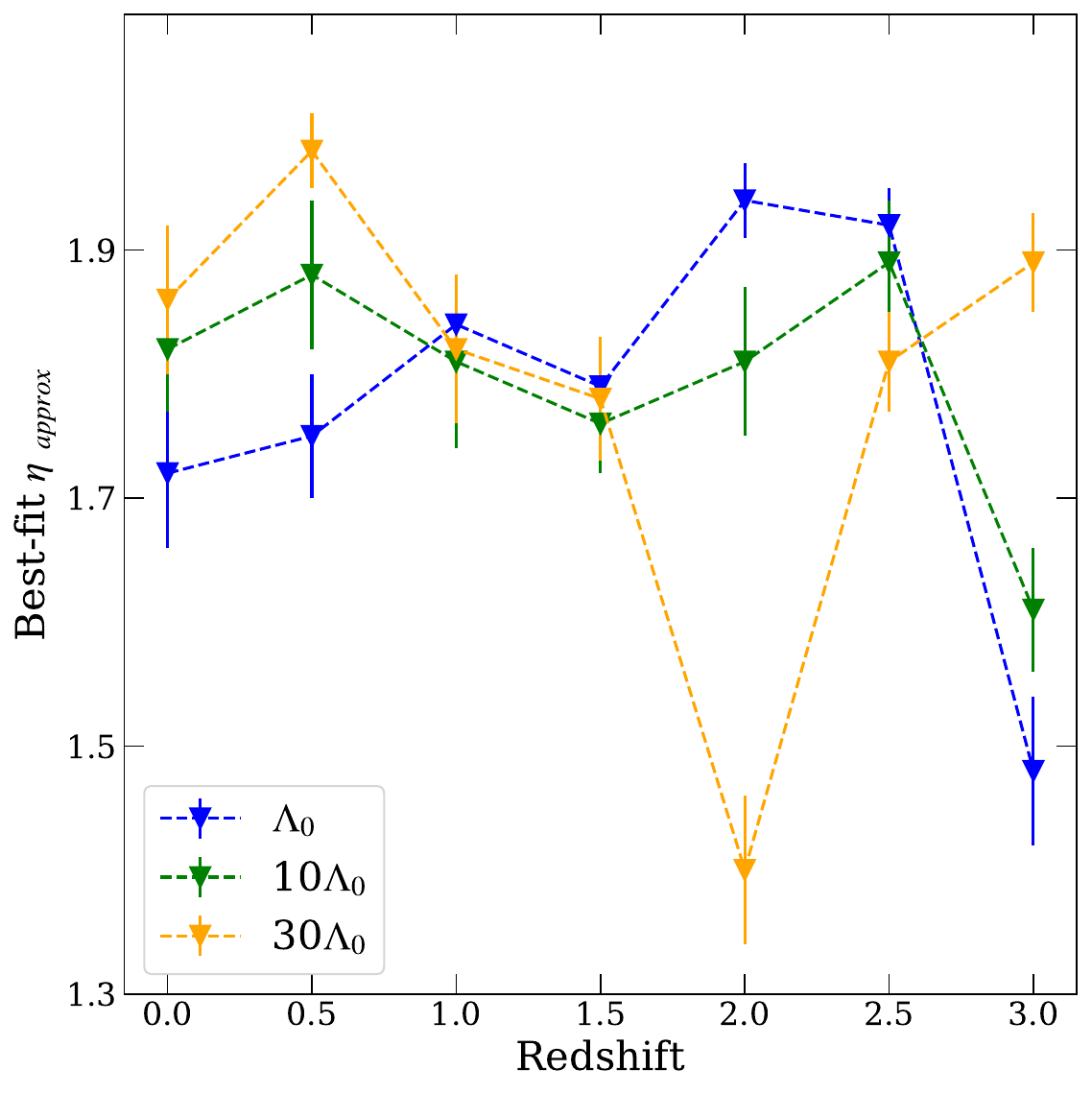}
    \includegraphics[width=0.42\textwidth]{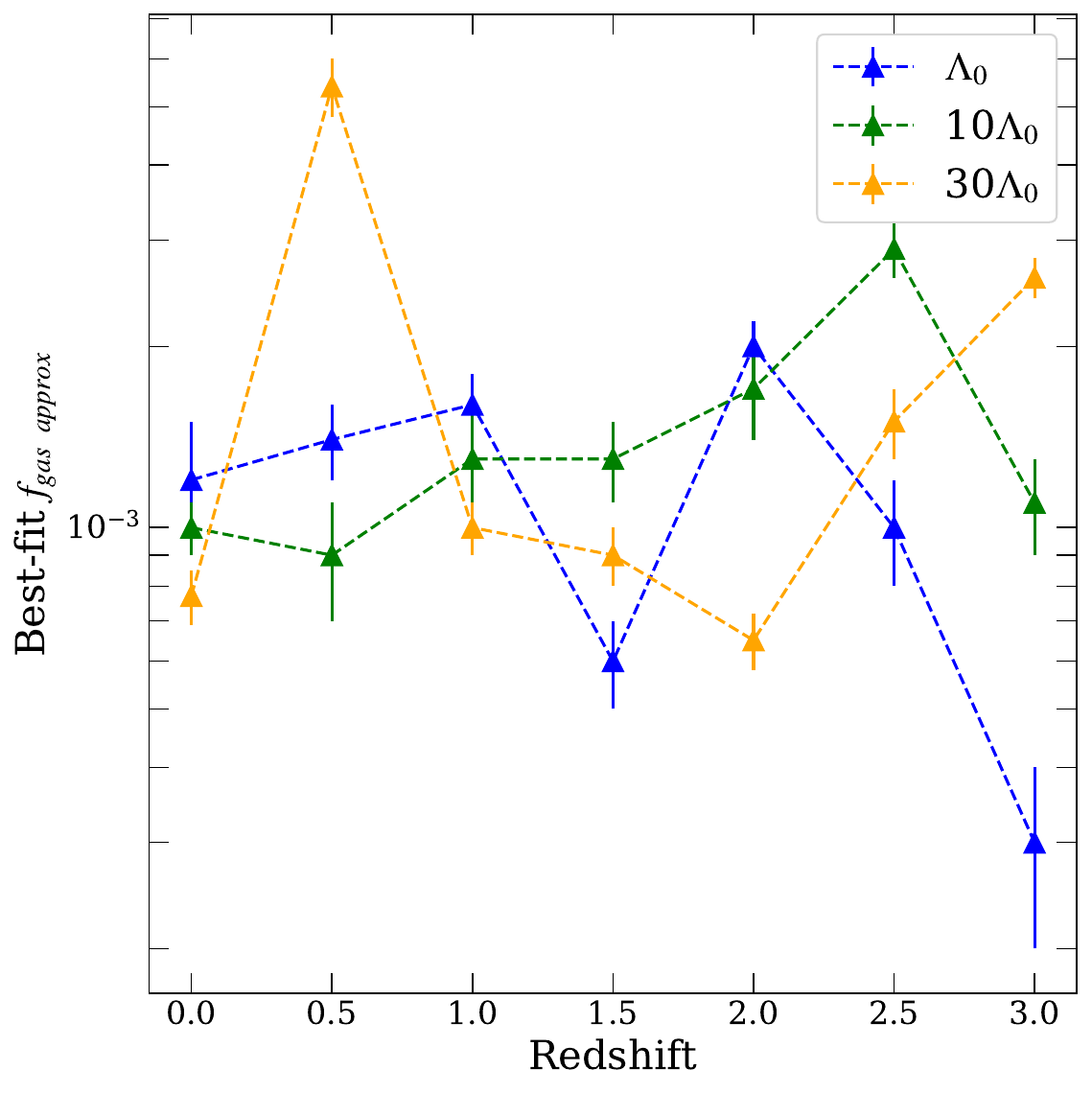} \\
    \includegraphics[width=0.42\textwidth]{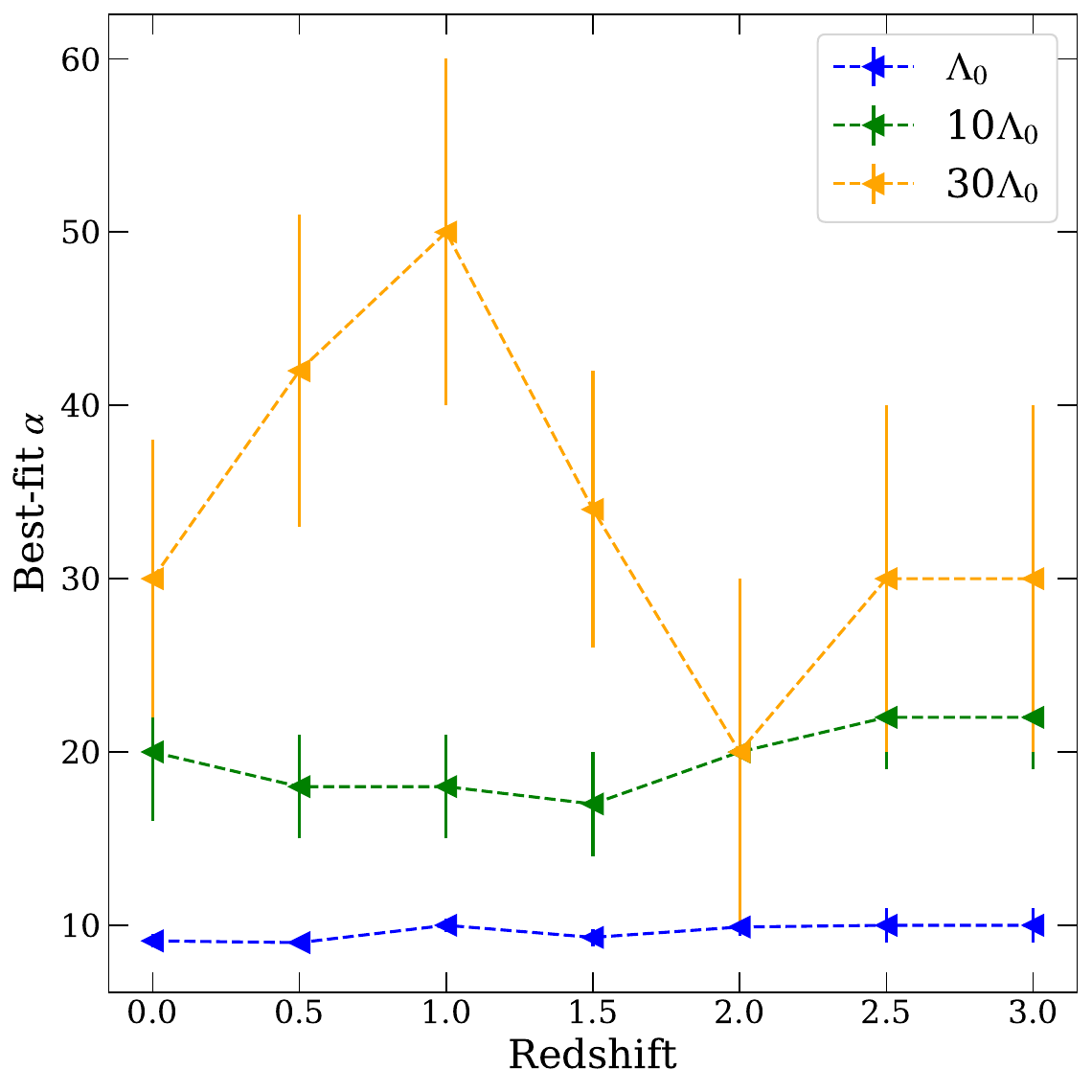}
    \includegraphics[width=0.42\textwidth]{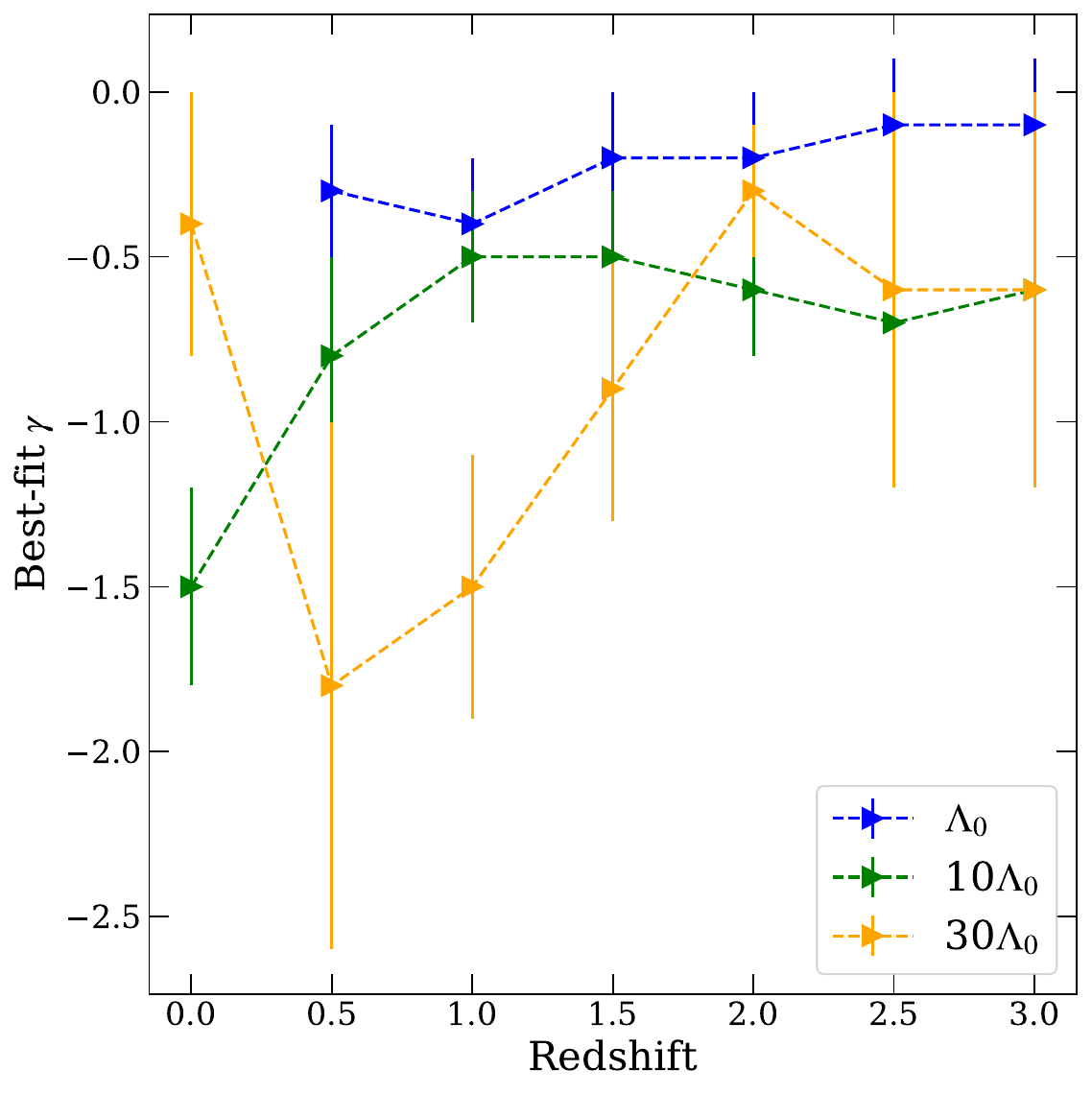}
    \caption{Best-fit parameter evolution with redshift taken from Table~\ref{tab:combinedparameters}.}
    \label{fig:2x2grid}
\end{figure*}

\begin{figure*}
    \centering
    \includegraphics[width=0.49\textwidth]{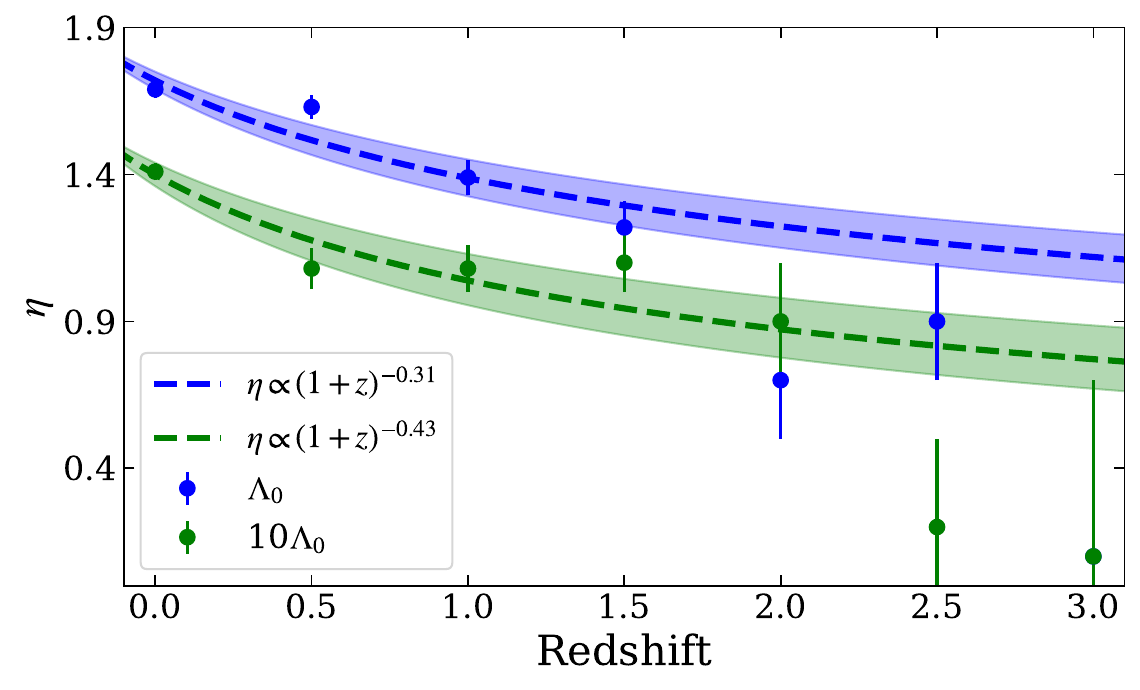}
    \includegraphics[width=0.49\textwidth]{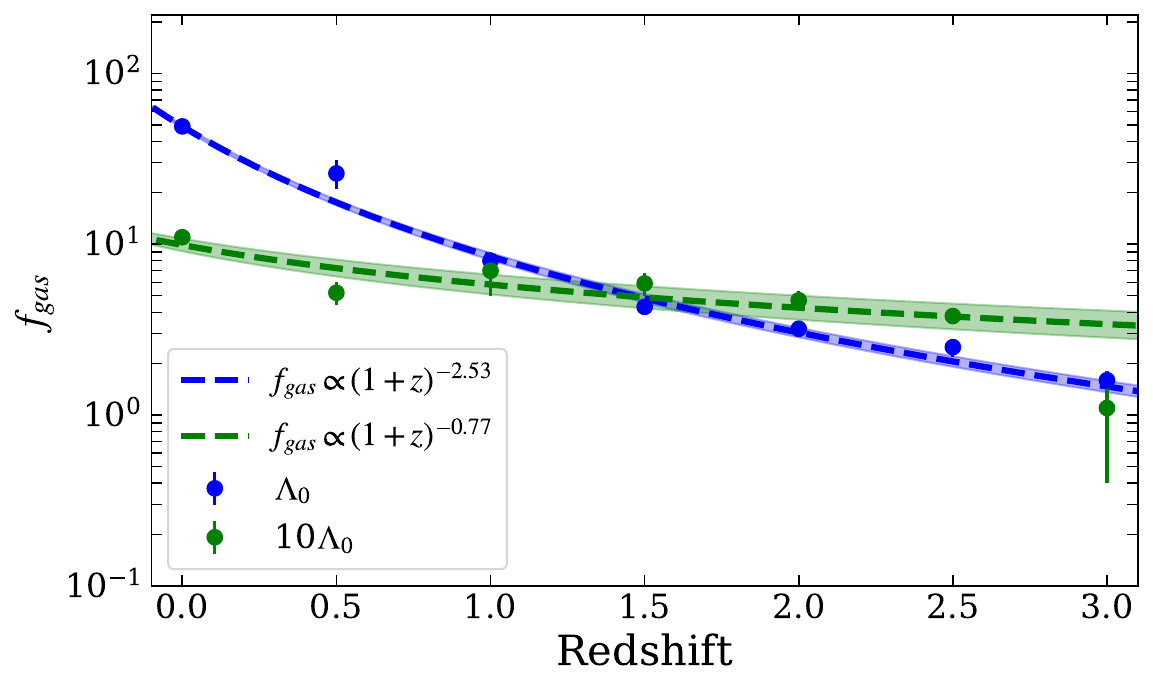} \\
    \caption{Best-fit $\eta$ and $f_{\text{gas}}/10^{-4}$ against redshift in the $\Lambda_0$ and 10$\Lambda_0$ simulations with a simple redshift dependence fit.}
    \label{fig:eta_fgas_z_correlation}
\end{figure*}

The best-fit values of the model parameters $\eta$ and $f_{\text{gas}}$ for the $\Lambda_0$ and $10\Lambda_0$ simulations exhibit clear redshift evolution, as shown in Table~\ref{tab:combinedparameters} and the top panels of Fig.~\ref{fig:2x2grid}. Specifically, $\eta$ and $f_{\text{gas}}$ decrease with redshift. A decreasing $\eta$ indicates that gas density declines more gradually with radial distance from the halo at higher redshifts. This suggests that over time, haloes tend to lose the gas which is closer to their centres, leading to a reduction in the size of their inner, gas-rich regions. Conversely, the gas fraction, $f_{\text{gas}}$, slowly increases with time, implying a growing fraction of gas mass within the virial radius—contrary to expectations given the increasing closure radii.

We suspect that this increase could be significant and may be linked to the unexpectedly small best-fit values of $f_{\text{gas}}$ found in our analysis. However, the underlying reason for this remains unclear. It is possible that the $f_{\text{gas}}$ parameter used here does not correspond directly to the true gas fraction. Some of the approximations made in our the formalism (see Section~\ref{sec:newmodel}), such as assuming the gas remains at the virial temperature, may affect the normalisation of the closure radius -- baryon mass fraction relationship, hence the value of $f_{\rm gas}$  derived numerically. 

The parameter errors for $\eta$ and $f_{\text{gas}}$ increase with redshift, indicating reduced model accuracy at earlier times. This is expected, as haloes at $z > 1$ generally have notably poorer defined closure radii, as shown by their increased errors in Fig.~\ref{fig:3x3grid}. The trends in $\eta$ and $f_{\text{gas}}$ are closely correlated: as $\eta$ increases, $f_{\text{gas}}$ typically also increases, and vice versa. Both parameters either diverge from or converge toward zero with changing redshift. A similar relationship is observed in the $30\Lambda_0$ simulation, though high scatter and larger errors make it unclear whether the parameters converge or diverge. 

It is clear that $\eta$ and $f_{\text{gas}}$ retain a degree of redshift evolution, suggesting that our new model has not fully captured the entire redshift dependence of the closure radius explicitly. However, we can easily show that fitting a simple function which is proportional to some power of $(1+z)$ provides good fits. This is shown in Fig.~\ref{fig:eta_fgas_z_correlation} for $\Lambda_0$ and 10$\Lambda_0$ (where there seems to be a clear monotonic redshift dependence).

Fig.~\ref{fig:eta_fgas_z_correlation} shows a fairly good fit for a simple redshift relation for both parameters, but especially $f_{\text{gas}}$, in both simulations. At higher redshift, $z > 2$, the $\eta$ relationship does not fit well within one error for either simulation; however, this could be due to most haloes not being properly relaxed or had time for dark energy or feedback to redistribute their baryons at those times. While the overall redshift evolution of the parameters is well described by the fits, the $(1+z)$ exponent is not constant between cosmologies, and was just an arbitrary fitting parameter with no physical intuition into what its value should be for either parameter. Nonetheless, one could substitute these redshift relations into equation~(\ref{eq:Rc_total_eta}) and possibly acquire a (complicated) fully redshift explicit relation in our new model. However, introducing further empirical, arbitrary fits may cause us to lose the physical intuition into how the closure radius evolves with time, and forcing such a function is not our aim.

\subsubsection{A linear fit to the EAGLE simulations}
\label{sec:approxsolutionresults}

While our model provides a good fit to the simulations considered, in most cases the closure radius -- baryon mass fraction relationship appears to exhibit a linear trend. This was indeed the observation that led \cite{Ayromlou_2023} to describe such correlation with a linear fit. It is therefore tempting to consider a simpler, linear formula to represent the numerical results, as long as this is also rooted in physical principles.

If we look back at equation~\eqref{eq:Rc_total}, we notice that the first factor on the RHS is of order 0.01 – 0.1 depending on the exact value of $n$ (which, as discussed in Section~\ref{sec:newmodel}, must be negative), since $\Delta$ is of order 100. Considering that all other factors are of order 0.1 – 1, it follows that, in general, the first term in the RHS of equation~(\ref{eq:Rc_total}) is much smaller than one, so that we can safely make the following approximation:
\begin{equation}
\label{eq:Rc_approx}
\begin{aligned}
    \frac{R_{\text{closure}}}{R_{200}} - 1 \approx \frac{n}{2+3n} 
    \left( 1 + \frac{1}{n} \right)^{\frac{1}{n}} 
R    \left( 1 + \frac{\Omega_{m0}}{\Omega_{\Lambda 0}}(1+z)^3 \right)^{\frac{1}{n}} \\
      \left( \frac{\Delta}{2} \right)^{\frac{1}{n}} \frac{f_{\text{b-cosmic}}}{f_{\text{gas}}} \left( 1 - \frac{f_{\text{b-halo}}}{f_{\text{b-cosmic}}} \right) \, .
\end{aligned}
\end{equation}

This approximate form of the model is linear with $1 - f_{\text{b-halo}}/f_{\text{b-cosmic}}$ and we now compare it with our numerical results and the predictions of our model in its most general incarnation (equation~\ref{eq:Rc_total}). For the same reasons explained in Section~\ref{sec:newmodel}, we replace $n$ with $\eta = -2/n$ in equation~\eqref{eq:Rc_approx} before seeking the best-fit values of the model parameters. To distinguish them from the $f_{\rm gas}$ and $\eta$ parameters obtained by fitting equation~\eqref{eq:Rc_total} to the simulations, we will call them $f_{\rm gas-approx}$ and $\eta_{\rm approx}$, respectively.

The linear fits are shown as the purple lines in Fig.~\ref{fig:3x3grid}. These results are also tabulated in Table~\ref{tab:combinedparameters}. Surprisingly, they show that $\eta_{\text{approx}}$ does not vary much - if at all, and a similar case could be made for $f_{\text{gas-approx}}$ not varying with redshift, but overall having a larger error.

Since the approximate model parameters appear to be constant (see also their panels in Fig.~\ref{fig:2x2grid}), it seems reasonable to calculate their average value in each cosmology. The results from Table~\ref{tab:combinedparameters} yield average values in the following cosmologies between $0 < z < 3$ of $\Lambda_0: \eta_{\text{approx}} = 1.83 \pm 0.09, f_{\text{gas-approx}} = 0.0013 \pm 0.0005$ (omitting $z=3$). $10\Lambda_0: \eta_{\text{approx}} = 1.83 \pm 0.05, f_{\text{gas-approx}} = 0.0017 \pm 0.007$ (omitting $z=3$). $30\Lambda_0: \eta_{\text{approx}} = 1.86 \pm 0.07, f_{\text{gas-approx}} = 0.002 \pm 0.002$ (omitting $z=2$). Across all three cosmologies and twenty one snapshots, this has an average of: $\eta_{\text{approx}} = 1.8 \pm 0.1, f_{\text{gas-approx}} = 0.002 \pm 0.001$, with a low standard deviation, showing a high level of constancy overall.

An $\eta = 1.8$ corresponds to $n = -1.1$, which is physically feasible. Instead, $f_{\text{gas}} = 0.002$ is quite small and harder to justify if this represents the true gas fraction. But we once again suggest that this is due to $f_{\text{gas}}$ here being a proxy for the actual gas fraction and its absolute value not corresponding to the physical gas mass fraction within the virialised halo.

The observation that our fit parameters, which initially seem to vary strongly with redshift and dark energy, become fairly precise constants under the linear approximation suggests that the unaccounted for dependencies of $\eta$ and $f_{\text{gas}}$ on $z$ and $\Lambda$ are largely second-order effects. This is an interesting result, indicating that the primary closure radius trend can be accurately captured based only on halo mass or radius for each cosmology.

\subsection{Dark energy impact on the closure radius}
\label{sec:darkenergyimpact}

Examining the fitted values of $\eta$ and $f_{\rm gas}$ from our full model, we find clear trends in Table~\ref{tab:combinedparameters}, Fig.~\ref{fig:3x3grid}, and particularly the top panels of Fig.~\ref{fig:2x2grid}: $\eta$ is consistently lower in simulations with larger $\Lambda$ across most redshifts. Since lower values of $\eta$ correspond to more extended gas density profiles, this indicates that baryons are distributed over larger distances from halo centres in cosmologies with higher dark energy densities. Consequently, haloes of comparable mass exhibit larger closure radii as $\Lambda$ increases.

We also find that $f_{\rm gas}$ decreases systematically with increasing $\Lambda$, indicating that haloes contain a smaller fraction of their baryons within $R_{200}$ in higher $\Lambda$ cosmologies. Importantly, we do not interpret this as dark energy directly ejecting gas from haloes. Rather, feedback processes are responsible for transporting baryons to large radii, while the presence of a larger cosmological constant suppresses the subsequent growth of structure and may reduce the efficiency with which this material is reaccreted. Within this picture, increasing $\Lambda$ leads to more extended baryon distributions and larger closure radii. Our model successfully captures these trends and reproduces the behaviour observed in the 1, 10, and 30 $\Lambda_0$ simulations.

In regard to the approximate form of our model, the relatively constant values of $\eta_{\text{approx}}$ and $f_{\text{gas-approx}}$ across redshift and cosmology are noteworthy, as it implies that first-order differences in closure radii may arise solely from variations in the density parameters $\Omega_m$ and $\Omega_\Lambda$ across simulations and redshifts, rather than direct changes in the gas behaviour. This leads to a universal relation for the closure radius, depending only on $z$, $R_{200}$, $\Omega_{m0}$, $\Omega_{\Lambda0}$, and $f_{\text{b-halo}}$. By setting $\eta_{\text{approx}} = 1.8$ and $f_{\text{gas-approx}} = 0.002$, the closure radius trend for haloes in each cosmology at any $0 < z < 3$ can be determined using equation~(\ref{eq:Rc_approx}). This approach requires only potentially observable quantities of redshift, virial radius (or mass), and halo baryon fraction, since now we impose $\eta_{\text{approx}}$ and $f_{\text{gas-approx}}$ remain constant. The approximate model is thus more versatile, predicting a constant radial gas density profile beyond the virial radius with a fixed slope given by $\eta_{\text{approx}} \sim 1.8$.

\begin{figure*}
    \centering
    \includegraphics[width=\textwidth]{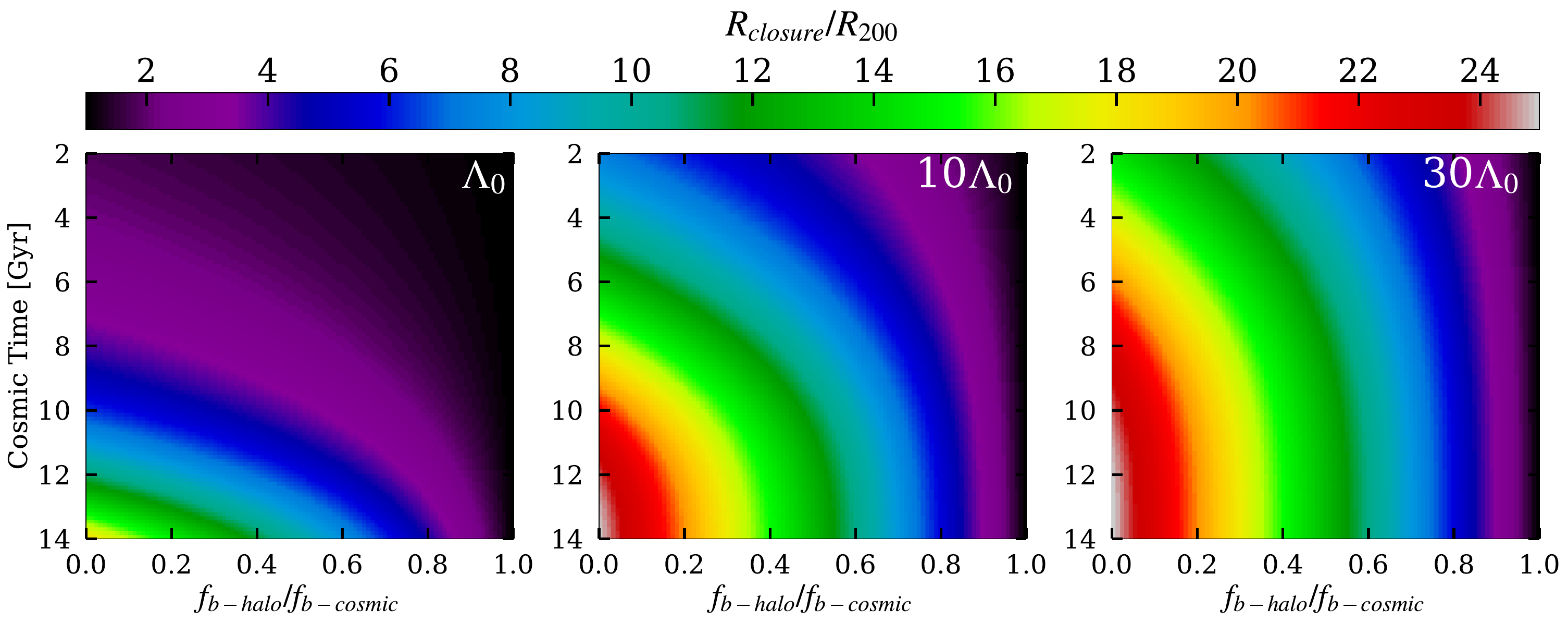}
    \caption{2D Histograms showing how the normalised closure radius varies for haloes depending on their cosmology, cosmic time and baryon fraction from our approximate model, equation~(\ref{eq:Rc_approx}) assuming constant $\eta_{\text{approx}} = 1.8$ and $f_{\text{gas-approx}} = 0.002$.} 
    \label{fig:contour_xt}
\end{figure*}

We can also show what constant $\eta_{\text{approx}}$ and $f_{\text{gas-approx}}$ predict for the normalised closure radii of haloes with different halo baryon fraction in each cosmology over cosmic time. This is shown in Fig.~\ref{fig:contour_xt}. This figure is plotted showing a variety of cosmic times instead of redshift to avoid inaccurate comparison of closure radii where significantly different amounts of cosmic time have passed. This demonstrates that our model successfully predicts the observed trends in the normalised closure radius, including the expected positive correlation with both elapsed cosmic time and dark energy. Consequently, our model indicates that in universes with a higher cosmological constant, closure radii can exceed tens of virial radii within just a few Gyr for haloes with modest baryon fractions. Naturally, in such universes, the increased dominance of dark energy also results in fewer baryons collapsing into haloes in the first place. This leads to a larger fraction of haloes with lower $f_{\text{b-halo}}$, which in turn produces even larger closure radii. Thus, our model, when assuming fixed $\eta_{\text{approx}}$ and $f_{\text{gas-approx}}$, effectively captures the complex interplay between cosmology and astrophysical feedback in driving the dispersal of halo baryons far beyond the virial radius.

These results also have wider implications for anthropic arguments \citep{barrow1991anthropic, Sorini_2024_cosmo}. In universes with larger $\Lambda$, haloes are rapidly characterised by extremely large closure radii, meaning that a substantial fraction of baryons is diffuse well into the IGM and is unable to be retained into haloes and form stars \citep{Oh_2021}. Since star formation is a prerequisite for stellar nucleosynthesis, which in turn produces the heavier elements necessary for planetary formation and life, such universes would be far less hospitable to the emergence of complex chemistry and biological systems. Consequently, the relationship between dark energy and the ability of haloes to retain their baryons could be an important factor in shaping the conditions for habitability. Thus, our model (which describes the relationship between dark energy and baryonic retention in haloes) may impose a constraint on the conditions necessary for the formation of observers. Analogous plots to Fig.~\ref{fig:contour_xt} for a wider range of cosmologies could therefore serve as a useful illustration of how the presence of dark energy influences the retention of baryons within haloes and, ultimately, the potential for life-supporting environments.

\section{Comparison with previous work}
\label{sec:previous_work}

\subsection{The A23 formula}
\label{sec:A23_model_limiting_case}

Previously \citet{Ayromlou_2023} proposed a simple formula for the closure radius:

\begin{equation}
\frac{R_{\text{closure}}}{R_{200}} - 1 = \alpha (1+z)^\gamma \left(1 - \frac{f_{\text{b}}(<R_{200})}{f_{\text{b-cosmic}}}\right) \, ,
\label{eq:A23model}
\end{equation}
where $\alpha$ and $\gamma$ are free parameters to be calibrated against simulations. The former is a normalisation constant that quantifies the factor by which the closure radius exceeds $R_{200}$ at $z = 0$. Meanwhile, $\gamma$ quantifies the redshift dependence of the closure radius; a more negative $\gamma$ indicates a faster growth of the closure radius with decreasing redshift. For convenience, \cite{Ayromlou_2023} incorporated the free parameters into the redshift-dependent function $\beta(z) = \alpha (1+z)^{\gamma}$. Equation~(\ref{eq:A23model}) will be referred to as the A23 formula from here on.

Moreover, \citet{Ayromlou_2023} evaluate their formula within a $\Lambda_0$ cosmology only, and until now, its validity and accuracy in universes with $\Lambda \neq \Lambda_0$ must be assumed -- which may not be valid. Furthermore, \citet{Ayromlou_2023} showed that their formula provides a good fit primarily for haloes with $M_{200} > 10^{13} M_{\odot}$ in the EAGLE, IllustrisTNG, and SIMBA simulations. These considerations suggest that while their formula is valuable, its predictive power for exploring the impact of dark energy on halo baryon distributions may be limited, particularly since higher $\Lambda$ universes are expected to contain fewer massive haloes in the mass regime where their fit has been calibrated.

The A23 formula provides an empirical description of the closure radius relation that has been shown to fit several simulation suites successfully. Our goal is not to replace this approach, but rather to explore whether some of its behaviour can be understood within a simple physically motivated framework. In contrast to the A23 parametrisation, our model introduces quantities related to the gas distribution and halo baryon content, providing a possible physical interpretation of the trends seen in the closure radius. However, these parameters are themselves calibrated using simulations, and their universality across different galaxy formation models remains to be tested.

As we show in Fig.~\ref{fig:3x3grid} (also shown as the purple dashed lines, as at fixed redshift our approximate linear model has the same form as the A23 fit and is therefore exactly identical to it.), the A23 formula can provide good fits for lower mass haloes, $M_{200} > 10^{11} M_{\odot}$, across various cosmologies within the redshift range $0 < z < 3$ in EAGLE, making it valuable for quantifying the effect of varying dark energy on the evolution of the closure radius and halo baryon distribution.

\begin{figure*}
    \centering
    \includegraphics[width=0.33\textwidth]{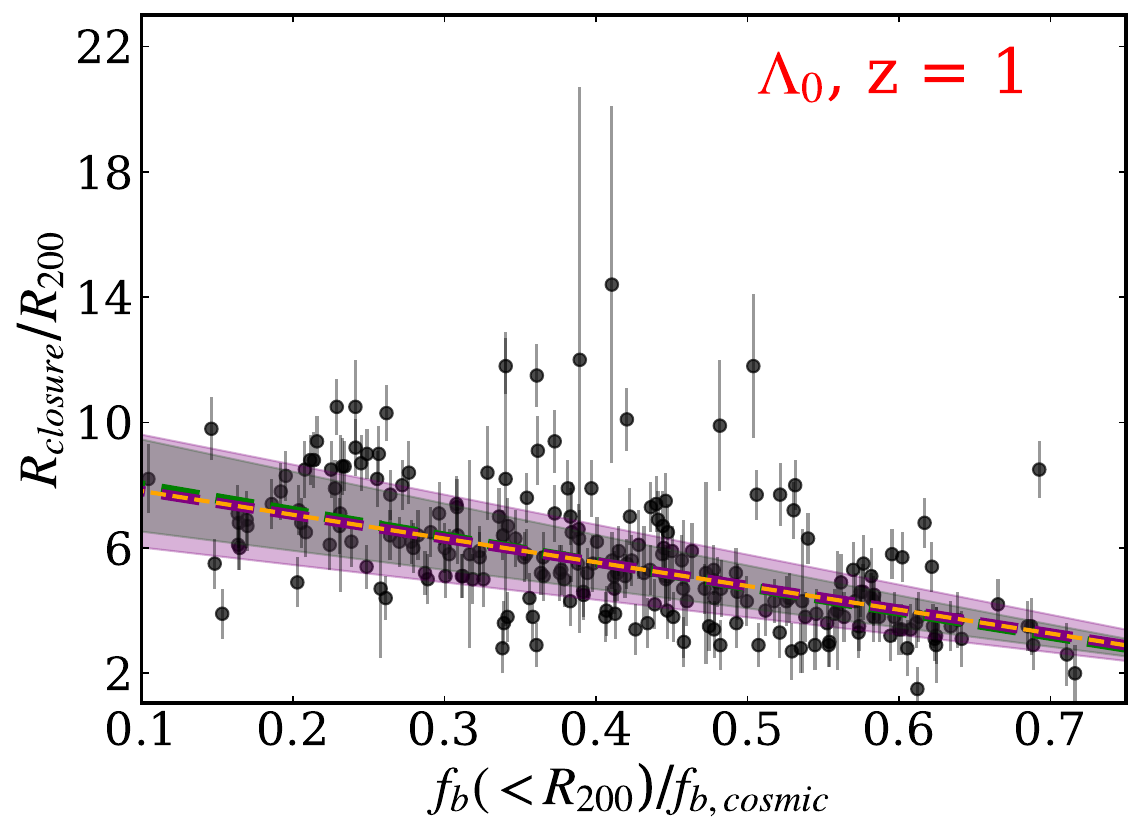}
    \includegraphics[width=0.33\textwidth]{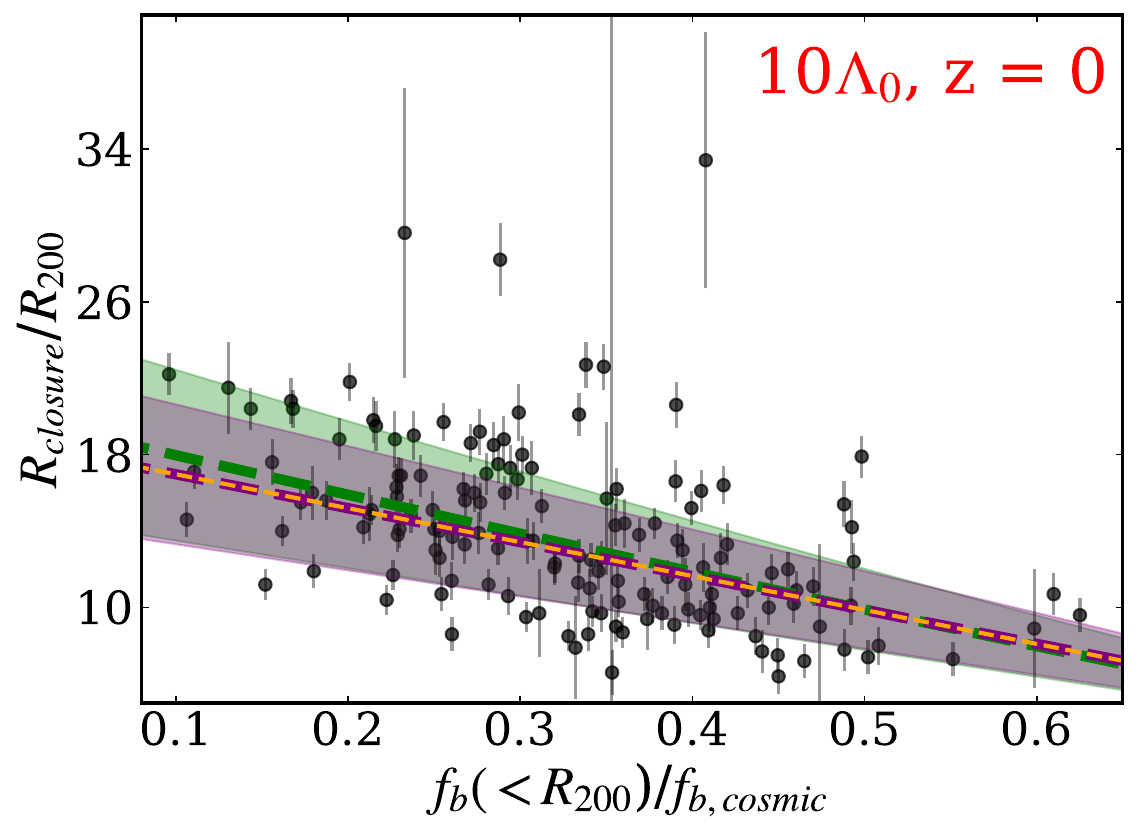}
    \includegraphics[width=0.33\textwidth]{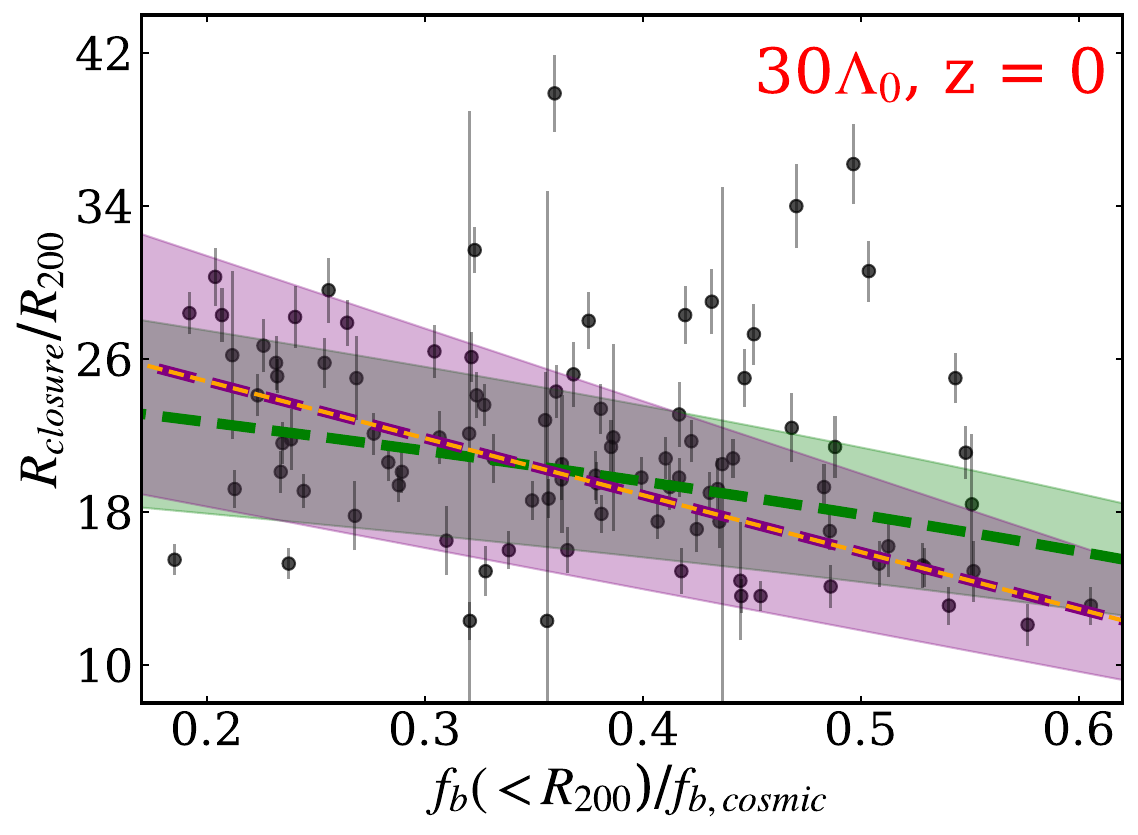}
    \caption{Relative closure radii against halo baryon fraction for three example snapshots. Green lines show the best-fit full solution of the new model, equation~(\ref{eq:Rc_total_eta}), purple lines the  best-fit A23 formula, equation~(\ref{eq:A23model}), and orange lines show the best-fit approximation of the new model, equation~(\ref{eq:Rc_approx}), which has a tight best-fit $\eta = 1.8$ and $f_{\text{gas}} = 0.002$ across all cosmologies and redshift. Our approximation agrees completely with the A23 formula in all snapshots (with just three plotted here as examples), as they are both linear at fixed redshift.}
    \label{fig:threelimitplots}
\end{figure*}

As discussed, at fixed redshift, our approximation (equation~(\ref{eq:Rc_approx})) agrees with the A23 formula for the functional form of the closure radius-virial radius relation -- although, as expected, they differ with changing $z$. This is shown explicitly in Fig.~\ref{fig:threelimitplots} with the A23 formula (fit just to each snapshot) in purple and the approximate model with fixed $\eta_{\text{approx}}$ and $f_{\text{gas-approx}}$ in dashed orange.

\subsubsection{A23 formula fits}

Since the A23 formula shares the exact functional form as our approximate model at fixed redshift, its fits are also shown by the purple lines in Fig.~\ref{fig:3x3grid}, and lead to similar conclusions. Namely, this formula effectively captures the overall trend by approximating the closure radius–halo baryon fraction relation as a straight line. While it provides a good fit across most redshift snapshots and cosmologies, it is outperformed (through having worse chi-squared values) by the full model, which is more flexible and better suited to capturing non-linear trends evident in many snapshots.

Fig.~\ref{fig:2x2grid} also illustrates the redshift evolution of the fitted $\alpha$ and $\gamma$ across the 1, 10, and 30 $\Lambda_0$ simulations. Notably, $\alpha$ appears largely invariant with redshift in the 1 and 10$\Lambda_0$, and potentially 30$\Lambda_0$ simulation as well, though the larger uncertainties make the latter less certain. A similar argument could be made for $\gamma$, but the consistently large uncertainties across all simulations make its constancy less definitive than that of $\alpha$. If these parameters are indeed redshift independent, it would align with our findings from fitting $\eta_{\text{approx}}$ and $f_{\text{gas-approx}}$ in our approximate model. Since both are linear at fixed redshift, it is unsurprising that their fit parameters exhibit similar constancy. At first consideration, this might suggest that the $\alpha(1+z)^\gamma$ relation fully encapsulates redshift evolution in the closure radius relation. However, our full model indicates that best-fit parameters do vary with redshift, implying that this simple scaling does not capture the entire picture. Nonetheless, the agreement between the A23 formula and our approximate model highlights their effectiveness in providing a reasonable first-order fit to the closure radius distribution, with the added advantage that they may only need to be calibrated at a single redshift snapshot per simulation to capture the overall trend.

\begin{figure}
    \includegraphics[width=\columnwidth]{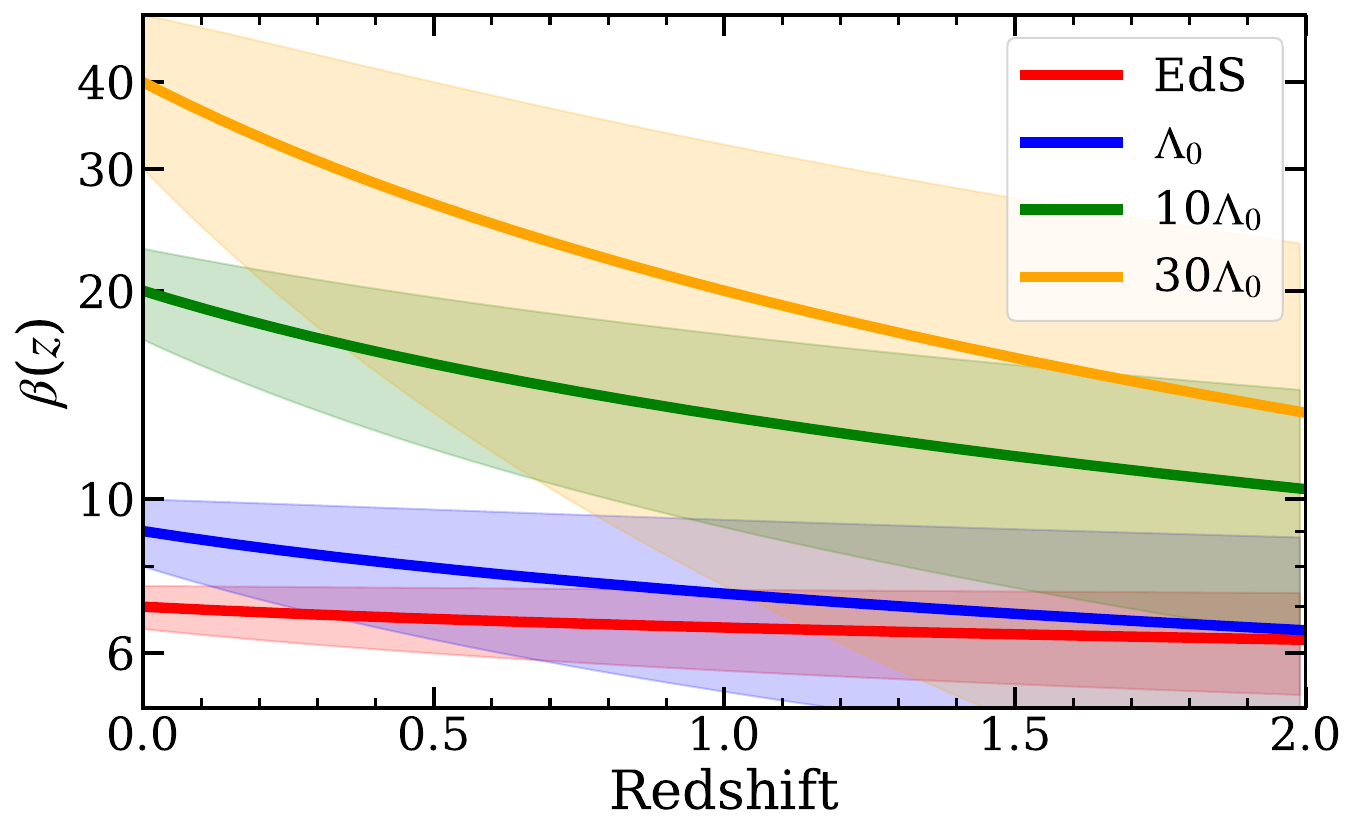}
    \caption{Best fit $\beta(z)=\alpha(1+z)^\gamma$ and error from each simulation for $\log_{10}\left(M_{200}/M_{\odot}\right) >$ 11. This approximates the factor by which $R_{\text{closure}}$ exceeds $R_{200}$ at different $z$. The errors on the parameters also increase with $\Lambda$, indicating the A23 formula breaks down in high $\Lambda$ universes.}
    \label{fig:bestfitbeta}
\end{figure}

Fig.~\ref{fig:2x2grid} further reveals that $\alpha$ tends to be larger in simulations with higher $\Lambda$, indicating an increased normalised closure radius, $R_{\text{closure}} / R_{200}$ — a trend that aligns with expectations. Additionally, $\gamma$ generally becomes more negative as $\Lambda$ increases, suggesting a faster growth of the closure radius with redshift, which is again consistent with physical intuition. However, in order to more comprehensively compare the effect of changing $\Lambda$ on the A23 formula parameters (which are instead somewhat constant in $z$), we follow the approach by \citealt{Ayromlou_2023} by calculating the constant average best-fit parameters $\alpha$ and $\gamma$ through fitting $\beta(z)$ to every simulation ($0 < z < 2$), except from $100 \Lambda_0$. These results are presented in Table~\ref{tab:alphagammatable}. Additionally, to compare the redshift closure radius trend in each cosmology within their formalism, each best-fit $\beta(z)$ function is also plotted in Fig.~\ref{fig:bestfitbeta}.

\begin{table}
    \centering
    \begin{tabular}{|c|c|c|}
        \hline
        $\Lambda / \Lambda_0$ & $\alpha$ & $\gamma$ \\
        \hline
        0 (EdS) & 7.0 $\pm$ 0.5 & -0.1 $\pm$ 0.1 \\
        \hline
        1 & 9 $\pm$ 1 & -0.3 $\pm$ 0.1 \\
        \hline
        10 & 20 $\pm$ 3 & -0.6 $\pm$ 0.3 \\
        \hline
        30 & 40 $\pm$ 10 & -1.0 $\pm$ 0.7 \\
        \hline
    \end{tabular}
    \caption{Best-fit $\alpha$ and $\gamma$ (averaged over all snapshots in the redshift range 0-2) in each 25 Mpc simulation with varying $\Lambda$. $\alpha$ becomes larger, quantifying how much further baryons are pushed from haloes, and $\gamma$ becomes more negative, quantifying the rate at which more baryons are pushed far from haloes over time, with increasing dark energy.}
    \label{tab:alphagammatable}
\end{table}

These results quantify the extent to which the factor $R_{\text{closure}} / R_{200}$ increases across the available simulations due to changes in $\Lambda$ and indicate a positive correlation between $\alpha$ and $\Lambda$, aligning with theoretical expectations.

Table~\ref{tab:alphagammatable} also confirms that the average $\gamma$ becomes more negative with heightened $\Lambda$. This implies that $R_{\text{closure}}$ expands more rapidly with decreasing redshift (increasing time) in universes with greater vacuum energy. The physical reason likely stems from higher $\Lambda$ universes (10 and 30 $\Lambda_0$) entering vacuum-dominated states before $0 < z < 2$, Fig.~\ref{fig:timeevolution}. Consequently, spacetime expansion is already accelerating, and as time progresses, spacetime expands by a greater factor compared to universes with less vacuum energy. Consequently, its effect of pushing non-gravitationally bound baryons further from haloes intensifies. Thus, $R_{\text{closure}}$ increases more rapidly with decreasing $z$, a trend quantified by the decrease in $\gamma$.

These trends can be understood through our model too if one considers that $\eta$ is typically lower than 1.7 in its best fit values in Table~\ref{tab:combinedparameters} in the larger $\Lambda$ simulations. As discussed previously, $0 < \eta < 1.7$ is monotonic in causing the relative closure radius to increase. Hence, larger dark energy causes the gas density profile to be more stretched from the centre, thus causing more baryons to lie beyond the virial radius. This was discussed more in Section~\ref{sec:newmodelresults}, thus the results from applying our new model corroborate these findings from applying the A23 formula.

Another observation is the notable increase in parameter uncertainties, as displayed in Table~\ref{tab:alphagammatable}, indicating heightened error when applying the A23 formula to more extreme cosmologies with elevated $\Lambda$. This formula was not designed to specifically address the effects of vacuum energy, exacerbated by the tendency of higher $\Lambda$ universes to exhibit fewer haloes at higher masses (Fig.~\ref{fig:HMF}) with less well-defined closure radii. Consequently, this results in significantly poorer minimum chi-squared values when applying this formula, leading to substantial increases in error ranges. This is unsurprising and could impact conclusions. For instance, while $\gamma$ appears to become more negative with increased $\Lambda$, the magnitude of error on $\gamma$ is often of the same order of $\gamma$ itself, raising doubts about the certainty of the declining $\gamma$ trend. It is plausible that $\gamma$ could remain constant, such as $\gamma = -0.2$, across all cosmologies, aligning somewhat with most findings. Thus, drawing accurate conclusions about $\gamma$ is constrained by error size. Improvements could involve running larger simulations, as the effect of $\Lambda$ is most prevalent on the largest scales. Conducting simulations with intermediate $\Lambda$ values for broader cosmology comparison, and simulations with more particles leading to smoother individual halo baryon distributions and lower closure radii errors, would likely improve accuracy in such analyses.

Comparing the results of the EdS universe to the $\Lambda_0$ universe provides valuable information into the specific impact of the vacuum energy on halo baryon distribution in our Universe. The distinction between EdS and $\Lambda_0$ is particularly important, reflecting differences solely attributed to varying $\Lambda$. Remembering all simulations share identical initial conditions and strength of subgrid physics models for feedback, ensuring any disparities originate solely from changing $\Lambda$. Table~\ref{tab:alphagammatable} indicates a slight difference in $\gamma$ with almost overlapping error ranges, but a statistically significant increase in the best-fit $\alpha$ beyond one error when incorporating $\Lambda_0$ effects from the EdS simulation. This correlates to a rise in the average ratio of $R_{\text{closure}}$ to $R_{200}$ (from 7.0 to 9) due to the dark energy presence in our Universe. This result quantifies the impact of dark energy relative to other processes. Assuming astrophysical feedback and gas pressure as the sole contributors to baryon dispersion beyond $R_{200}$ in the EdS universe (yielding $\alpha = 7.0$), including dark energy $\Lambda = \Lambda_0$ results in a marginal increase in $\alpha$. Consequently, dark energy's role in expelling baryons from haloes (post cosmic-noon, $z < 2$) is estimated to be of order 30\% of that attributed to astrophysical feedback and gas pressure ($2 / 7.0 \sim 30\%$). Thus, a key takeaway is that while dark energy plays a lesser role compared to established feedback mechanisms, it significantly contributes to pushing baryons far from haloes and contributes to the missing baryon problem \citep{2010ApJ...708L..14M} in our Universe, as inferred from applying the approximate form of our model and the A23 formula within an EAGLE suite of hydrodynamical simulations. 

Our conclusion gains credibility as both EdS and $\Lambda_0$ simulations exhibited the lowest chi-squared values and smaller best-fit parameter uncertainties out of all simulations, indicating the model accurately captures physics in these simulations. A difference of one sigma is notable, but further study is necessary to confirm this conclusion. Future improved simulations (as previously mentioned) or testing whether this result arises in other simulation suites would be beneficial for testing the robustness of our conclusions regarding the impact of cosmological parameters, under different models of galaxy formation.

One impact of not having a constant $\eta$ and $f_{\text{gas}}$ with redshift is now we cannot construct an exact analogous $\beta(z) = \alpha (1+z)^\gamma$ function directly from our model to compare to. This was extra motivation to investigate the approximate solution to our model which does yield a straight line at fixed redshift and was then found to yield best-fit parameter values which appear constant with redshift. The approximate model is in full agreement with the A23 formula at fixed redshift, when comparing normalised closure radii to normalised halo baryon fractions, Fig.~\ref{fig:threelimitplots}, explaining why the approximate model parameters fit constant values within a single cosmology as $\alpha$ and $\gamma$ also do not appear to vary. However, the surprising result is that $\eta_{\text{approx}}$ and $f_{\text{gas-approx}}$ also remain constant across different cosmologies, suggesting that $\alpha$ and $\gamma$ are highly sensitive to changes in cosmological parameters, while $\eta_{\text{approx}}$ and $f_{\text{gas-approx}}$ are mildly independent of cosmology. This makes $\eta_{\text{approx}}$ and $f_{\text{gas-approx}}$ the more robust choice for calculating model closure radii.

\subsubsection{High redshift limit}

We can also show that the approximate model can be forced to take the exact same functional form as the A23 formula even with variable $z$ by considering different limits.

In the high-$z$ limit, $(1 + z) \gg 1$ causes equation~(\ref{eq:Rc_approx}) take the exact functional shape of the A23 formula, equation~(\ref{eq:A23model}), with the following relations for $\alpha$ and $\gamma$:

\begin{equation}
\label{eq:alphahighz}
\alpha(z) = \frac{n(z)}{2+3n(z)} \left(1+ \frac{1}{n(z)}\right)^{\frac{1}{n(z)}} \left(\frac{\Delta}{2}\right)^{\frac{1}{n(z)}} \left(\frac{\Omega_{m0}}{\Omega_{\Lambda 0}} \right)^{\frac{1}{n(z)}} \frac{f_{\text{b-cosmic}}}{f_{\text{gas}}(z)} ,
\end{equation}

\begin{equation}
\label{eq:gammahighz}
\gamma(z) = \frac{3}{n(z)} .
\end{equation}

\subsubsection{Low redshift, $\Lambda \sim \Lambda_0$ limit}

In the low-$z$ limit we have $z \ll 1$ and so $(1 + z)^3 \sim 1 + 3z$. Coincidentally, for our universe, $\Lambda = \Lambda_0$, we find $\Omega_{\Lambda0} /3 \Omega_{m0} + 1/3 = 1.058$, and so we can safely approximate it to unity. This again leads to a function with the same form as the A23 formula with the following expressions for $\alpha$ and $\gamma$:

\begin{equation}
\label{eq:alphalowz}
\alpha(z) = \frac{n(z)}{2+3n(z)} \left(1+ \frac{1}{n(z)}\right)^{\frac{1}{n(z)}} \left(\frac{\Delta}{2}\right)^{\frac{1}{n(z)}} \left(3\frac{\Omega_{m0}}{\Omega_{\Lambda 0}}\right)^{\frac{1}{n(z)}} \frac{f_{\text{b-cosmic}}}{f_{\text{gas}}(z)} ,
\end{equation}

\begin{equation}
\label{eq:gammalowz}
\gamma(z) = \frac{1}{n(z)} .
\end{equation}

We conclude that the A23 formula represents a limiting case of our full closure radius model. However, since both $f_{\text{gas}}(z)$ and $\eta(z)$ may exhibit redshift dependence — as demonstrated in Section~\ref{sec:newmodelresults}, there is no direct correspondence between our model and their $\beta(z)$ function, as the redshift dependencies differ fundamentally. Consequently, it would be incorrect to use these relations between $\alpha$, $\gamma$, and the parameters $\eta$ and $f_{\text{gas}}$ to convert between the two models while expecting equivalent values.

In the A23 formula, all redshift dependence is assumed to be encapsulated by the $(1+z)^\gamma$ term, implying that $\alpha$ and $\gamma$ remain constant with redshift and fully describe the evolution through their $\beta$ function: $\beta(z) = \alpha (1+z)^\gamma$. In contrast, our formalism imposes no prior assumption on the redshift dependence of $\eta(z)$ and $f_{\text{gas}}(z)$. Thus, unless one can fully disentangle the redshift dependence of both $\eta(z)$ and $f_{\text{gas}}(z)$ and demonstrate that it precisely follows the form of $\beta(z)$, any direct conversion between the two parameter sets should be avoided, as they do not correspond exactly in either limit. Nevertheless, we include these limits here as a matter of theoretical interest, as it shows that for our Universe the A23 function can be derived from our model, both at high and low redshift.

\subsubsection{Testing redshift limits}

Earlier, we derived expressions for $\alpha$ and $\gamma$ in terms of $\eta_{\text{approx}}$ and $f_{\text{gas-approx}}$ under specific limits, as given in equations~\eqref{eq:alphahighz}-\eqref{eq:gammalowz}. Here, we explicitly evaluate the accuracy of the low- and high-$z$ limits of the approximate solution. For $\eta = 1.8$, the corresponding $n = -1.1$. In the low-$z$ limit, where $\Lambda \sim \Lambda_0$, we expect $\gamma \sim 1/n$. Substituting $n = -1.1$ yields $\gamma = -0.9$. However, comparing this theoretical value with the average $\gamma$ values in low-$z$ snapshots ($z < 1$) from Table~\ref{tab:combinedparameters}, we find significant disagreement, with $\gamma$ closer to -0.35. Furthermore, our previous analysis of the A23 formula across $0 < z < 3$ indicates that $\gamma$ remains nearly constant at $\sim -0.3$. Thus, the low-$z$ limit does not appear to be a reliable approximation. For the high-$z$ limit, where $\gamma = 3/n$, using $n = -1.1$ gives $\gamma = -2.7$, which similarly fails to match any best-fit $\gamma$ values in the snapshots. Calculations of $\alpha$ in each limit show similar inconsistencies. These results suggest that neither the low- nor high-$z$ limits provide concurring parameter values when substituting the approximate model directly.

Directly fitting for $\eta_{\text{approx}}$ and $f_{\text{gas-approx}}$ in the limits instead reveals that $\eta$ remains largely invariant with redshift and cosmology, suggesting it is likely a constant. This aligns with the approximate model  discussed previously. Across the redshift range $0 < z < 3$, the average fitted value is $\eta = 1.96 \pm 0.03$, derived from both low- and high-$z$ limits (which surprisingly agree closely with each other within this uncertainty). While slightly more negative than the $\eta_{\text{approx}} = 1.8 \pm 0.1$ found earlier, its value remains consistent in magnitude and constancy. Using this value of $\eta$, we can estimate $\gamma$ in the high- and low-$z$ cases, yielding $\gamma_{\text{low-z}} \sim -0.9$ and $\gamma_{\text{high-z}} \sim -2.9$. Again, comparing these results to Tables~\ref{tab:alphagammatable} and~\ref{tab:combinedparameters} reveals discrepancies. Specifically, $\gamma$ is closer to zero in low-redshift, low-$\Lambda$ snapshots, decreasing to approximately -1 in higher-$\Lambda$ simulations, with only a weak dependence on redshift. While the limits predict $\gamma$ to be more negative than observed, they do correctly suggest that $\gamma$ becomes more negative with increasing $\Lambda$ (consistent with the high-$z$ limit assuming a large $\Lambda$ value). Thus, while the trends align, the exact parameter values differ.

The high- and low-redshift limits for $f_{\text{gas}}$ do not predict constancy. Instead, $f_{\text{gas}}$ typically decreases slowly with increasing redshift, taking values roughly ten times larger than the $f_{\text{gas-approx}}$ values found previously. The trend of $f_{\text{gas}}$ increasing over time agrees with the full model, as shown in Fig.~\ref{fig:eta_fgas_z_correlation}. This indicates that $\alpha$ is also weakly correlated with time/redshift. Furthermore, the values of $f_{\text{gas}}$ obtained from fitting the low-$z$ limit are consistently slightly larger in every snapshot, which further suggests that $f_{\text{gas}}$ increases over time. This is consistent with expectations, as the low-$z$ limit is most accurate at later times.

Determining the utility of the redshift limits for comparing our parameters, $\eta$ and $f_{\text{gas}}$, with those from \citet{Ayromlou_2023}, $\alpha$ and $\gamma$, is challenging. Nonetheless, the trends largely align with the redshift evolution of all parameters, suggesting that $\alpha$ and $\gamma$ are not constant across cosmologies as assumed, though this change is obscured by significant errors in the $\alpha-\gamma$ parameter space. These limits demonstrate that the A23 functional form can emerge from our analytic framework under specific assumptions and limiting cases. This connection provides a useful link between the empirical parametrisation of A23 and the physically motivated quantities appearing in our model.

Furthermore, the corresponding $\alpha$ and $\gamma$ found from substituting these values of $\eta$ and $f_{\text{gas}}$ into equations~\eqref{eq:alphahighz}-\eqref{eq:gammalowz} do not agree with those found from fitting the A23 formula in  Table~\ref{tab:alphagammatable}. The reason for this was already discussed when these limits were derived in Section~\ref{sec:A23_model_limiting_case} and is because the A23 formula imposes a total redshift dependence of the form $(1+z)^\gamma$ on the closure radius relation, however as we have shown in our model, the real redshift dependence is more complex, with $\eta$ and $f_{\text{gas}}$ themselves also having redshift dependence (Fig.~\ref{fig:2x2grid}), giving the total closure radius relation a complicated form, with no easy (or more likely correct) way to factorise into the A23 redshift relation form. One could force $\eta$ and $f_{\text{gas}}$ to have simple $(1+z)^b$ relations for different $b$, e.g. Fig.~\ref{fig:eta_fgas_z_correlation}, and redefine our $\gamma'$ to match that of A23, $\gamma$, such as $\gamma = \gamma' - b_\eta - b_{f_{\text{gas}}}$. This gives $\gamma$ values closer to A23, but this method lacks physical intuition and only purpose is to force a relation similar to a previous empirical formula which is not the purpose of this paper. 

Our model reproduces the closure radius trends observed in the simulations with comparable or better accuracy to the existing parametrisations, while additionally providing a framework in which quantities such as the gas density profile slope and halo gas fraction can be related to the resulting closure radius.

Furthermore, our model can be applied across cosmologies with different $\Lambda$ within EAGLE simulations, redshift and agrees with the A23 formula under certain approximations at fixed redshift, and also provides physical meaning to the factors affecting the baryon dispersion of haloes, such as how the gas density profile index, $\eta$, and $f_{\text{gas}}$ being related to the true gas fraction of haloes, affects the closure radius.

A limitation of our present study is that the model has only been calibrated and tested using the EAGLE simulation suite considered here. Since closure radii are ultimately influenced by baryonic feedback processes, the fitted values of $\eta$ and $f_{\rm gas}$ may differ in simulations employing substantially different feedback prescriptions. Assessing the robustness of this model across alternative simulation suites remains an important avenue for future work.

\subsection{Implications for anthropic reasoning}
\label{sec:Further_implications}

Our formalism extends the cosmic star formation history model from \citet{Sorini_2021}, which was also shown to effectively describe the evolution of baryon mass fractions in haloes. In our derivation, we examined just the simple power-law gas density distribution as a solution to the low-density limit of equation~(\ref{eq:generalODE}). Recently, \citet{Sorini_2024} demonstrated that a power-law baryon density profile provides an excellent fit for haloes within the mass and redshift range considered here within SIMBA hydrodynamical simulations, extending out to the virial radius. Our results now confirm that this assumption also leads to accurate closure radius fits, showing that the power-law gas density model not only describes baryon distribution within haloes but also captures the scale at which baryons are expelled far beyond the virial radius and conventional halo boundaries.

We emphasise that our model is designed to be robust across a wide range of potential values for $\Lambda$, making it independent of specific cosmological parameter choices. This adaptability is crucial, as it allows us to quantify baryon evacuation trends without being tied to a particular cosmology. Moreover, since our model successfully captures closure radii trends across vastly different dark energy densities suggests a deeper, underlying physical mechanism governing baryon expulsion, rather than one finely tuned to our specific Universe. This has intriguing anthropic implications - if structure formation and baryon retention are strongly influenced by $\Lambda$, then our model provides a framework to explore the limits of habitability in alternate cosmologies. The ability to describe these trends with minimal cosmological assumptions also suggests that the closure radius can serve as a fundamental tracer of halo evolution across a broad range of universes, reinforcing the idea that the formation and distribution of baryons follow a predictable behaviour regardless of the specific value of $\Lambda$.

In the context of other previous work, \citet{Salcido_2018} and \citet{Barnes_2018} investigated the impact of varying $\Lambda$ within these specific EAGLE simulations, demonstrating that at late times, dark energy suppresses star formation by a small but measurable amount, thus providing an initial indication that $\Lambda$ influences baryonic astrophysics in our own Universe. Our study extends this by quantitatively assessing its role in baryon evacuation from haloes, showing that its impact reaches approximately $30\%$ of that caused by astrophysical feedback processes. Additionally, \citet{salcido2020feedback} developed an analytical model linking the growth of stellar mass in galaxies to the growth of their host dark matter haloes. Their model is governed by baryonic astrophysics regulating star formation, whereas our work investigates the subsequent removal of those baryons from haloes due to stellar feedback. Both studies highlight the critical role baryons play in shaping halo evolution over cosmological timescales. Similarly, \citet{oh2020calibration} examined feedback effects within ENZO simulations, showing that feedback energy significantly influences halo baryon content. Our work expands upon this by demonstrating how baryon retention and expulsion are further modulated by cosmological parameters, particularly $\Lambda$. Furthermore, \citet{BK_2022} explored the effects of varying $\Lambda$ on halo and IGM evolution, and star formation history. Their work indicated that higher $\Lambda$ values lead to reduced star formation rates in haloes, which aligns with our results suggesting that this suppression may stem from the increased efficiency of $\Lambda$ in stripping baryons from haloes, supported by our observed trend of increasing closure radii with $\Lambda$. Also the work by \citet{Sorini_2024_cosmo} reinforces the idea of a cosmological constant-halo baryon content connection. By calculating the impact of changing $\Lambda$ on cosmic star formation efficiency, they estimate the likelihood of generating observers in different cosmological models, hence deriving the posterior distribution on the observed value of $\Lambda$. The anthropic arguments regarding how the baryonic content of haloes influences potential habitability from these works align closely with the same implications explored in our study.

\section{Conclusions}
\label{sec:conclusions}

In this work we presented a new analytical model for the closure radius (quantifying the distance from a halo where the enclosed baryon fraction returns to the cosmic level), with the goal of determining the effects of astrophysical feedback and dark energy on the baryon distribution within and around haloes across different masses and redshifts. We began by showing that under the assumption of spherically symmetric haloes with constant feedback-induced outflows, the radial gas density profile can be described by a power law. This solution enabled us to analytically determine the closure radius, as a function of the baryon mass fraction enclosed within the halo, through equation~\eqref{eq:Rc_total_eta}. The resulting relationship depends on two astrophysical parameters (the radial gas density profile power-law slope, $\eta$, and the halo gas fraction, $f_{\text{gas}}$), the background cosmological model (specifically, $\Omega_{m 0}$, $\Omega_{\Lambda 0}$, and $f_{\text{b-cosmic}}$), and redshift.

We tested our framework against a suite of hydrodynamical simulations based on the EAGLE galaxy formation model, and spanning different cosmological parameters. We specifically examined the influence of the cosmological constant ($\Lambda$) in expelling baryons from haloes. Additionally, we compared the impact of dark energy to that of well-established astrophysical feedback mechanisms in redistributing baryons throughout the IGM. The main conclusions of this study are as follows:

\begin{itemize}

    \item The numerical closure radius -- baryon mass fraction relationship is well reproduced by our model over a 
    wide range in the cosmological constant (up to 30 times the observed value), and redshift interval $0<z<3$ (Figure~\ref{fig:3x3grid}). 

    \item In our formalism, the radial gas density profile power-law slope, $\eta$, and gas fraction of haloes, $f_{\text{gas}}$, depend on redshift, $z$, as $\propto (1+z)^b$ with an index $b$ which varies with $\Lambda$. We find that $\eta$ increases with decreasing redshift, showing that the halo gas density distribution changes over time as haloes lose gas from their inner regions (Figures~\ref{fig:2x2grid}-\ref{fig:eta_fgas_z_correlation}).

    \item Increasing $\Lambda$ causes the best fit $\eta$ to decrease, corresponding to a faster decrease in halo baryon density with radial distance from the halo centre. Moreover, we see evidence for $f_{\text{gas}}$ also decreasing with increasing $\Lambda$; however this trend is not as prominent (Figure~\ref{fig:contour_xt}).

    \item Fitting a first-order linear approximation of our new model showed that the radial gas density profile power-law slope, $\eta_{\text{approx}}$, and gas fraction of haloes, $f_{\text{gas-approx}}$, are constant across $0 < z < 3$. These best-fit values are $\eta_{\text{approx}} = 1.8 \pm 0.1$ and $f_{\text{gas-approx}} = 0.002 \pm 0.001$. Due to this constancy of $\eta_{\text{approx}}$ and $f_{\text{gas-approx}}$ between different redshift and values of $\Lambda$, this means that differences in closure radii between different cosmologies is primarily caused by differences in cosmological parameters, and not changes to the thermodynamic properties of the gas within haloes itself.

    \item A linear approximation of our model is suitable to describe the numerical results at the observed value of the cosmological constant, but fails for higher $\Lambda$ values. Such approximation resembles the dependence of the closure radius on the enclosed baryon mass fraction within haloes suggested by \citet{Ayromlou_2023}. Our model therefore explains the physical origin of the empirical \cite{Ayromlou_2023} relation, and constitutes a generalisation under different cosmological parameters.

    \item Comparing best-fit parameters between different cosmologies shows that $\Lambda$ significantly influences baryon distribution in high $\Lambda \geq 10\Lambda_0$ universes, whereas in our $\Lambda = \Lambda_0$ Universe, its impact is relatively small, around 30\% of that from astrophysical feedback mechanisms, yet still statistically significant (Table~\ref{tab:alphagammatable}, Fig.~\ref{fig:bestfitbeta}).

\end{itemize}

A primary limitation of this work lies in the large uncertainties associated with best-fit parameters, particularly evident in high $\Lambda$ simulations. These uncertainties hinder precise quantification of the impact of dark energy on the baryon distribution, thus limiting the robustness of conclusions drawn from the findings. The high uncertainties result from using 25 Mpc simulations with only $2 \times 376^3$ particles. Both the full and linear models are most accurate for higher mass haloes, necessitating a high mass threshold and resulting in fewer haloes suitable for chi-squared analysis (shown by the HMF, Fig.~\ref{fig:HMF}) or inaccurate fitting of low-mass haloes. Future studies should prioritise testing the reproducibility of these results in larger simulations with increased particle counts to improve parameter uncertainties. 
Additionally, future investigations should assess the efficacy of our closure radius model. This could involve validating its applicability across various other cosmological hydrodynamical simulations with different or varied astrophysical feedback prescriptions e.g., SIMBA \citep{Dav__2019}, IllustrisTNG \citep{Springel_2017}, Horizon-AGN \citep{2014MNRAS.444.1453D}. 

These results also complement ongoing and future efforts in observational astronomy to constrain the distribution of baryons in the Universe. The current generation of space-based integral field spectrographs, such as JWST NIRSpec \citep{Jakobsen_2022, boker2023orbit} and MIRI/MRS \citep{wells2015mid, argyriou2023jwst}, along with upcoming ground-based instruments like ELT HARMONI \citep{https://doi.org/10.18727/0722-6691/5215} and METIS \citep{2014SPIE.9147E..21B}, possess the resolving power to spatially map galaxies and their spectra at excellent spatial resolution out to the redshifts considered in this study. An intriguing avenue for future research would be to estimate the closure radius of real galaxies to compare the observed trends with those predicted by our model and other simulations. Our model already provides robust predictions for the closure radius as a function of halo mass, enclosed baryon fraction within the virial radius, cosmology, and redshift, in turn offering guidance for observational studies. 

In the context of a recent thread of works on rigorous tests of anthropic reasoning \citep{Carter_1974} through our understanding of galaxy formation in different cosmological models \citep[e.g.][]{Barnes_2018, BK_2022, Sorini_2024_cosmo}, our work quantifies how the abundance of dark energy depletes haloes of their baryons. In turn, this can severely limit the star formation rate, thereby reducing likelihood of the emergence of planetary systems and, complex systems such as intelligent life. We leave a deeper study on this matter for future work.

\section*{Acknowledgements}

OV is supported by a Science and Technology Facilities Council (STFC) studentship No.\ ST/Y509474/1. SB and DS are supported by the UK Research and Innovation (UKRI) Future Leaders Fellowship [grant number MR/V023381/1 and UKRI2044]. 
This work used the DiRAC@Durham facility managed by the Institute for Computational Cosmology on behalf of the STFC DiRAC HPC Facility (\url{www.dirac.ac.uk}). The equipment was funded by BEIS capital funding via STFC capital grants ST/K00042X/1, ST/P002293/1, ST/R002371/1 and ST/S002502/1, Durham University and STFC operations grant ST/R000832/1. DiRAC is part of the National e-Infrastructure.

\section*{Data Availability}

The primary data products (snapshots, halo catalogues) of the fiducial EAGLE simulation ($\Lambda = \Lambda_0$) are publicly available\footnote{\url{https://eagle.strw.leidenuniv.nl/wordpress/index.php/eagle-simulations-public-database/}}. The derived data, as well as the codes used for analysis and for creating relevant plots are available on \url{https://github.com/OVeenema/Veenema-et-al.-2026-closure-radius}.



\bibliographystyle{mnras}
\bibliography{references} 

@ARTICLE{Mcalpine,
       author = {{McAlpine}, S. and others},
        title = "{The EAGLE simulations of galaxy formation: Public release of halo and galaxy catalogues}",
      journal = {Astronomy and Computing},
         year = 2016,
        month = apr,
       volume = {15},
        pages = {72-89},
          doi = {10.1016/j.ascom.2016.02.004},
archivePrefix = {arXiv},
       eprint = {1510.01320},
 primaryClass = {astro-ph.GA},
       adsurl = {https://ui.adsabs.harvard.edu/abs/2016A&C....15...72M}
}

@article{Ayromlou_2023,
    author = {Ayromlou, Mohammadreza and Nelson, Dylan and Pillepich, Annalisa},
    title = "{Feedback reshapes the baryon distribution within haloes, in halo outskirts, and beyond: the closure radius from dwarfs to massive clusters}",
    journal = {Monthly Notices of the Royal Astronomical Society},
    volume = {524},
    number = {4},
    pages = {5391-5410},
    year = {2023},
    month = {07},
    issn = {0035-8711},
    doi = {10.1093/mnras/stad2046},
    url = {https://doi.org/10.1093/mnras/stad2046},
}

@ARTICLE{Sorini_2022,
       author = {{Sorini}, Daniele and {Dav{\'e}}, Romeel and {Cui}, Weiguang and {Appleby}, Sarah},
        title = "{How baryons affect haloes and large-scale structure: a unified picture from the SIMBA simulation}",
      journal = {\mnras},
         year = 2022,
        month = oct,
       volume = {516},
       number = {1},
        pages = {883-906},
          doi = {10.1093/mnras/stac2214},
archivePrefix = {arXiv},
       eprint = {2111.13708},
 primaryClass = {astro-ph.GA},
       adsurl = {https://ui.adsabs.harvard.edu/abs/2022MNRAS.516..883S}
}

@ARTICLE{Barnes_2018,
       author = {{Barnes}, Luke A. and others},
        title = "{Galaxy formation efficiency and the multiverse explanation of the cosmological constant with EAGLE simulations}",
      journal = {\mnras},
         year = 2018,
        month = jul,
       volume = {477},
       number = {3},
        pages = {3727-3743},
          doi = {10.1093/mnras/sty846},
archivePrefix = {arXiv},
       eprint = {1801.08781},
 primaryClass = {astro-ph.CO},
       adsurl = {https://ui.adsabs.harvard.edu/abs/2018MNRAS.477.3727B}
}

@ARTICLE{Schaye,
       author = {{Schaye}, Joop and others},
        title = "{The EAGLE project: simulating the evolution and assembly of galaxies and their environments}",
      journal = {\mnras},
         year = 2015,
        month = jan,
       volume = {446},
       number = {1},
        pages = {521-554},
          doi = {10.1093/mnras/stu2058},
archivePrefix = {arXiv},
       eprint = {1407.7040},
 primaryClass = {astro-ph.GA},
       adsurl = {https://ui.adsabs.harvard.edu/abs/2015MNRAS.446..521S}
}

@ARTICLE{2007ARA&A..45..221B,
       author = {{Bregman}, Joel N.},
        title = "{The Search for the Missing Baryons at Low Redshift}",
      journal = {\araa},
         year = 2007,
        month = sep,
       volume = {45},
       number = {1},
        pages = {221-259},
          doi = {10.1146/annurev.astro.45.051806.110619},
archivePrefix = {arXiv},
       eprint = {0706.1787},
 primaryClass = {astro-ph},
       adsurl = {https://ui.adsabs.harvard.edu/abs/2007ARA&A..45..221B}
}

@ARTICLE{2012ApJ...759...23S,
       author = {{Shull}, J. Michael and {Smith}, Britton D. and {Danforth}, Charles W.},
        title = "{The Baryon Census in a Multiphase Intergalactic Medium: 30\% of the Baryons May Still be Missing}",
      journal = {\apj},
         year = 2012,
        month = nov,
       volume = {759},
       number = {1},
          eid = {23},
        pages = {23},
          doi = {10.1088/0004-637X/759/1/23},
 primaryClass = {astro-ph.CO},
       adsurl = {https://ui.adsabs.harvard.edu/abs/2012ApJ...759...23S}
}

@ARTICLE{2010ApJ...708L..14M,
       author = {{McGaugh}, Stacy S. and {Schombert}, James M. and {de Blok}, W.~J.~G. and {Zagursky}, Matthew J.},
        title = "{The Baryon Content of Cosmic Structures}",
      journal = {\apjl},
         year = 2010,
        month = jan,
       volume = {708},
       number = {1},
        pages = {L14-L17},
          doi = {10.1088/2041-8205/708/1/L14},
archivePrefix = {arXiv},
       eprint = {0911.2700},
 primaryClass = {astro-ph.CO},
       adsurl = {https://ui.adsabs.harvard.edu/abs/2010ApJ...708L..14M}
}

@article{Lacey_1994,
   title={Merger rates in hierarchical models of galaxy formation – II. Comparison with N-body simulations},
   volume={271},
   ISSN={1365-2966},
   url={http://dx.doi.org/10.1093/mnras/271.3.676},
   DOI={10.1093/mnras/271.3.676},
   number={3},
   journal={Monthly Notices of the Royal Astronomical Society},
   publisher={Oxford University Press (OUP)},
   author={Lacey, Cedric and Cole, Shaun},
   year={1994},
   month=dec, pages={676–692} }

@article{article,
author = {Coles, Peter and Lucchin, Francesco},
year = {1995},
month = {01},
pages = {},
title = {Cosmology, The Origin and Evolution of Cosmic Structure},
volume = {-1},
journal = {Chichester: Wiley, |c1995}
}

@article{aghanim2020planck,
  title={Planck 2018 results-VI. Cosmological parameters},
  author={Aghanim, Nabila and others},
  journal={Astronomy \& Astrophysics},
  volume={641},
  pages={A6},
  year={2020},
  publisher={EDP sciences}
}

@ARTICLE{Jenkins2010,
       author = {{Jenkins}, Adrian},
        title = "{Second-order Lagrangian perturbation theory initial conditions for resimulations}",
      journal = {\mnras},
         year = 2010,
        month = apr,
       volume = {403},
       number = {4},
        pages = {1859-1872},
          doi = {10.1111/j.1365-2966.2010.16259.x},
archivePrefix = {arXiv},
       eprint = {0910.0258},
 primaryClass = {astro-ph.CO},
       adsurl = {https://ui.adsabs.harvard.edu/abs/2010MNRAS.403.1859J}
}

@article{Crain_2015,
   title={The EAGLE simulations of galaxy formation: calibration of subgrid physics and model variations},
   volume={450},
   ISSN={0035-8711},
   url={http://dx.doi.org/10.1093/mnras/stv725},
   DOI={10.1093/mnras/stv725},
   number={2},
   journal={Monthly Notices of the Royal Astronomical Society},
   publisher={Oxford University Press (OUP)},
   author={Crain, Robert A. and others},
   year={2015},
   month=apr, pages={1937–1961} }

@article{Springel_2001,
   title={Populating a cluster of galaxies - I. Results at \fontshape{it}{z}=0},
   volume={328},
   ISSN={1365-2966},
   url={http://dx.doi.org/10.1046/j.1365-8711.2001.04912.x},
   DOI={10.1046/j.1365-8711.2001.04912.x},
   number={3},
   journal={Monthly Notices of the Royal Astronomical Society},
   publisher={Oxford University Press (OUP)},
   author={Springel, Volker and White, Simon D. M. and Tormen, Giuseppe and Kauffmann, Guinevere},
   year={2001},
   month=dec, pages={726–750} }

@article{Salcido_2018,
   title={The impact of dark energy on galaxy formation. What does the future of our Universe hold?},
   volume={477},
   ISSN={1365-2966},
   url={http://dx.doi.org/10.1093/mnras/sty879},
   DOI={10.1093/mnras/sty879},
   number={3},
   journal={Monthly Notices of the Royal Astronomical Society},
   publisher={Oxford University Press (OUP)},
   author={Salcido, Jaime and others},
   year={2018},
   month=apr, pages={3744–3759} }

@article{Austin_2023,
doi = {10.3847/2041-8213/ace18d},
url = {https://dx.doi.org/10.3847/2041-8213/ace18d},
year = {2023},
month = {jul},
publisher = {The American Astronomical Society},
volume = {952},
number = {1},
pages = {L7},
author = {Duncan Austin and others},
title = {A Large Population of Faint 8 &lt; z &lt; 16 Galaxies Found in the First JWST NIRCam Observations of the NGDEEP Survey},
journal = {The Astrophysical Journal Letters}
}

@article{Oh_2021,
   title={Evolving beyond z=0: insights about the future of stars and the intergalactic medium},
   volume={507},
   ISSN={1365-2966},
   url={http://dx.doi.org/10.1093/mnras/stab2473},
   DOI={10.1093/mnras/stab2473},
   number={4},
   journal={Monthly Notices of the Royal Astronomical Society},
   publisher={Oxford University Press (OUP)},
   author={Oh, Boon Kiat and Peacock, John A and Khochfar, Sadegh and Smith, Britton D},
   year={2021},
   month=sep, pages={5432–5450} }

@ARTICLE{Khrykin_2024,
       author = {{Khrykin}, Ilya S. and {Ata}, Metin and {Lee}, Khee-Gan and {Simha}, Sunil and {Huang}, Yuxin and {Prochaska}, J. Xavier and {Tejos}, Nicolas and {Bannister}, Keith W. and {Cooke}, Jeff and {Day}, Cherie K. and {Deller}, Adam and {Glowacki}, Marcin and {Gordon}, Alexa C. and {James}, Clancy W. and {Marnoch}, Lachlan and {Shannon}, Ryan. M. and {Zhang}, Jielai and {Bernales-Cortes}, Lucas},
        title = "{FLIMFLAM DR1: The First Constraints on the Cosmic Baryon Distribution from Eight Fast Radio Burst Sight Lines}",
      journal = {\apj},
         year = 2024,
        month = oct,
       volume = {973},
       number = {2},
          eid = {151},
        pages = {151},
          doi = {10.3847/1538-4357/ad6567},
archivePrefix = {arXiv},
       eprint = {2402.00505},
 primaryClass = {astro-ph.GA},
       adsurl = {https://ui.adsabs.harvard.edu/abs/2024ApJ...973..151K}
}

@article{Hernquist_2003,
   title={An analytical model for the history of cosmic star formation},
   volume={341},
   ISSN={1365-2966},
   url={http://dx.doi.org/10.1046/j.1365-8711.2003.06499.x},
   DOI={10.1046/j.1365-8711.2003.06499.x},
   number={4},
   journal={Monthly Notices of the Royal Astronomical Society},
   publisher={Oxford University Press (OUP)},
   author={Hernquist, L. and Springel, V.},
   year={2003},
   month=jun, pages={1253–1267} }

@article{Yao_2012,
doi = {10.1088/0004-637X/746/2/166},
url = {https://dx.doi.org/10.1088/0004-637X/746/2/166},
year = {2012},
month = {feb},
publisher = {The American Astronomical Society},
volume = {746},
number = {2},
pages = {166},
author = {Yangsen Yao and J. Michael Shull and Q. Daniel Wang and Webster Cash},
title = {DETECTING THE WARM–HOT INTERGALACTIC MEDIUM THROUGH X-RAY ABSORPTION LINES},
journal = {The Astrophysical Journal}
}

@article{Grego_2001,
doi = {10.1086/320443},
url = {https://dx.doi.org/10.1086/320443},
year = {2001},
month = {may},
publisher = {},
volume = {552},
number = {1},
pages = {2},
author = {Laura Grego and others},
title = {Galaxy Cluster Gas Mass Fractions from Sunyaev-Zeldovich
Effect Measurements: Constraints on
ΩM},
journal = {The Astrophysical Journal}
}

@article{Lewis_2000,
   title={Efficient Computation of Cosmic Microwave Background Anisotropies in Closed Friedmann‐Robertson‐Walker Models},
   volume={538},
   ISSN={1538-4357},
   url={http://dx.doi.org/10.1086/309179},
   DOI={10.1086/309179},
   number={2},
   journal={The Astrophysical Journal},
   publisher={American Astronomical Society},
   author={Lewis, Antony and Challinor, Anthony and Lasenby, Anthony},
   year={2000},
   month=aug, pages={473–476} }

@article{Macquart_2020,
   title={A census of baryons in the Universe from localized fast radio bursts},
   volume={581},
   ISSN={1476-4687},
   url={http://dx.doi.org/10.1038/s41586-020-2300-2},
   DOI={10.1038/s41586-020-2300-2},
   number={7809},
   journal={Nature},
   publisher={Springer Science and Business Media LLC},
   author={Macquart, J.-P. and others},
   year={2020},
   month=may, pages={391–395} }

@article{Springel_2017,
   title={First results from the IllustrisTNG simulations: matter and galaxy clustering},
   volume={475},
   ISSN={1365-2966},
   url={http://dx.doi.org/10.1093/mnras/stx3304},
   DOI={10.1093/mnras/stx3304},
   number={1},
   journal={Monthly Notices of the Royal Astronomical Society},
   publisher={Oxford University Press (OUP)},
   author={Springel, Volker and others},
   year={2017},
   month=dec, pages={676–698} }

@article{Dav__2019,
   title={simba: Cosmological simulations with black hole growth and feedback},
   volume={486},
   ISSN={1365-2966},
   url={http://dx.doi.org/10.1093/mnras/stz937},
   DOI={10.1093/mnras/stz937},
   number={2},
   journal={Monthly Notices of the Royal Astronomical Society},
   publisher={Oxford University Press (OUP)},
   author={Davé, Romeel and others},
   year={2019},
   month=apr, pages={2827–2849} }

@ARTICLE{1991ApJ...379...52W,
       author = {{White}, Simon D.~M. and {Frenk}, Carlos S.},
        title = "{Galaxy Formation through Hierarchical Clustering}",
      journal = {\apj},
         year = 1991,
        month = sep,
       volume = {379},
        pages = {52},
          doi = {10.1086/170483},
       adsurl = {https://ui.adsabs.harvard.edu/abs/1991ApJ...379...52W}
}

@article{Cen_1999,
doi = {10.1086/306949},
url = {https://dx.doi.org/10.1086/306949},
year = {1999},
month = {mar},
publisher = {},
volume = {514},
number = {1},
pages = {1},
author = {Renyue Cen and Jeremiah P. Ostriker},
title = {Where Are the Baryons?},
journal = {The Astrophysical Journal}
}

@article{Mota_2004,
   title={On the spherical collapse model in dark energy cosmologies},
   volume={421},
   ISSN={1432-0746},
   url={http://dx.doi.org/10.1051/0004-6361:20041090},
   DOI={10.1051/0004-6361:20041090},
   number={1},
   journal={Astronomy &amp; Astrophysics},
   publisher={EDP Sciences},
   author={Mota, D. F. and C. van de Bruck},
   year={2004},
   month=jun, pages={71–81} }

@BOOK{1980lssu.book.....P,
       author = {{Peebles}, P.~J.~E.},
        title = "{The large-scale structure of the universe}",
         year = 1980,
       adsurl = {https://ui.adsabs.harvard.edu/abs/1980lssu.book.....P}
}

@article{Haider_2016,
   title={Large-scale mass distribution in the Illustris simulation},
   volume={457},
   ISSN={1365-2966},
   url={http://dx.doi.org/10.1093/mnras/stw077},
   DOI={10.1093/mnras/stw077},
   number={3},
   journal={Monthly Notices of the Royal Astronomical Society},
   publisher={Oxford University Press (OUP)},
   author={Haider, M. and Steinhauser, D. and Vogelsberger, M. and Genel, S. and Springel, V. and Torrey, P. and Hernquist, L.},
   year={2016},
   month=feb, pages={3024–3035} }

@article{Pintos-Castro_2019,
doi = {10.3847/1538-4357/ab14ee},
url = {https://dx.doi.org/10.3847/1538-4357/ab14ee},
year = {2019},
month = {may},
publisher = {The American Astronomical Society},
volume = {876},
number = {1},
pages = {40},
author = {I. Pintos-Castro and H. K. C. Yee and A. Muzzin and L. Old and G. Wilson},
title = {The Evolution of the Quenching of Star Formation in Cluster Galaxies since z ∼ 1},
journal = {The Astrophysical Journal}
}

@article{10.1111/j.1365-2966.2008.14191.x,
    author = {Wiersma, Robert P. C. and Schaye, Joop and Smith, Britton D.},
    title = "{The effect of photoionization on the cooling rates of enriched, astrophysical plasmas}",
    journal = {Monthly Notices of the Royal Astronomical Society},
    volume = {393},
    number = {1},
    pages = {99-107},
    year = {2009},
    month = {01},
    issn = {0035-8711},
    doi = {10.1111/j.1365-2966.2008.14191.x},
    url = {https://doi.org/10.1111/j.1365-2966.2008.14191.x},
    eprint = {https://academic.oup.com/mnras/article-pdf/393/1/99/3777521/mnras0393-0099.pdf},
}

@ARTICLE{1985ApJ...292..371D,
       author = {{Davis}, M. and {Efstathiou}, G. and {Frenk}, C.~S. and {White}, S.~D.~M.},
        title = "{The evolution of large-scale structure in a universe dominated by cold dark matter}",
      journal = {\apj},
         year = 1985,
        month = may,
       volume = {292},
        pages = {371-394},
          doi = {10.1086/163168},
       adsurl = {https://ui.adsabs.harvard.edu/abs/1985ApJ...292..371D}
}

@article{Schaye_2007,
   title={On the relation between the Schmidt and Kennicutt-Schmidt star formation laws and its implications for numerical simulations: Schmidt and Kennicutt-Schmidt laws},
   volume={383},
   ISSN={1365-2966},
   url={http://dx.doi.org/10.1111/j.1365-2966.2007.12639.x},
   DOI={10.1111/j.1365-2966.2007.12639.x},
   number={3},
   journal={Monthly Notices of the Royal Astronomical Society},
   publisher={Oxford University Press (OUP)},
   author={Schaye, Joop and Dalla Vecchia, Claudio},
   year={2007},
   month=dec, pages={1210–1222} }

@article{10.1046/j.1365-8711.2003.06207.x,
    author = {Springel, Volker and Hernquist, Lars},
    title = "{The history of star formation in a Λ cold dark matter universe}",
    journal = {Monthly Notices of the Royal Astronomical Society},
    volume = {339},
    number = {2},
    pages = {312-334},
    year = {2003},
    month = {02},
    issn = {0035-8711},
    doi = {10.1046/j.1365-8711.2003.06207.x},
    url = {https://doi.org/10.1046/j.1365-8711.2003.06207.x},
    eprint = {https://academic.oup.com/mnras/article-pdf/339/2/312/3231433/339-2-312.pdf},
}

@article{10.1111/j.1365-2966.2005.09655.x,
    author = {Springel, Volker},
    title = "{The cosmological simulation code gadget-2}",
    journal = {Monthly Notices of the Royal Astronomical Society},
    volume = {364},
    number = {4},
    pages = {1105-1134},
    year = {2005},
    month = {12},
    issn = {0035-8711},
    doi = {10.1111/j.1365-2966.2005.09655.x},
    url = {https://doi.org/10.1111/j.1365-2966.2005.09655.x},
    eprint = {https://academic.oup.com/mnras/article-pdf/364/4/1105/18657201/364-4-1105.pdf},
}

@article{rigby2023science,
  title={The science performance of JWST as characterized in commissioning},
  author={Rigby, Jane and Perrin, Marshall and McElwain, Michael and Kimble, Randy and Friedman, Scott and Lallo, Matt and Doyon, Ren{\'e} and Feinberg, Lee and Ferruit, Pierre and Glasse, Alistair and others},
  journal={Publications of the Astronomical Society of the Pacific},
  volume={135},
  number={1046},
  pages={048001},
  year={2023},
  publisher={IOP Publishing}
}

@ARTICLE{2014MNRAS.444.1453D,
       author = {{Dubois}, Y. and {Pichon}, C. and {Welker}, C. and {Le Borgne}, D. and {Devriendt}, J. and {Laigle}, C. and {Codis}, S. and {Pogosyan}, D. and {Arnouts}, S. and {Benabed}, K. and {Bertin}, E. and {Blaizot}, J. and {Bouchet}, F. and {Cardoso}, J. -F. and {Colombi}, S. and {de Lapparent}, V. and {Desjacques}, V. and {Gavazzi}, R. and {Kassin}, S. and {Kimm}, T. and {McCracken}, H. and {Milliard}, B. and {Peirani}, S. and {Prunet}, S. and {Rouberol}, S. and {Silk}, J. and {Slyz}, A. and {Sousbie}, T. and {Teyssier}, R. and {Tresse}, L. and {Treyer}, M. and {Vibert}, D. and {Volonteri}, M.},
        title = "{Dancing in the dark: galactic properties trace spin swings along the cosmic web}",
      journal = {\mnras},
         year = 2014,
        month = oct,
       volume = {444},
       number = {2},
        pages = {1453-1468},
          doi = {10.1093/mnras/stu1227},
archivePrefix = {arXiv},
       eprint = {1402.1165},
 primaryClass = {astro-ph.CO},
       adsurl = {https://ui.adsabs.harvard.edu/abs/2014MNRAS.444.1453D}
}

@article{Jakobsen_2022,
   title={The Near-Infrared Spectrograph (NIRSpec) on theJames WebbSpace Telescope: I. Overview of the instrument and its capabilities},
   volume={661},
   ISSN={1432-0746},
   url={http://dx.doi.org/10.1051/0004-6361/202142663},
   DOI={10.1051/0004-6361/202142663},
   journal={Astronomy &amp; Astrophysics},
   publisher={EDP Sciences},
   author={Jakobsen, P. and Ferruit, P. and Alves de Oliveira, C. and Arribas, S. and Bagnasco, G. and Barho, R. and Beck, T. L. and Birkmann, S. and Böker, T. and Bunker, A. J. and Charlot, S. and de Jong, P. and de Marchi, G. and Ehrenwinkler, R. and Falcolini, M. and Fels, R. and Franx, M. and Franz, D. and Funke, M. and Giardino, G. and Gnata, X. and Holota, W. and Honnen, K. and Jensen, P. L. and Jentsch, M. and Johnson, T. and Jollet, D. and Karl, H. and Kling, G. and Köhler, J. and Kolm, M.-G. and Kumari, N. and Lander, M. E. and Lemke, R. and López-Caniego, M. and Lützgendorf, N. and Maiolino, R. and Manjavacas, E. and Marston, A. and Maschmann, M. and Maurer, R. and Messerschmidt, B. and Moseley, S. H. and Mosner, P. and Mott, D. B. and Muzerolle, J. and Pirzkal, N. and Pittet, J.-F. and Plitzke, A. and Posselt, W. and Rapp, B. and Rauscher, B. J. and Rawle, T. and Rix, H.-W. and Rödel, A. and Rumler, P. and Sabbi, E. and Salvignol, J.-C. and Schmid, T. and Sirianni, M. and Smith, C. and Strada, P. and te Plate, M. and Valenti, J. and Wettemann, T. and Wiehe, T. and Wiesmayer, M. and Willott, C. J. and Wright, R. and Zeidler, P. and Zincke, C.},
   year={2022},
   month=may, pages={A80} }

@article{wells2015mid,
  title={The mid-infrared instrument for the james webb space telescope, vi: The medium resolution spectrometer},
  author={Wells, Martyn and Pel, J-W and Glasse, Alistair and Wright, GS and Aitink-Kroes, Gabby and Azzollini, Ruym{\'a}n and Beard, Steven and Brandl, BR and Gallie, Angus and Geers, VC and others},
  journal={Publications of the Astronomical Society of the Pacific},
  volume={127},
  number={953},
  pages={646},
  year={2015},
  publisher={IOP Publishing}
}

@article{argyriou2023jwst,
  title={JWST MIRI flight performance: the medium-resolution spectrometer},
  author={Argyriou, Ioannis and Glasse, Alistair and Law, David R and Labiano, Alvaro and {\'A}lvarez-M{\'a}rquez, Javier and Patapis, Polychronis and Kavanagh, Patrick J and Gasman, Danny and Mueller, Michael and Larson, Kirsten and others},
  journal={Astronomy \& Astrophysics},
  volume={675},
  pages={A111},
  year={2023},
  publisher={EDP Sciences}
}

@article{boker2023orbit,
  title={In-orbit performance of the near-infrared spectrograph NIRSpec on the James Webb Space Telescope},
  author={B{\"o}ker, T and Beck, TL and Birkmann, SM and Giardino, G and Keyes, C and Kumari, N and Muzerolle, J and Rawle, T and Zeidler, P and Abul-Huda, Y and others},
  journal={Publications of the Astronomical Society of the Pacific},
  volume={135},
  number={1045},
  pages={038001},
  year={2023},
  publisher={IOP Publishing}
}

@article{https://doi.org/10.18727/0722-6691/5215,
  doi = {10.18727/0722-6691/5215},
  
  url = {http://doi.eso.org/10.18727/0722-6691/5215},
  
  author = {Thatte, Niranjan and Tecza, Matthias and Schnetler, Hermine and Neichel, Benoit and Melotte, Dave and Fusco, Thierry and Ferraro-Wood, Vanessa and Clarke, Fraser and Bryson, Ian and O’Brien, Kieran and Mateo, Mario and Garcia Lorenzo, Begoña and Evans, Chris and Bouché, Nicolas and Arribas, Santiago and Consortium, The HARMONI},
  
  title = {HARMONI: the ELT’s First-Light Near-infrared and Visible Integral Field Spectrograph},
  
  journal = {Published in The Messenger vol. 182},
  
  volume = {pp. 7-12},
  
  pages = {March 2021.},
  
  publisher = {European Southern Observatory (ESO)},
  
  year = {2021},
  
  copyright = {Copyright European Southern Observatory}
}

@INPROCEEDINGS{2014SPIE.9147E..21B,
       author = {{Brandl}, Bernhard R. and {Feldt}, Markus and {Glasse}, Alistair and {Guedel}, Manuel and {Heikamp}, Stephanie and {Kenworthy}, Matthew and {Lenzen}, Rainer and {Meyer}, Michael R. and {Molster}, Frank and {Paalvast}, Sander and {Pantin}, Eric J. and {Quanz}, Sascha P. and {Schmalzl}, Eva and {Stuik}, Remko and {Venema}, Lars and {Waelkens}, Christoffel},
        title = "{METIS: the mid-infrared E-ELT imager and spectrograph}",
    booktitle = {Ground-based and Airborne Instrumentation for Astronomy V},
         year = 2014,
       editor = {{Ramsay}, Suzanne K. and {McLean}, Ian S. and {Takami}, Hideki},
       series = {Society of Photo-Optical Instrumentation Engineers (SPIE) Conference Series},
       volume = {9147},
        month = aug,
          eid = {914721},
        pages = {914721},
          doi = {10.1117/12.2056468},
archivePrefix = {arXiv},
       eprint = {1409.3087},
 primaryClass = {astro-ph.IM},
       adsurl = {https://ui.adsabs.harvard.edu/abs/2014SPIE.9147E..21B}
}

@article{barrow1991anthropic,
  title={The anthropic cosmological principle},
  author={Barrow, John D and Tipler, Frank J},
  journal={Di{\'a}logo filos{\'o}fico},
  number={19},
  pages={119--120},
  year={1991},
  publisher={Di{\'a}logo filos{\'o}fico}
}

@article{weinberg1987anthropic,
  title={Anthropic bound on the cosmological constant},
  author={Weinberg, Steven},
  journal={Physical Review Letters},
  volume={59},
  number={22},
  pages={2607},
  year={1987},
  publisher={APS}
}

@article{glazebrook2024massive,
  title={A massive galaxy that formed its stars at z≈ 11},
  author={Glazebrook, Karl and Nanayakkara, Themiya and Schreiber, Corentin and Lagos, Claudia and Kawinwanichakij, Lalitwadee and Jacobs, Colin and Chittenden, Harry and Brammer, Gabriel and Kacprzak, Glenn G and Labbe, Ivo and others},
  journal={Nature},
  volume={628},
  number={8007},
  pages={277--281},
  year={2024},
  publisher={Nature Publishing Group UK London}
}

@article{Sorini_2024,
   title={The impact of feedback on the evolution of gas density profiles from galaxies to clusters: a universal fitting formula from the Simba suite of simulations},
   volume={7},
   ISSN={2565-6120},
   url={http://dx.doi.org/10.33232/001c.126621},
   DOI={10.33232/001c.126621},
   journal={The Open Journal of Astrophysics},
   publisher={Maynooth University},
   author={Sorini, Daniele and Bose, Sownak and Davé, Romeel and Anglés-Alcázar, Daniel},
   year={2024},
   month=dec }

@article{salcido2020feedback,
  title={How feedback shapes galaxies: an analytic model},
  author={Salcido, Jaime and Bower, Richard G and Theuns, Tom},
  journal={Monthly Notices of the Royal Astronomical Society},
  volume={491},
  number={4},
  pages={5083--5100},
  year={2020},
  publisher={Oxford University Press}
}

@article{oh2020calibration,
  title={Calibration of a star formation and feedback model for cosmological simulations with Enzo},
  author={Oh, Boon Kiat and Smith, Britton D and Peacock, John A and Khochfar, Sadegh},
  journal={Monthly Notices of the Royal Astronomical Society},
  volume={497},
  number={4},
  pages={5203--5219},
  year={2020},
  publisher={Oxford University Press}
}

@ARTICLE{IllustrisTNG2018,
       author = {{Pillepich}, Annalisa and {Springel}, Volker and {Nelson}, Dylan and
         {Genel}, Shy and {Naiman}, Jill and {Pakmor}, R{\"u}diger and
         {Hernquist}, Lars and {Torrey}, Paul and {Vogelsberger}, Mark and
         {Weinberger}, Rainer and {Marinacci}, Federico},
        title = "{Simulating galaxy formation with the IllustrisTNG model}",
      journal = {\mnras},
         year = 2018,
        month = jan,
       volume = {473},
       number = {3},
        pages = {4077-4106},
          doi = {10.1093/mnras/stx2656},
archivePrefix = {arXiv},
       eprint = {1703.02970},
 primaryClass = {astro-ph.GA},
       adsurl = {https://ui.adsabs.harvard.edu/abs/2018MNRAS.473.4077P}
}

@ARTICLE{Dolag_2016,
       author = {{Dolag}, K. and {Komatsu}, E. and {Sunyaev}, R.},
        title = "{SZ effects in the Magneticum Pathfinder simulation: comparison with the Planck, SPT, and ACT results}",
      journal = {\mnras},
         year = 2016,
        month = dec,
       volume = {463},
       number = {2},
        pages = {1797-1811},
          doi = {10.1093/mnras/stw2035},
archivePrefix = {arXiv},
       eprint = {1509.05134},
 primaryClass = {astro-ph.CO},
       adsurl = {https://ui.adsabs.harvard.edu/abs/2016MNRAS.463.1797D}
}

@ARTICLE{Lim_2021,
       author = {{Lim}, S.~H. and {Barnes}, D. and {Vogelsberger}, M. and {Mo}, H.~J. and {Nelson}, D. and {Pillepich}, A. and {Dolag}, K. and {Marinacci}, F.},
        title = "{Properties of the ionized CGM and IGM: tests for galaxy formation models from the Sunyaev-Zel'dovich effect}",
      journal = {\mnras},
         year = 2021,
        month = jul,
       volume = {504},
       number = {4},
        pages = {5131-5143},
          doi = {10.1093/mnras/stab1172},
archivePrefix = {arXiv},
       eprint = {2007.11583},
 primaryClass = {astro-ph.GA},
       adsurl = {https://ui.adsabs.harvard.edu/abs/2021MNRAS.504.5131L}
}

@ARTICLE{Davies_2019,
       author = {{Davies}, Jonathan J. and {Crain}, Robert A. and {McCarthy}, Ian G. and {Oppenheimer}, Benjamin D. and {Schaye}, Joop and {Schaller}, Matthieu and {McAlpine}, Stuart},
        title = "{The gas fractions of dark matter haloes hosting simulated {\ensuremath{\sim}}L$^{{\ensuremath{\star}}}$ galaxies are governed by the feedback history of their black holes}",
      journal = {\mnras},
         year = 2019,
        month = may,
       volume = {485},
       number = {3},
        pages = {3783-3793},
          doi = {10.1093/mnras/stz635},
archivePrefix = {arXiv},
       eprint = {1810.07696},
 primaryClass = {astro-ph.GA},
       adsurl = {https://ui.adsabs.harvard.edu/abs/2019MNRAS.485.3783D}
}

@ARTICLE{Davies_2021,
       author = {{Davies}, Jonathan J. and {Crain}, Robert A. and {Pontzen}, Andrew},
        title = "{Quenching and morphological evolution due to circumgalactic gas expulsion in a simulated galaxy with a controlled assembly history}",
      journal = {\mnras},
         year = 2021,
        month = jan,
       volume = {501},
       number = {1},
        pages = {236-253},
          doi = {10.1093/mnras/staa3643},
archivePrefix = {arXiv},
       eprint = {2006.13221},
 primaryClass = {astro-ph.GA},
       adsurl = {https://ui.adsabs.harvard.edu/abs/2021MNRAS.501..236D}
}

@ARTICLE{Davies_2022,
       author = {{Davies}, Jonathan J. and {Pontzen}, Andrew and {Crain}, Robert A.},
        title = "{Galaxy mergers can initiate quenching by unlocking an AGN-driven transformation of the baryon cycle}",
      journal = {\mnras},
         year = 2022,
        month = sep,
       volume = {515},
       number = {1},
        pages = {1430-1443},
          doi = {10.1093/mnras/stac1742},
archivePrefix = {arXiv},
       eprint = {2203.08157},
 primaryClass = {astro-ph.GA},
       adsurl = {https://ui.adsabs.harvard.edu/abs/2022MNRAS.515.1430D}
}

@ARTICLE{Davies_2020,
       author = {{Davies}, Jonathan J. and {Crain}, Robert A. and {Oppenheimer}, Benjamin D. and {Schaye}, Joop},
        title = "{The quenching and morphological evolution of central galaxies is facilitated by the feedback-driven expulsion of circumgalactic gas}",
      journal = {\mnras},
         year = 2020,
        month = jan,
       volume = {491},
       number = {3},
        pages = {4462-4480},
          doi = {10.1093/mnras/stz3201},
archivePrefix = {arXiv},
       eprint = {1908.11380},
 primaryClass = {astro-ph.GA},
       adsurl = {https://ui.adsabs.harvard.edu/abs/2020MNRAS.491.4462D}
}

@ARTICLE{Tollet_2019,
       author = {{Tollet}, {\'E}douard and {Cattaneo}, Andrea and {Macci{\`o}}, Andrea V. and {Dutton}, Aaron A. and {Kang}, Xi},
        title = "{NIHAO XIX: how supernova feedback shapes the galaxy baryon cycle}",
      journal = {\mnras},
         year = 2019,
        month = may,
       volume = {485},
       number = {2},
        pages = {2511-2531},
          doi = {10.1093/mnras/stz545},
archivePrefix = {arXiv},
       eprint = {1902.03888},
 primaryClass = {astro-ph.GA},
       adsurl = {https://ui.adsabs.harvard.edu/abs/2019MNRAS.485.2511T}
}

@ARTICLE{Borrow_2020,
       author = {{Borrow}, Josh and {Angl{\'e}s-Alc{\'a}zar}, Daniel and
         {Dav{\'e}}, Romeel},
        title = "{Cosmological baryon transfer in the SIMBA simulations}",
      journal = {\mnras},
         year = 2020,
        month = feb,
       volume = {491},
       number = {4},
        pages = {6102-6119},
          doi = {10.1093/mnras/stz3428},
archivePrefix = {arXiv},
       eprint = {1910.00594},
 primaryClass = {astro-ph.GA},
       adsurl = {https://ui.adsabs.harvard.edu/abs/2020MNRAS.491.6102B}
}

@ARTICLE{Appleby_2021,
       author = {{Appleby}, Sarah and {Dav{\'e}}, Romeel and {Sorini}, Daniele and {Storey-Fisher}, Kate and {Smith}, Britton},
        title = "{The low-redshift circumgalactic medium in SIMBA}",
      journal = {\mnras},
         year = 2021,
        month = oct,
       volume = {507},
       number = {2},
        pages = {2383-2404},
          doi = {10.1093/mnras/stab2310},
archivePrefix = {arXiv},
       eprint = {2102.10126},
 primaryClass = {astro-ph.GA},
       adsurl = {https://ui.adsabs.harvard.edu/abs/2021MNRAS.507.2383A}
}

@ARTICLE{Angelinelli_2022,
       author = {{Angelinelli}, M. and {Ettori}, S. and {Dolag}, K. and {Vazza}, F. and {Ragagnin}, A.},
        title = "{Mapping `out-of-the-box' the properties of the baryons in massive halos}",
      journal = {\aap},
         year = 2022,
        month = jul,
       volume = {663},
          eid = {L6},
        pages = {L6},
          doi = {10.1051/0004-6361/202244068},
archivePrefix = {arXiv},
       eprint = {2206.08382},
 primaryClass = {astro-ph.GA},
       adsurl = {https://ui.adsabs.harvard.edu/abs/2022A&A...663L...6A}
}

@ARTICLE{BK_2022,
       author = {{Oh}, Boon Kiat and {Peacock}, John A. and {Khochfar}, Sadegh and {Smith}, Britton D.},
        title = "{The fate of baryons in counterfactual universes}",
      journal = {\mnras},
         year = 2022,
        month = nov,
       volume = {517},
       number = {1},
        pages = {59-75},
          doi = {10.1093/mnras/stac2669},
archivePrefix = {arXiv},
       eprint = {2209.08783},
 primaryClass = {astro-ph.CO},
       adsurl = {https://ui.adsabs.harvard.edu/abs/2022MNRAS.517...59O}
}

@INPROCEEDINGS{Carter_1974,
       author = {{Carter}, B.},
        title = "{Large number coincidences and the anthropic principle in cosmology.}",
    booktitle = {Confrontation of Cosmological Theories with Observational Data},
    series = {IAU Symposium},
         year = 1974,
       editor = {{Longair}, M.~S.},
       volume = {63},
        month = jan,
        pages = {291-298},
       adsurl = {https://ui.adsabs.harvard.edu/abs/1974IAUS...63..291C}
}

@ARTICLE{Sorini_2024_cosmo,
       author = {{Sorini}, Daniele and {Peacock}, John A. and {Lombriser}, Lucas},
        title = "{The impact of the cosmological constant on past and future star formation}",
      journal = {\mnras},
         year = 2024,
        month = dec,
       volume = {535},
       number = {2},
        pages = {1449-1474},
          doi = {10.1093/mnras/stae2236},
archivePrefix = {arXiv},
       eprint = {2411.07301},
 primaryClass = {astro-ph.CO},
       adsurl = {https://ui.adsabs.harvard.edu/abs/2024MNRAS.535.1449S}
}

@ARTICLE{Sorini_2021,
       author = {{Sorini}, Daniele and {Peacock}, John A.},
        title = "{Extended Hernquist-Springel formalism for cosmic star formation}",
      journal = {\mnras},
         year = 2021,
        month = dec,
       volume = {508},
       number = {4},
        pages = {5802-5824},
          doi = {10.1093/mnras/stab2845},
archivePrefix = {arXiv},
       eprint = {2109.01146},
 primaryClass = {astro-ph.GA},
       adsurl = {https://ui.adsabs.harvard.edu/abs/2021MNRAS.508.5802S}
}

@article{dave2012analytic,
  title={An analytic model for the evolution of the stellar, gas and metal content of galaxies},
  author={Dav{\'e}, Romeel and Finlator, Kristian and Oppenheimer, Benjamin D},
  journal={Monthly Notices of the Royal Astronomical Society},
  volume={421},
  number={1},
  pages={98--107},
  year={2012},
  publisher={Blackwell Publishing Ltd Oxford, UK}
}

@ARTICLE{Madau1996,
       author = {{Madau}, Piero and {Ferguson}, Henry C. and {Dickinson}, Mark E. and {Giavalisco}, Mauro and {Steidel}, Charles C. and {Fruchter}, Andrew},
        title = "{High-redshift galaxies in the Hubble Deep Field: colour selection and star formation history to z\raisebox{-0.5ex}\textasciitilde4}",
      journal = {\mnras},
         year = 1996,
        month = dec,
       volume = {283},
       number = {4},
        pages = {1388-1404},
          doi = {10.1093/mnras/283.4.1388},
archivePrefix = {arXiv},
       eprint = {astro-ph/9607172},
 primaryClass = {astro-ph},
       adsurl = {https://ui.adsabs.harvard.edu/abs/1996MNRAS.283.1388M}
}

@ARTICLE{Ellis1997,
       author = {{Ellis}, Richard S.},
        title = "{Faint Blue Galaxies}",
      journal = {\araa},
         year = 1997,
        month = jan,
       volume = {35},
        pages = {389-443},
          doi = {10.1146/annurev.astro.35.1.389},
archivePrefix = {arXiv},
       eprint = {astro-ph/9704019},
 primaryClass = {astro-ph},
       adsurl = {https://ui.adsabs.harvard.edu/abs/1997ARA&A..35..389E}
}

@ARTICLE{wayland2025calibrating,
       author = {{Wayland}, Amy and {Alonso}, David and {Zennaro}, Matteo},
        title = "{Calibrating baryonic effects in cosmic shear with external data in the LSST era}",
      journal = {\mnras},
         year = 2025,
        month = oct,
       volume = {543},
       number = {2},
        pages = {1518-1534},
          doi = {10.1093/mnras/staf1541},
archivePrefix = {arXiv},
       eprint = {2506.11943},
 primaryClass = {astro-ph.CO},
       adsurl = {https://ui.adsabs.harvard.edu/abs/2025MNRAS.543.1518W}
}

@ARTICLE{naidu2025cosmic,
       author = {{Naidu}, Rohan P. and {Oesch}, Pascal A. and {Brammer}, Gabriel and {Weibel}, Andrea and {Li}, Yijia and {Matthee}, Jorryt and {Chisolm}, John and {Pollock}, Clara L. and {Heintz}, Kasper E. and {Johnson}, Benjamin D. and {Shen}, Xuejian and {Hviding}, Raphael E. and {Leja}, Joel and {Tacchella}, Sandro and {Ganguly}, Arpita and {Witten}, Callum and {Atek}, Hakim and {Belli}, Siro and {Bose}, Sownak and {Bouwens}, Rychard and {Dayal}, Pratika and {Decarli}, Roberto and {de Graaff}, Anna and {Fudamoto}, Yoshinobu and {Giovinazzo}, Emma and {Greene}, Jenny E. and {Illingworth}, Garth and {Inoue}, Akio K. and {Kane}, Sarah G. and {Labbe}, Ivo and {Leonova}, Ecaterina and {Marques-Chaves}, Rui and {Meyer}, Roman A. and {Nelson}, Erica J. and {Roberts-Borsani}, Guido and {Schaerer}, Daniel and {Simcoe}, Robert A. and {Stefanon}, Mauro and {Sugahara}, Yuma and {Toft}, Sune and {van der Wel}, Arjen and {van Dokkum}, Pieter and {Walter}, Fabian and {Watson}, Darrach and {Weaver}, John R. and {Whitaker}, Katherine E.},
        title = "{A Cosmic Miracle: A Remarkably Luminous Galaxy at zspec = 14.44 Confirmed with JWST}",
      journal = {The Open Journal of Astrophysics},
         year = 2026,
        month = jan,
       volume = {9},
        pages = {56033},
          doi = {10.33232/001c.156033},
archivePrefix = {arXiv},
       eprint = {2505.11263},
 primaryClass = {astro-ph.GA},
       adsurl = {https://ui.adsabs.harvard.edu/abs/2026OJAp....956033N}
}

@article{hu2023hubble,
  title={Hubble tension: The evidence of new physics},
  author={Hu, Jian-Ping and Wang, Fa-Yin},
  journal={Universe},
  volume={9},
  number={2},
  pages={94},
  year={2023},
  publisher={MDPI}
}

@article{bargiacchi2023tensions,
  title={Tensions with the flat $\Lambda$CDM model from high-redshift cosmography},
  author={Bargiacchi, G and Dainotti, MG and Capozziello, S},
  journal={Monthly Notices of the Royal Astronomical Society},
  volume={525},
  number={2},
  pages={3104--3116},
  year={2023},
  publisher={Oxford University Press}
}

@article{adil2024s,
  title={S 8 increases with effective redshift in $\Lambda$CDM cosmology},
  author={Adil, Shahnawaz A and Akarsu, {\"O}zg{\"u}r and Malekjani, Mohammad and {\'O} Colg{\'a}in, E and Pourojaghi, Saeed and Sen, Anjan A and Sheikh-Jabbari, MM},
  journal={Monthly Notices of the Royal Astronomical Society: Letters},
  volume={528},
  number={1},
  pages={L20--L26},
  year={2024},
  publisher={Oxford University Press}
}

@article{monaco2014semi,
  title={A semi-analytic model comparison: testing cooling models against hydrodynamical simulations},
  author={Monaco, Pierluigi and Benson, Andrew J and De Lucia, Gabriella and Fontanot, Fabio and Borgani, Stefano and Boylan-Kolchin, Michael},
  journal={Monthly Notices of the Royal Astronomical Society},
  volume={441},
  number={3},
  pages={2058--2077},
  year={2014},
  publisher={Oxford University Press}
}

@article{dale2015modelling,
  title={The modelling of feedback in star formation simulations},
  author={Dale, James E},
  journal={New Astronomy Reviews},
  volume={68},
  pages={1--33},
  year={2015},
  publisher={Elsevier}
}

@ARTICLE{chisari2019modelling,
       author = {{Chisari}, Nora Elisa and {Mead}, Alexander J. and {Joudaki}, Shahab and {Ferreira}, Pedro G. and {Schneider}, Aurel and {Mohr}, Joseph and {Tr{\"o}ster}, Tilman and {Alonso}, David and {McCarthy}, Ian G. and {Martin-Alvarez}, Sergio and {Devriendt}, Julien and {Slyz}, Adrianne and {van Daalen}, Marcel P.},
        title = "{Modelling baryonic feedback for survey cosmology}",
      journal = {The Open Journal of Astrophysics},
         year = 2019,
        month = jun,
       volume = {2},
       number = {1},
          eid = {4},
        pages = {4},
          doi = {10.21105/astro.1905.06082},
archivePrefix = {arXiv},
       eprint = {1905.06082},
 primaryClass = {astro-ph.CO},
       adsurl = {https://ui.adsabs.harvard.edu/abs/2019OJAp....2E...4C}
}

@article{bondi1952spherically,
  title={On spherically symmetrical accretion},
  author={Bondi, HJ1952MNRAS},
  journal={Monthly Notices of the Royal Astronomical Society},
  volume={112},
  number={2},
  pages={195--204},
  year={1952},
  publisher={Oxford University Press Oxford, UK}
}

@article{Bigwood:2024,
    author = "Bigwood, L. and others",
    collaboration = "DES",
    title = "{Weak lensing combined with the kinetic Sunyaev\textendash{}Zel\textquoteright{}dovich effect: a study of baryonic feedback}",
    eprint = "2404.06098",
    archivePrefix = "arXiv",
    primaryClass = "astro-ph.CO",
    reportNumber = "DES-2024-0827, FERMILAB-PUB-24-0130-PPD",
    doi = "10.1093/mnras/stae2100",
    journal = "Mon. Not. Roy. Astron. Soc.",
    volume = "534",
    number = "1",
    pages = "655--682",
    year = "2024"
}

@ARTICLE{LaPosta:2024,
       author = {{La Posta}, Adrien and {Alonso}, David and {Chisari}, Nora Elisa and {Ferreira}, Tassia and {Garc{\'\i}a-Garc{\'\i}a}, Carlos},
        title = "{Insights on gas thermodynamics from the combination of x-ray and thermal Sunyaev-Zel'dovich data cross correlated with cosmic shear}",
      journal = {\prd},
         year = 2025,
        month = aug,
       volume = {112},
       number = {4},
          eid = {043525},
        pages = {043525},
          doi = {10.1103/m77z-w7pl},
archivePrefix = {arXiv},
       eprint = {2412.12081},
 primaryClass = {astro-ph.CO},
       adsurl = {https://ui.adsabs.harvard.edu/abs/2025PhRvD.112d3525L}
}

@article{Reischke:2023,
       author = {{Reischke}, Robert and {Neumann}, Dennis and {Bertmann}, Klara Antonia and {Hagstotz}, Steffen and {Hildebrandt}, Hendrik},
        title = "{Calibrating baryonic feedback with weak lensing and fast radio bursts}",
      journal = {arXiv e-prints},
         year = 2023,
        month = sep,
          eid = {arXiv:2309.09766},
        pages = {arXiv:2309.09766},
          doi = {10.48550/arXiv.2309.09766},
archivePrefix = {arXiv},
       eprint = {2309.09766},
primaryClass = {astro-ph.CO},
       adsurl = {https://ui.adsabs.harvard.edu/abs/2023arXiv230909766R}
}

@ARTICLE{Sorini_2026,
       author = {{Sorini}, Daniele and {Bose}, Sownak and {Denison}, Mathilda and {Dav{\'e}}, Romeel},
        title = "{Interpretable machine learning of halo gas density profiles: a sensitivity analysis of cosmological hydrodynamical simulations}",
      journal = {The Open Journal of Astrophysics},
         year = 2026,
        month = jun,
       volume = {9},
        pages = {63762},
          doi = {10.33232/001c.163762},
archivePrefix = {arXiv},
       eprint = {2512.09021},
 primaryClass = {astro-ph.GA},
       adsurl = {https://ui.adsabs.harvard.edu/abs/2026OJAp....963762S}
}

@ARTICLE{Sorini_2025,
       author = {{Sorini}, Daniele and {Bose}, Sownak and {Pakmor}, R{\"u}diger and {Hernquist}, Lars and {Springel}, Volker and {Hadzhiyska}, Boryana and {Hern{\'a}ndez-Aguayo}, C{\'e}sar and {Kannan}, Rahul},
        title = "{The impact of baryons on the internal structure of dark matter haloes from dwarf galaxies to superclusters in the redshift range 0 < z < 7}",
      journal = {\mnras},
         year = 2025,
        month = jan,
       volume = {536},
       number = {1},
        pages = {728-751},
          doi = {10.1093/mnras/stae2613},
archivePrefix = {arXiv},
       eprint = {2409.01758},
 primaryClass = {astro-ph.CO},
       adsurl = {https://ui.adsabs.harvard.edu/abs/2025MNRAS.536..728S}
}

@ARTICLE{2020SciPy-NMeth,
  author  = {Virtanen, Pauli and Gommers, Ralf and Oliphant, Travis E. and
            Haberland, Matt and Reddy, Tyler and Cournapeau, David and
            Burovski, Evgeni and Peterson, Pearu and Weckesser, Warren and
            Bright, Jonathan and {van der Walt}, St{\'e}fan J. and
            Brett, Matthew and Wilson, Joshua and Millman, K. Jarrod and
            Mayorov, Nikolay and Nelson, Andrew R. J. and Jones, Eric and
            Kern, Robert and Larson, Eric and Carey, C J and
            Polat, {\.I}lhan and Feng, Yu and Moore, Eric W. and
            {VanderPlas}, Jake and Laxalde, Denis and Perktold, Josef and
            Cimrman, Robert and Henriksen, Ian and Quintero, E. A. and
            Harris, Charles R. and Archibald, Anne M. and
            Ribeiro, Ant{\^o}nio H. and Pedregosa, Fabian and
            {van Mulbregt}, Paul and {SciPy 1.0 Contributors}},
  title   = {{{SciPy} 1.0: Fundamental Algorithms for Scientific
            Computing in Python}},
  journal = {Nature Methods},
  year    = {2020},
  volume  = {17},
  pages   = {261--272},
  adsurl  = {https://rdcu.be/b08Wh},
  doi     = {10.1038/s41592-019-0686-2},
}

@ARTICLE{Wayland2026,
       author = {{Wayland}, Amy and {Alonso}, David and {Reischke}, Robert},
        title = "{Probing baryonic feedback with fast radio bursts: joint analyses with cosmic shear and galaxy clustering}",
      journal = {\mnras},
         year = 2026,
        month = apr,
       volume = {547},
       number = {4},
          eid = {stag557},
        pages = {stag557},
          doi = {10.1093/mnras/stag557},
archivePrefix = {arXiv},
       eprint = {2602.12174},
 primaryClass = {astro-ph.CO},
       adsurl = {https://ui.adsabs.harvard.edu/abs/2026MNRAS.547ag557W}
}

@article{akino2022hsc,
  title={HSC-XXL: Baryon budget of the 136 XXL groups and clusters},
  author={Akino, Daichi and Eckert, Dominique and Okabe, Nobuhiro and Sereno, Mauro and Umetsu, Keiichi and Oguri, Masamune and Gastaldello, Fabio and Chiu, I-Non and Ettori, Stefano and Evrard, August E and others},
  journal={Publications of the Astronomical Society of Japan},
  volume={74},
  number={1},
  pages={175--208},
  year={2022},
  publisher={Oxford University Press}
}

@article{allen2002cosmological,
  title={Cosmological constraints from the X-ray gas mass fraction in relaxed lensing clusters observed with Chandra},
  author={Allen, SW and Schmidt, RW and Fabian, AC},
  journal={Monthly Notices of the Royal Astronomical Society},
  volume={334},
  number={2},
  pages={L11--L15},
  year={2002},
  publisher={Blackwell Science Ltd}
}

@article{simionescu2011baryons,
  title={Baryons at the edge of the X-ray--brightest galaxy cluster},
  author={Simionescu, Aurora and Allen, Steven W and Mantz, Adam and Werner, Norbert and Takei, Yoh and Morris, R Glenn and Fabian, Andrew C and Sanders, Jeremy S and Nulsen, Paul EJ and George, Matthew R and others},
  journal={Science},
  volume={331},
  number={6024},
  pages={1576--1579},
  year={2011},
  publisher={American Association for the Advancement of Science}
}

@article{mccarthy2016bahamas,
  title={The BAHAMAS project: calibrated hydrodynamical simulations for large-scale structure cosmology},
  author={McCarthy, Ian G and Schaye, Joop and Bird, Simeon and Le Brun, Amandine M C},
  journal={Monthly Notices of the Royal Astronomical Society},
  pages={stw2792},
  year={2016},
  publisher={Oxford University Press}
}

@article{dolag2017distribution,
  title={Distribution and evolution of metals in the magneticum simulations},
  author={Dolag, Klaus and Mevius, Emilio and Remus, Rhea-Silvia},
  journal={Galaxies},
  volume={5},
  number={3},
  pages={35},
  year={2017},
  publisher={MDPI}
}

@article{schaye2023flamingo,
  title={The FLAMINGO project: cosmological hydrodynamical simulations for large-scale structure and galaxy cluster surveys},
  author={Schaye, Joop and Kugel, Roi and Schaller, Matthieu and Helly, John C and Braspenning, Joey and Elbers, Willem and McCarthy, Ian G and Van Daalen, Marcel P and Vandenbroucke, Bert and Frenk, Carlos S and others},
  journal={Monthly Notices of the Royal Astronomical Society},
  volume={526},
  number={4},
  pages={4978--5020},
  year={2023},
  publisher={Oxford University Press}
}

@article{villaescusa2023camels,
  title={The CAMELS project: public data release},
  author={Villaescusa-Navarro, Francisco and Genel, Shy and Angl{\'e}s-Alc{\'a}zar, Daniel and Perez, Lucia A and Villanueva-Domingo, Pablo and Wadekar, Digvijay and Shao, Helen and Mohammad, Faizan G and Hassan, Sultan and Moser, Emily and others},
  journal={The Astrophysical Journal Supplement Series},
  volume={265},
  number={2},
  pages={54},
  year={2023},
  publisher={The American Astronomical Society}
}

@article{mead2016accurate,
  title={Accurate halo-model matter power spectra with dark energy, massive neutrinos and modified gravitational forces},
  author={Mead, Alexander J and Heymans, Catherine and Lombriser, Lucas and Peacock, John A and Steele, Olivia I and Winther, Hans A},
  journal={Monthly Notices of the Royal Astronomical Society},
  volume={459},
  number={2},
  pages={1468--1488},
  year={2016},
  publisher={Oxford University Press}
}

@article{schneider2025baryonification,
  title={Baryonification: an alternative to hydrodynamical simulations for cosmological studies},
  author={Schneider, Aurel and Kova{\v{c}}, Michael and Bucko, Jozef and Nicola, Andrina and Reischke, Robert and Giri, Sambit K and Teyssier, Romain and Tr{\"o}ster, Tilman and Refregier, Alexandre and Schaller, Matthieu and others},
  journal={Journal of Cosmology and Astroparticle Physics},
  volume={2025},
  number={12},
  pages={043},
  year={2025},
  publisher={IOP Publishing}
}

@article{arico2021simultaneous,
  title={Simultaneous modelling of matter power spectrum and bispectrum in the presence of baryons},
  author={Arico, Giovanni and Angulo, Raul E and Hern{\'a}ndez-Monteagudo, Carlos and Contreras, Sergio and Zennaro, Matteo},
  journal={Monthly Notices of the Royal Astronomical Society},
  volume={503},
  number={3},
  pages={3596--3609},
  year={2021},
  publisher={Oxford University Press}
}

@article{schneider2015new,
  title={A new method to quantify the effects of baryons on the matter power spectrum},
  author={Schneider, Aurel and Teyssier, Romain},
  journal={Journal of Cosmology and Astroparticle Physics},
  volume={2015},
  number={12},
  pages={049--049},
  year={2015}
}

@article{Shaw2010,
  title={IMPACT OF CLUSTER PHYSICS ON THE SUNYAEV--ZEL'DOVICH POWER SPECTRUM},
  author={Shaw, Laurie D and Nagai, Daisuke and Bhattacharya, Suman and Lau, Erwin T},
  journal={The Astrophysical Journal},
  volume={725},
  number={2},
  pages={1452--1465},
  year={2010},
  publisher={The American Astronomical Society}
}

@ARTICLE{Sunseri2026,
       author = {{Sunseri}, James and {Amon}, Alexandra and {Dunkley}, Jo and {Battaglia}, Nicholas and {Ferraro}, Simone and {Hadzhiyska}, Boryana and {Guachalla}, Bernardita Ried and {Schaan}, Emmanuel},
        title = "{Disentangling the halo: joint model for measurements of the kinetic Sunyaev─Zeldovich effect and galaxy─galaxy lensing}",
      journal = {\mnras},
         year = 2026,
        month = feb,
       volume = {546},
       number = {2},
          eid = {stag031},
        pages = {stag031},
          doi = {10.1093/mnras/stag031},
archivePrefix = {arXiv},
       eprint = {2505.20413},
 primaryClass = {astro-ph.CO},
       adsurl = {https://ui.adsabs.harvard.edu/abs/2026MNRAS.546ag031S}
}

@ARTICLE{Costello2026,
       author = {{Costello}, Emily E. and {McCarthy}, Ian G. and {Salcido}, Jaime and {Helly}, John C. and {McGibbon}, Robert J. and {Schaller}, Matthieu and {Schaye}, Joop},
        title = "{FLAMINGO: tracing the co-evolution of hot gas and black holes in galaxy groups and clusters}",
      journal = {\mnras},
         year = 2026,
        month = jul,
       volume = {549},
       number = {4},
          eid = {stag945},
        pages = {stag945},
          doi = {10.1093/mnras/stag945},
archivePrefix = {arXiv},
       eprint = {2510.17980},
 primaryClass = {astro-ph.CO},
       adsurl = {https://ui.adsabs.harvard.edu/abs/2026MNRAS.549ag945C}
}

@ARTICLE{Nagamine_2004,
       author = {{Nagamine}, Kentaro and {Loeb}, Abraham},
        title = "{Future evolution of the intergalactic medium in a universe dominated by a cosmological constant}",
      journal = {\na},
         year = 2004,
        month = oct,
       volume = {9},
       number = {8},
        pages = {573-583},
          doi = {10.1016/j.newast.2004.03.003},
archivePrefix = {arXiv},
       eprint = {astro-ph/0310505},
 primaryClass = {astro-ph},
       adsurl = {https://ui.adsabs.harvard.edu/abs/2004NewA....9..573N}
}

@ARTICLE{Stevens_2017,
       author = {{Stevens}, Adam R.~H. and {Lagos}, Claudia del P. and {Contreras}, Sergio and {Croton}, Darren J. and {Padilla}, Nelson D. and {Schaller}, Matthieu and {Schaye}, Joop and {Theuns}, Tom},
        title = "{How to get cool in the heat: comparing analytic models of hot, cold, and cooling gas in haloes and galaxies with EAGLE}",
      journal = {\mnras},
         year = 2017,
        month = may,
       volume = {467},
       number = {2},
        pages = {2066-2084},
          doi = {10.1093/mnras/stx243},
archivePrefix = {arXiv},
       eprint = {1608.04389},
 primaryClass = {astro-ph.GA},
       adsurl = {https://ui.adsabs.harvard.edu/abs/2017MNRAS.467.2066S}
}




\appendix

\section{Complete halo gas density solutions}
\label{app:solutions}

In this appendix, we extend the formalism of our closure radius model. While this broader theoretical framework was not tested in the main report, since the simpler formulations were shown to be accurate, we include it for the sake of completeness, with the potential for future application.

Previously, we only considered a power-law solution for the gas density profile far from the halo to derive $R_{\text{closure}}$, to get an exact solution when integrating $\rho_{\rm gas}$ out to $R_{\text{closure}}$. However the gas density profile, $\rho_{\rm gas} = Ar^{-\eta}$ is just a special case of the full solution of (\ref{eq:generalODE}) with the RHS equal to $\Lambda c^2 / 4 \pi G$ in this limit, which can be solved for exactly. This can be done by rewriting (\ref{eq:generalODE}) as a function of $Y = \rho_{\rm gas}^{n-1}\frac{d\rho_{\rm gas}}{dr}$ - the repeated term in square brackets since the product rule was not fully expanded. After some rearrangement, this becomes:

\begin{equation}
Y' + \frac{2}{r}Y = \frac{\Lambda c^2}{w_n (n+1)} = K .
\label{eq:farlimit}
\end{equation}

Remembering that we ignore the $\rho_{\rm gas}$ term on the RHS as we are only considering the limit $\rho_{\rm gas} \ll \rho_{\text{crit}}$ or $r \gg R_{200}$. Several constants have been collected into a new constant, $K$. (\ref{eq:farlimit}) is a 1st order ODE which can be solved by multiplying through by the integrating factor, $r^2$. Collecting the LHS into a derivative of a product and integrating yields:

\begin{equation}
r^2 Y = \frac{K}{3} r^3 + c_1 .
\label{eq:farlimitgeneralsolution}
\end{equation}

Where $c_1$ is a constant of integration. Then substituting the definition of $Y$ as a function of $\rho_{\rm gas}$ into (\ref{eq:farlimitgeneralsolution}) and rearranging leads to the following separable 1st order ODE:

\begin{equation}
\frac{d\rho_{\rm gas}}{dr} = \left(\frac{K}{3}r +c_1 r^{-2}\right) \rho_{\rm gas}^{1-n}.
\label{eq:farlimitgeneralsolution2}
\end{equation}

Which can be easily solved to give the exact density profile in this limit:

\begin{equation}
\frac{\rho_{\rm gas}^n}{n} = \frac{K}{6} r^2 -c_1 r^{-1} + c_2 .
\label{eq:farlimitdensity}
\end{equation}

Where $c_2$ is another constant of integration. From here, it is simple to see that the previous assumed power-law is just a special case of the full density profile where $c_1 = c_2 = 0$. Unfortunately, (\ref{eq:farlimitdensity}) cannot be integrated out to $R_{\text{closure}}$ to derive an analytical solution in full generality. Instead, (\ref{eq:farlimitdensity}) must be  simplified. This can be done through employing boundary conditions (BCs) on $\rho_{\rm gas}$. Since this solution is only be valid far from the halo, a condition on $\rho_{\rm gas}(r \rightarrow 0)$ is not suitable. One potential BC is requiring that $\rho_{\rm gas} \rightarrow \rho_{\text{IGM}}$ at and beyond the closure radius as we are only considering a single halo in a homogeneous universe. Another BC is requiring that $\frac{d\rho_{\rm gas}}{dr} < 0$, i.e. the gas density is decreasing in the range of interest. This is supported by hydrodynamical simulation data and intuition that the density should not increase the further one goes far away from the halo -- but as discussed in the main text, should not be taken as a hard requirement. Finally, another BC is that $Y' \rightarrow K$ far from the halo - evident from (\ref{eq:farlimit}). Investigations into the relative size of different terms around $R_{200}$ also hints towards $c_1 = 0$ otherwise that term would unrealistically dominate far from the halo. Therefore, for now, it seems sensible to set $c_1 = 0$ - but not $c_2$. 

Requiring then that $\rho_{\rm gas}(R_{\text{closure}}) = \rho_{\text{IGM}}$ and $c_1 = 0$ leads to:

\begin{equation}
\rho_{\text{IGM}} = \left(\frac{Kn R_{\text{closure}}^2}{6} + c_2 n\right)^{\frac{1}{n}} .
\label{eq:IGMdensity}
\end{equation}

$c_2$ can be found from (\ref{eq:IGMdensity}), allowing the density profile far from the halo to be written as:

\begin{equation}
\rho_{\rm gas}(r) = \left[ \frac{Kn}{6} (r^2 - R_{\text{closure}}^2) + \rho_{\text{IGM}}^n \right]^\frac{1}{n} .
\label{eq:farlimitfullsolution}
\end{equation}

This is the full gas density profile far from the halo in universes with $\Lambda > 0$ and could be integrated numerically to more accurately arrive at the closure radius than the model in (\ref{eq:generalintegral2}) and (\ref{eq:Asolution}). (\ref{eq:farlimitfullsolution}) can be differentiated and examined to see where $\frac{d \rho_{\rm gas}}{dr} < 0$ (BC) constraining that $n < -1$ which does not support the assumption that $n = 2/3$ made earlier. However, to understand the gas density near and just beyond the virial radius requires going back to (\ref{eq:generalODE}) and considering the other limit where $\rho_{\rm gas} \gg \Lambda c^2 / 4 \pi G$. This is also the full gas density profile in the EdS universe beyond $R_{200}$. In this limit, (\ref{eq:generalODE}) becomes:

\begin{equation}
\frac{-w_n (n+1) f_{\rm gas}}{4 \pi G} \left[\frac{2}{r} \rho_{\rm gas}^{n-1} \frac{d \rho_{\rm gas}}{dr} + \frac{d}{dr} \left(\rho_{\rm gas}^{n-1} \frac{d \rho_{\rm gas}}{dr}\right)\right] = \rho_{\rm gas}.
\label{eq:nearlimitODE}
\end{equation}

This cannot be easily solved completely unlike the other limit, however the simple power-law is a solution. Substituting this into (\ref{eq:nearlimitODE}) to work out the $\eta-n$ relation gives $\eta = 2/(1-m)$, where $n$ has been re-indexed to $m$ for this solution since it does not have to be the same $n$ as the solution far from the halo as the gas may be behaving differently at different ranges. $A$ can be similarly worked out, giving $\rho_{\rm gas}$ for $r - R_{200} \ll R_{200}$ (or everywhere in EdS) as $\rho_*(r)$:

\begin{equation}
\rho_*(r) = \left[ \frac{-2w_m (1-m) (3m+1) f_{\rm gas}}{4 \pi G (1+m)^2} \right]^{\frac{1}{-m-1}} r^\frac{2}{-m-1}.
\label{eq:nearlimitsolution}
\end{equation}

Similarly, requiring $\frac{d \rho_*}{dr} < 0$ leads to the condition: $-1/3 < m < 1$, which is closer to the $n = 2/3$ assumption than before, however still not in agreement. Finally, the two gas density profiles, (\ref{eq:farlimitfullsolution}) and (\ref{eq:nearlimitsolution}) can be used to integrate out to the closure radius. Assuming they are equal at some distance, $r*$, (\ref{eq:nearlimitsolution}) can be used to integrate between $R_{200}$ and $r*$, and (\ref{eq:farlimitfullsolution}) to integrate from $r*$ to $R_{\text{closure}}$, allowing one to calculate the closure radius for any $m$ and $n$. $r*$ can be found by equating $\rho$ (\ref{eq:farlimitfullsolution}) to $\rho_*$ (\ref{eq:nearlimitsolution}), but does not have a simple form.

The equations in this more general form of the model are complicated, making it unlikely that an analytical expression for the closure radius can be obtained. Therefore, numerical integration is now necessary. Without an exact analytical solution, it becomes challenging to deduce the influence of various variables on the closure radius without tedious numerical exploration of effects of their variations. This formalism is still in development and refinement for future publication. However, the simplified model outlined in Section~\ref{sec:newmodel} is adequate for fitting to simulation data and represents a significant advancement over previous work.


\bsp	
\label{lastpage}
\end{document}